\documentclass[12pt]{article}
\usepackage{amsfonts,amssymb,amsmath}
\usepackage{epsfig}
\usepackage{float}

\newcommand{\RR}{\mathbb R}
\newcommand{\CC}{\mathbb C}
\renewcommand{\Re}{\mathop{\mathrm{Re}}}
\renewcommand{\Im}{\mathop{\mathrm{Im}}}
\newcommand{\res}{\mathop{\mathrm{res}}}

\newtheorem{lemma}{Lemma}
\newtheorem{remark}{Remark}

\newcommand{\LL}{{\cal L}}
\newcommand{\bigzero}{\mbox{\Large $0$}}
\newcommand{\bigP}[1]{\mbox{\Large $P_{#1}$}}

\newcommand{\NN}{{\mathbb N}}

\newcommand{\ZZ}{{\mathbb Z}}

\newcommand{\beq}{\begin{equation}}
\newcommand{\eeq}{\end{equation}}
\newcommand{\ba}{\begin{array}}
\newcommand{\ea}{\end{array}}
\newcommand{\bea}{\begin{eqnarray}}
\newcommand{\eea}{\end{eqnarray}}
\newcommand{\eps}{{\epsilon}}

\usepackage{graphicx}
\usepackage[usenames,dvipsnames]{color}
\usepackage{transparent}

\DeclareMathAlphabet{\mathpzc}{OT1}{pzc}{m}{it}
\newcommand{\rr}{\mathpzc{R}}
\begin{document}

\begin{center}
{\bf The finite gap method and the analytic description of the exact rogue wave recurrence in the periodic NLS Cauchy problem. 1}  
\vskip 10pt
{\it P. G. Grinevich $^{1,3}$ and P. M. Santini $^{2,4}$}

\vskip 10pt

\vskip 10pt

{\it 
$^1$ L.D. Landau Institute for Theoretical Physics, pr. Akademika Semenova 1a, 
Chernogolovka, 142432, Russia, and \\
Lomonosov Moscow State University, Faculty of Mechanics and Mathematics, Russia, 119991, Moscow, GSP-1, 1 Leninskiye Gory, Main Building, and 
Moscow Institute of Physics and Technology, 9 Institutskiy per., Dolgoprudny, Moscow Region, 141700, Russia

\smallskip

$^2$ Dipartimento di Fisica, Universit\`a di Roma "La Sapienza", and \\
Istituto Nazionale di Fisica Nucleare, Sezione di Roma, 
Piazz.le Aldo Moro 2, I-00185 Roma, Italy}

\vskip 10pt

$^{3}$e-mail:  {\tt pgg@landau.ac.ru}\\
$^{4}$e-mail:  {\tt paolo.santini@roma1.infn.it}
\vskip 10pt

{\today}

\end{center}
\begin{abstract}
The focusing Nonlinear Schr\"odinger (NLS) equation is the simplest universal model describing the modulation instability (MI) of quasi monochromatic waves in weakly nonlinear media, considered the main physical mechanism for the appearence of rogue (anomalous) waves (RWs) in Nature. In this paper we study, using the finite gap method, the NLS Cauchy problem for periodic initial perturbations of the unstable background solution of NLS exciting just one of the unstable modes. We distinguish two cases. In the case in which only the corresponding unstable gap is theoretically open, the solution describes an exact deterministic alternate recurrence of linear and nonlinear stages of MI, and the nonlinear RW stages are described by the 1-breather Akhmediev solution, whose parameters, different at each RW appearence, are always given in terms of the initial data through elementary functions. If the number of unstable modes is $>1$, this uniform in $t$ dynamics is sensibly affected by perturbations due to numerics and/or real experiments, provoking $O(1)$ corrections to the result. In the second case in which more than one unstable gap is open, a detailed investigation of all these gaps is necessary to get a uniform in $t$ dynamics, and this study is postponed to a subsequent paper. It is however possible to obtain the elementary description of the first nonlinear stage of MI, given again by the Akhmediev 1-breather solution, and how perturbations due to numerics and/or real experiments can affect this result. Since the solution of the Cauchy problem is given in terms of different elementary functions in different time intervals, obviously matching in the corresponding overlapping regions, an alternative approach, based on matched asymptotic expansions, is suggested and presented in a separate paper in which the RW recurrence, in the case of a finite number of unstable modes and of a generic initial perturbation exciting all of them, is again described in term of elementary functions. 
\end{abstract}

\section{Introduction}

The self-focusing Nonlinear Schr\"odinger (NLS) equation 
\beq\label{NLS}
i u_t +u_{xx}+2 |u|^2 u=0, \ \ u=u(x,t)\in\CC
\eeq
is a universal model in the description of the propagation of a quasi monochromatic wave in a weakly nonlinear medium; in particular, it is relevant in deep water \cite{Zakharov}, in nonlinear optics \cite{Solli,Bortolozzo,PMContiADelRe}, in Langmuir waves in a plasma \cite{Sulem}, and in the theory of attracting Bose-Einstein condensates \cite{Bludov}. It is well-known that its elementary solution
\beq\label{background0}
a \exp (2i |a|^2t), \ \ a\in \CC \ \mbox{and constant,}
\eeq
describing Stokes waves \cite{Stokes} in a water wave context, a state of constant light intensity in nonlinear optics, and a state of constant boson density in a Bose-Einstein condensate, is unstable under the perturbation of waves with sufficiently large wave length \cite{Talanov,BF,Zakharov,ZakharovOstro,Taniuti,Salasnich}, and this modulation instability (MI) is considered as the main cause for the formation of rogue (anomalous, extreme, freak) waves (RWs) in Nature \cite{HendersonPeregrine,Dysthe,Osborne,KharifPeli1,KharifPeli2,Onorato2}. 

The integrable nature \cite{ZakharovShabat} of the NLS equation allows one to construct solutions corresponding to perturbations of the background by degenerating finite-gap solutions \cite{Its,BBEIM,Krichever2,Krichever3}, when the spectral curve becomes rational, or, more directly, using classical Darboux \cite{Matveev0,Ercolani}, Dressing \cite{ZakharovShabatdress,ZakharovMikha} techniques. Among these basic solutions, we mention the Peregrine soliton \cite{Peregrine}, rationally localized  in $x$ and $t$ over the background (\ref{background0}), the so-called Kuznetsov \cite{Kuznetsov} - Ma \cite{Ma} soliton, exponentially localized in space over the background and periodic in time; the so-called Akhmediev breather \cite{Akhmed1,Akhmed2}, periodic in $x$ and exponentially localized in time over the background (\ref{background0}). A more general one soliton solution over the background (\ref{background0}) can be found, f.i., in \cite{Its, ZakharovGelash1}, corresponding to a spectral parameter in general position. These solutions have also been generalized to the case of multi-soliton solutions, describing their nonlinear interaction, see, f.i., \cite{DGKMatv,Its,Hirota,Akhm6,ZakharovGelash2}. We remark that the Peregrine solitons are homoclinic, describing RWs appearing apparently from nowhere and desappearing in the future, while the multisoliton solution of Akhmediev type is almost homoclinic, returning to the original background up to a multiplicative phase factor. Generalizations of these solutions to the case of integrable multicomponent NLS equations have also been found \cite{BDegaCW,DegaLomb}. 

Concerning the NLS Cauchy problems in which the initial condition consists of a perturbation of the exact background (\ref{background0}), if such a perturbation is localized, then slowly modulated periodic oscillations described by the elliptic solution of (\ref{NLS}) play a relevant role in the longtime regime \cite{Biondini1,Biondini2}. If the initial perturbation is $x$-periodic, numerical experiments and qualitative considerations indicate that the solutions of (\ref{NLS}) exhibit instead time recurrence \cite{Yuen1,Yuen2,Yuen3,Akhmed3,Simaeys,Kuznetsov2}, as well as numerically induced chaos \cite{AblowHerbst,AblowSchobHerbst,AblowHHShober}, in which the almost homoclinic solutions of Akhmediev type seem to play a relevant role \cite{CaliniEMcShober,CaliniShober1,CaliniShober2}. There are reports of experiments in which the Peregrine and the Akhmediev solitons were observed \cite{CHA_observP,KFFMDGA_observP,Yuen3,Tulin}, but no analytic proof of their relevance, within generic Cauchy problems associated with NLS, has been given so far, to the best of our knowledge, prior to the present work. 

In this paper we apply the finite gap method \cite{Novikov,Its2,Krichever} to the solution of the NLS Cauchy problem on the segment $[0,L]$, with periodic boundary conditions, and we consider, as initial condition, a generic, smooth, periodic, zero average, small perturbation of the background solution (\ref{background0}) (or, better, of its simplified form 
\beq\label{background}
u_0(x,t)=e^{2it},
\eeq
obtained setting, without loss of generality, $a=1$, having used the scaling symmetry of NLS):
\beq\label{Cauchy}
\ba{l}
u(x,0)=1+\eps(x), \\ \eps(x+L)=\eps(x), \ \ \ \ 
||\eps(x)||_{\infty}=\eps \ll 1, \ \ \ \ \int\limits_{0}^L\eps(x)dx =0 .
\ea
\eeq
It is well-known that, in this Cauchy problem, the MI is due to the fact that, expanding the initial perturbation in Fourier components:
\beq\label{Fourier}
\eps(x)=\sum\limits_{j\ge 1}\left(c_j e^{i k_j x}+c_{-j} e^{-i k_j x}\right), \ \ k_j=\frac{2\pi}{L}j , \ \ |c_j |=O(\eps), 
\eeq
and defining $N\in\NN^+$ through the inequalities
\beq\label{def_N}
\frac{L}{\pi}-1<N<\frac{L}{\pi}, \ \ \pi < L,
\eeq 
the first $N$ modes $k_j,~k_{-j}=-k_j,~1\le j \le N$, are unstable, since they give rise to exponentially growing and decaying waves of amplitudes $O(\eps e^{\pm \sigma_j t})$, where the growing factors $\sigma_j$ are defined by
\beq\label{def_ampl}
\sigma_j=k_j\sqrt{4-k^2_j}, \ \ 1\le j \le N,
\eeq 
while the remaining modes give rise to oscillations of amplitude $O(\eps e^{\pm i \omega_j t})$, where 
\beq
\omega_j=k_j\sqrt{k^2_j -4}, \ \ j>N, 
\eeq 
and therefore are stable. We have in mind the following {\it qualitative recurrence scenario}, associated with the Cauchy problem (\ref{Cauchy}). 

The exponentially growing waves become $O(1)$ at times 
\beq
T_j=O({\sigma_j}^{-1}|\log~\eps|), \ \ \ 1\le j \le N ,
\eeq
when one enters the second asymptotic stage: the nonlinear stage of MI. In this second time interval one expects the generation of a transient, $O(1)$, coherent structure, described by a soliton - like solution of NLS over the unstable background (\ref{background}), the 
so-called RW. Such a RW will have an internal structure, due to the nonlinear interaction between the $N$ unstable modes, fully described by the integrable NLS theory. Due again to MI, this RW is expected to be destroyed in a finite time interval, and one enters the third asymptotic stage, characterized, like the first one, by the background plus an $O(\eps)$ perturbation, and described again by the NLS theory linearized around the background. This second linear stage is expected, due again to MI, to give rise to the formation of a second RW (the second nonlinear stage of MI). This procedure should iterate forever, in the integrable NLS model, giving rise to the generation of an infinite sequence of RWs. 

Therefore one is expected to be dealing with the following basic, deterministic open problems.\\ 
\vskip 2pt
For a given generic initial condition of the type (\ref{Cauchy}), how to predict: 
\begin{enumerate}
\item the ``generation time'' of the first RW; 
\item  the ``recurrence times'' measuring the time intervals between two consecutive RWs;
\item  the analytic form of this deterministic sequence of RWs.
\end{enumerate} 
\vskip 2pt

In this paper we concentrate on the situation in which the initial condition excites just one of the $N$ unstable modes, say the $n^{th}$ mode, $1\le n \le N$, and we discuss two cases: 
\begin{enumerate}
\item the case in which only the gap associated with the excited unstable mode is open; 
\item  the case in which a finite number $>1$ of gaps associated with unstable modes are open. 
\end{enumerate} 

In the first case, we show that the finite gap method ``theoretically'' provides a uniform in time, approximate solution of the above Cauchy problem in terms of Riemann $\theta$-functions, defined as an infinite sum of exponentials over a $g$-dimensional cartesian lattice, where the genus $g$ of the hyperelliptic Riemann surface is $2$. Since the 2 handles of this surface, generated by the $O(\eps)$ perturbation, are $O(\eps)$ thin, the constants appearing in the $\theta$ - function formula are all expressed in terms of the initial data via elementary functions, appearing as the coefficients of expansions involving $|\log~\eps |$ and powers of $\eps$. As a consequence, depending on the different time intervals we consider, corresponding to the linearized and the nonlinear stages of MI, the infinite sum takes its main contribution on different {\it finite} sets of lattice points; therefore {\it the solution of the Cauchy problem is ultimately described by different elementary functions in different asymptotic regions of the $t\ge 0$ semi-axis}. More precisely, we shall show that {\it the time evolution is described by an exact recurrence of linear and nonlinear stages of modulation instability; the nonlinear RW stages are characterized, at the leading $O(1)$, by a time sequence of RWs described by the 1-mode Akhmediev solution, whose $4$ free parameters change at each appearence, and are always expressed in terms of the initial data via elementary functions}. It turns out that, in this case, the dynamics turns out to be essentially periodic also in time (up to an overall $x$-translation of the profile and a multiplicative phase shift).  This periodicity becomes an ``exact recurrence'' when the number of open unstable gaps is $>1$, and this more general situation will be investigated in detail by the finite gap method in a subsequent paper. It important to remark, however, that this uniform in time theoretical result is very much affected by the above instabilities. Indeed, as we shall see in more detail in Section~\ref{sec:sec2.1}, perturbations associated with numerical experiments (due to non integrability properties of the numerical scheme and round-off error) and with real experiments (due to the fact that NLS is only a first approximation of reality and that a monochromatic initial profile is quasi monochromatic, in practice) open up all gaps, implying $O(1)$ corrections to the above evolution due to the unstable modes. Therefore the analytic and uniform in $t$ results of Section~\ref{sec:sec2.1} are presumably physically relevant only when $N=1$ ($\pi < L < 2\pi$). Analogously, {\it if we have $N>1$ unstable modes, a uniform in $t$ description of the RW recurrence in terms of elementary functions, not affected by the above instability problems, can be obtained when the initial perturbation excites ``democratically'' all the unstable modes}.

In the second case in which a finite ($>1$) number of unstable gaps is theoretically open, the application of the finite gap approach requires a detailed description of how all these unstable gaps open up, and this is presently under investigation. As we shall see in this paper, it is however possible, even without these informations, to have a theoretical prediction and analytic control on the first nonlinear stage of MI, the ``first appearance of the RW'', again described, at the leading order, by the 1-breather Akhmediev solution. Such a control can also be extended to the case in which perturbations like the round-off error play a relevant role (see section~\ref{sec:sec2.2}).

We remark that the first attempt to apply the finite gap method to solve the NLS Cauchy problem on the segment, for periodic perturbations of the background, was made in \cite{Tracy}; the fact that, in the $\theta$-function representation of the solution, different finite sets of lattice point are relevant in different time intervals was first observed there, but no connection was established between the initial data and the parameters of the $\theta$-function, and no description of the nonlinear stages of MI in terms of elementary functions was given.
\vskip 2pt

We also remark that, since, through our results, the solution of the Cauchy problem is described by different elementary functions in different time intervals of the positive time axis, and since these different representations obviously match in their overlapping time regions, these finite gap results naturally motivate the introduction of a matched asymptotic expansions (MAE) approach, presented in the papers \cite{GS2,GS3} and involving more elementary mathematical tools. The advantages of the finite gap approach are due to the fact that the $\theta$-function representation of the solution is uniform in time, and the analytic description of the nonlinear stages of MI (of the sequence of RWs) does not require any guess work. Such a guess work is instead needed in the MAE approach, when one has to select the proper nonlinear mode of NLS describing a certain nonlinear stage of MI and matching with the preceeding linear stage. In all the situations in which such a guess work is no problem, the MAE approach becomes very competitive, since it involves more elementary mathematics. For instance, a situation in which the MAE allows one to have a straightforward description of the RW recurrence is the case in which we have a finite number $N\ge 1$ of unstable modes and the generic initial perturbation excites all these modes.  In this case it is possible to show that {\it the exact RW recurrence is described analytically, at each appearance, by the $N$-breather solution of Akhmediev type, and the $3N+1$ free parameters of this solution, different at each appearance, are all expressed in terms of the initial perturbation through elementary functions} \cite{GS2,GS3}.  

We are also presently exploring the interesting case in which $N\gg 1$, and the situation in which the parameters appearing in the initial condition are random. This Cauchy problem has been investigated numerically in \cite{AgafontZakharov1}. At last, we mention that a nonlinear optics group of the Department of Physics of the University of Rome ``La Sapienza'' is presently testing, using the exact analytic results contained in this paper and in the paper \cite{GS2}, how well NLS describes RW recurrences in real nonlinear optics experiments.

The paper is organized as follows. In Section~\ref{sec:sec2} we summarize the main results of the paper. In Sections~\ref{sec:sec3} - \ref{sec:sec7} we show how to derive these results using the finite gap method.

\section{Results}
\label{sec:sec2}

In this section we summarize the main results obtained in this paper, introducing first the following convenient parameters associated with the $N$ unstable modes $1\le j \le N$:
\beq\label{def_alpha_beta_phi}
\ba{l}
\alpha_j=\overline{c_j}-e^{2i\phi_j}c_{-j}, \\ \beta_j=\overline{c_{-j}}-e^{-2i\phi_j}c_{j},\\
\phi_j=\arccos\left(\frac{\pi}{L}j \right)=\arccos\left(\frac{k_j}{2} \right),
\ea
\eeq
where $c_j, c_{-j}$ are the Fourier coefficients of the (unstable part of the) initial perturbation (see (\ref{Fourier})-(\ref{def_ampl})), implying that the wave numbers $k_j$ and the amplification factors $\sigma_j$ in (\ref{Fourier}),(\ref{def_ampl}) take the form
\beq\label{def_k1_sigma1}
k_j=2\cos(\phi_j), \ \ \sigma_j=2\sin(2\phi_j)
\eeq  
in terms of the angles $\phi_j$. 

We investigate the Cauchy problem on the segment $[0,L]$, with periodic boundary conditions, for the NLS equation (\ref{NLS}), under the hypothesis that the $O(\eps),~\eps\ll 1$ initial perturbation of the background (\ref{background}) excites just one of the $N$ unstable Fourier modes:
\beq\label{Cauchy1a}
\eps(x)=c_n e^{i k_n x}+c_{-n}e^{-i k_n x}, \ \ 1\le n \le N, \ \ |c_n|,|c_{-n}|=O(\eps),
\eeq
where $k_n$ is defined in (\ref{Fourier}), and where $N\in\NN^+$, defined by the inequalities (\ref{def_N}), is taken to be finite ($1/N\gg \eps$). Then only the double points $\lambda^{\pm}_{j}=\pm\sqrt{(\pi j/L)^2 -1},~j=ns,~s\in\NN^+$ of the spectrum of the associated Zakharov - Shabat spectral problem (\ref{eq:lp-x}), corresponding to the background $u_0(x,0)=1$, theoretically split into a pair of square root branch points, whose difference $\delta^{\pm}_j$ is $O(\eps)$ for the excited unstable mode $n$:
\beq\label{delta}
\delta^+_n=\frac{\sqrt{\alpha_n\beta_n}}{\lambda^+_n}+O(\eps^2), \ \ \lambda^+_n=i\sqrt{1-(\pi n/L)^2}, \ \ 
\delta^-_n=\overline{\delta^+_n}, 
\eeq 
and is $O(\eps^s)$ for the remaining infinitely many modes: $|\delta^{\pm}_{ns} |= O(\eps^s)$ (the simple but technical proof of this order estimate is postponed to a subsequent paper). Only the first $\tilde s$ of these modes are unstable, with $\tilde s$ defined by the inequalities
\beq
\frac{L}{\pi n}-1< \tilde s < \frac{L}{\pi n}.
\eeq
Then, if we delete from the spectrum the infinitely many gaps associated with the stable modes, reducing the problem to a finite gap theory, we make an $O(\eps^2)$ error or smaller, and therefore we do it. The corresponding Riemann surface is a genus $2\tilde s$ hyperelliptic curve, whose $2$ handles corresponding to $\lambda^{\pm}_n$ are $O(\eps)$ thin, while the remaining $2(\tilde s-1)$ handles are thinner and thinner. 

In this paper we consider two subcases: the case in which $\tilde s=1$, in which only the 2 gaps associated with the excited harmonic are ``theoretically'' open, and the case in which a finite number ($>1$) of unstable gaps are ``theoretically''  opened.

\subsection{The genus $2$ approximation and the uniform in $t$ description of the exact RW recurrence}
\label{sec:sec2.1}

If the initial condition excites the $n^{th}$ unstable mode as in (\ref{Cauchy1a}), and if $L/2\pi <n <L/\pi$, then the number of ``theoretically'' open gaps associated with unstable modes is $\tilde s=1$, and the dynamics is ``theoretically'' well described by the genus 2 approximation, in which only the 2 gaps associated with the excited harmonic are open. We say ``theoretically'', because of the intrinsic instabilities of the problem. Indeed: a) in any numerical experiment one uses non integrable numerical schemes approximating NLS; in addition, even if the numerical scheme were integrable, round off errors are not avoidable. All these facts cause the opening of basically all gaps and, due to the instability, no matter how small are the gaps associated with the unstable modes, they will cause $O(1)$ effects during the evolution (see also \cite{AblowHerbst,AblowSchobHerbst,CaliniEMcShober}). b) In physical phenomena involving weakly nonlinear quasi monochromatic waves, NLS is a first approximation of the reality, and higher order corrections have the effect of opening again all gaps, with $O(1)$ effects during the evolution caused by the unstable ones. At last, in a real experiment, the monochromatic initial perturbation (\ref{Cauchy1a}) is replaced by a quasi-monochromatic approximation of it, often with random coefficients, opening again all the gaps associated with the unstable modes, with $O(1)$ effects during on the evolution.  

All these considerations imply that the uniformly valid in time formulas of this section are expected to be in good agreement with numerical simulations and/or with real experiments only when we have just one unstable mode: $ N=1 \ \ \Leftrightarrow \ \ \pi < L <2\pi$. However, as we shall see in section~\ref{sec:sec2.2}, the formulas of this section should give an accurate description of the first appearance of the RW also in the case of a finite number of unstable modes.

The above considerations generalize to the case in which the initial perturbation excites more than one unstable mode. If such a perturbation is generic, exciting democratically all the $N$ unstable modes, the genus $2N$ approximation allows one to obtain a uniform in time description of the evolution, relevant also in numerical and real experiments, and the $N$-breather solution of Akhmediev type will play a basic role in such a description. If only a subset of the unstable modes are excited, then instabilities would limit the relevance of the theoretical results in the description of numerical and real experiments; however they should still provide a good analytic description of the first appearance of the RW. These generalizations will be discussed in detail in a forthcoming paper.        

With the precautions indicated above, we consider the initial perturbation (\ref{Cauchy1a}), with $L/2\pi <n <L/\pi$. 
In this case the approximate finite gap Riemann surface is a genus $2$ hyperelliptic curve whose $2$ handles, corresponding to the excited mode, are $O(\eps)$ thin. We recall that, in this case, the $\theta$-function is an infinite sum of exponentials over a 2-dimensional cartesian lattice \cite{Its2} (see formula (\ref{eq:pot1})). This representation is uniform in time but, since the parameters involved are expanded in $|\log\eps|$ and in powers of $\eps$, in different time intervals only different finite sets of lattice points give a non negligeable contribution. Therefore {\it the solution of the Cauchy problem is given in terms of different elementary functions in different time intervals} in the following way.   
 
 If $0\le t\le O(1)$, the $\theta$-functions appearing in the finite gap representation of the solution acquire an $O(1)$ contribution from the lattice point $(0,0)$ and $O(\eps)$ contributions from the lattice points $(-1,0)$,$(1,0)$,$(0,1)$,$(0,-1)$, while the contributions of the other lattice points are smaller and can be neglected. Correspondingly: 
\beq\label{reg_lin1}
\ba{l}
u(x,t)=e^{2it}\Big\{1+\frac{2}{\sigma_n}\Big[|\alpha_n |\cos\Big(k_n  (x-X^+_n)\Big)e^{\sigma_n t+i\phi_n}+ \\
|\beta_n |\cos\Big(k_n (x-X^-_n)\Big)e^{-\sigma_n t-i\phi_n}  \Big] \Big\}+O(\eps^2|\log\eps |),
\ea
\eeq
where $\alpha_n ,\beta_n ,\phi_n, k_n,\sigma_n$ are defined in (\ref{def_alpha_beta_phi}), (\ref{def_k1_sigma1}), 
and $X^{\pm}_n$, defined as 
\beq\label{def_X}
X^{+}_n=\frac{\arg(\alpha_n)-\phi_n+\pi/2}{k_n}, \ \ X^{-}_=\frac{-\arg(\beta_n)-\phi_n+\pi/2}{k_n},
\eeq   
are the positions of the maxima of the sinusoidal wave decomposition of the growing and decaying unstable modes.

This explicit expression of the solution coincides, as it has to be, with the solution of the NLS equation linearized about the background, and satisfying the initial condition. This is the first linear stage of MI, in which: \\
{\it the unstable part of the initial datum splits into exponentially growing and decaying waves, respectively the $\alpha_n$- and $\beta_n$-waves, each one carrying half of the information encoded into the unstable part of the initial datum}. 

\vskip 10pt

If $t=O(\sigma_n^{-1}|\log\eps |)$, the balance changes and the $\theta$ function is $O(1)$ at the four points $(0,0)$, $(0,1)$, $(-1,0)$, $(-1,1)$, while the contributions of the other lattice points are smaller and can be neglected; we are in the first nonlinear stage of MI (corresponding to the first appearence of the RW). More precisely, if $|t-T_n(|\alpha_n |)|\le O(1)$, where
\beq
T_n(\zeta)=\frac{1}{\sigma_n}\log\left(\frac{(\sigma_n)^2}{2\zeta} \right) ,
\eeq
then
\beq\label{reg_RW1}
u(x,t)=A(x,t;\phi_n,X^+_n,T_n(|\alpha_n |),2\phi_n)+O(\eps),
\eeq
where
\beq\label{Akhm1}
\ba{l}
A(x,t;\phi,X,T,\rho)\equiv e^{2it+i\rho}\frac{\cosh[\Sigma(\phi) (t-T)+2i\phi ]+\sin(\phi) \cos[K(\phi)(x-X)]}{\cosh[\Sigma(\phi) (t-T)]-\sin(\phi) \cos[K(\phi)(x-X)]}, \\
K(\phi)=2\cos\phi, \ \ \Sigma(\phi)=2\sin(2\phi)
\ea
\eeq
is the well-known Akhmediev 1-breather \cite{Akhmed1,Akhmed2}, exact solution of NLS for all values of the 4 real parameters $\phi,X,T,\rho$.

It follows that {\it the first RW appears in the time interval $|t-T_n(|\alpha_n |)|\le O(1)$ and is described by the the Akhmediev 1-breather, whose parameters are expressed in terms of the initial data through elementary functions}. Such a RW, appearing about the logarithmically large time  $T_n(|\alpha_n|)=O(\sigma^{-1}_n|\log\eps |)$, is exponentially localized in an $O(1)$ time interval over the background $u_0$, changing it by the multiplicative phase factor $e^{4 i\phi_n}$, since
\beq
A(x,t;\phi_n,X^+_n,T_n(|\alpha_n |),2\phi_n)\to e^{2it+i(2\phi_n\pm 2\phi_n)}, \ \ \mbox{as} \ \ t\to\pm\infty .
\eeq
We remark that the modulus of the first RW (\ref{first}) has its maxima at $t=T_n(|\alpha_n|)$ in the $n$ positions 
\beq\label{position_max}
X^+_n+\frac{L}{n}j, \ \ 0\le j\le n-1, \ \ \mbox{mod }L , 
\eeq
and the value of such maxima is
\beq\label{maxima_nN}
M(n,L)=1+2\sin\phi_n < 1+\sqrt{3} \sim 2.732 .
\eeq 
This upper bound, $2.732$ times the background amplitude, is consequence of the formula  
\beq
\sin\phi_n=\sqrt{1-(\pi n/L)^2}, \ \ \ \frac{L}{2\pi} <n <\frac{L}{\pi} ,
\eeq
and is obtained when $n$ is close to its lower bound $L/2\pi$. We also notice that the position $x=X^+_n$ of the maximum of the RW coincides with the position of the maximum of the growing sinusoidal wave of the linearized theory; this is due to the absence of nonlinear interactions with other unstable modes. We finally remark that, in the first appearance, the RW contains informations, at the leading order, only on half of the initial wave, the half associated with the $\alpha_n$-wave. 

\begin{figure}[H]
\centering
\includegraphics[width=14cm]{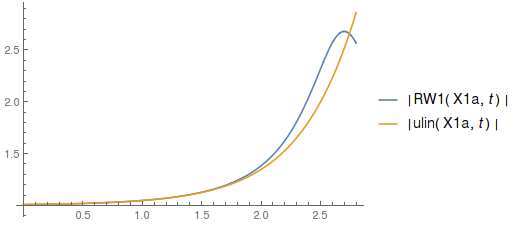} \\
\caption{\label{fig:fig1} Plotting, as function of $t$, of the moduli of the exponentially growing linearized solution (\ref{reg_lin1}) and of the first RW (\ref{reg_RW1}), for $L=5.78$ ($N=n=1$), evaluated at the maximum $x=X^+_1$, for $\alpha_1=0.011+0.005 i,~\beta_1=0.007+0.010 i$. They perfectly match in the overlapping region.}
\end{figure}

It is easy to verify that the two representations (\ref{reg_lin1}) and  (\ref{reg_RW1}) of the solution, valid respectively in the time intervals $0\le t \le O(1)$ and $|t-T_n(|\alpha_n |)|\le O(1)$, have the same behavior
\beq\label{overlapping1}
u(x,t)\sim e^{2it}\left(1+\frac{|\alpha_n |}{\sin 2\phi_n}e^{\sigma_n t+i\phi_n}\cos[k_n (x-X^+_n)]\right),
\eeq
in the intermediate region $O(1)\ll t \ll T_n(|\alpha_n |)$; therefore they match succesfully (see Fig.~\ref{fig:fig1}).

\vskip 5pt

The next balance takes place when $t=O(2(\sigma_n)^{-1}|\log\eps |)$, the second stage of linear MI, and the leading terms are obtained from the five points $(-1,1)$, $(0,1)$, $(-1,0)$, $(-1,2)$, $(-2,1)$. To be more precise, let $|t-T_p|\le O(1)$,
where
\beq\label{def_period}
T_p=T_n(|\alpha_n |)+T_n(|\beta_n |)=\frac{2}{\sigma_n}\log\left( \frac{\sigma^2_n}{2\sqrt{|\alpha_n\beta_n |}} \right)=
O(2 \sigma^{-1}_n |\log\eps |);
\eeq
then
\beq\label{reg_lin2}
\ba{l}
u(x,t)=e^{2it+4i\phi_n}\Big\{1+\frac{2}{\sigma_n}\Big[|\alpha_n|\cos\Big(k_n  (x-X^+_n-\Delta_x)\Big)e^{\sigma_n (t-T_p)+i\phi_n}+ \\
|\beta_n|\cos\Big(k_n (x-X^-_n-\Delta_x)\Big)e^{-\sigma_n (t-T_p)-i\phi_n}  \Big] \Big\}+O(\eps^2|\log\eps |),
\ea
\eeq
where
\beq
\Delta_x=X^{+}_n-X^{-}_n=\frac{\arg(\alpha_n\beta_n)}{k_n}.
\eeq
We observe that, evaluating (\ref{reg_lin2}) at $t=T_p$, one obtains
\beq
u(x,T_p)=e^{2iT_p+4i\phi_n}u(x-\Delta_x,0)+O(\eps^2|\log\eps |),
\eeq
implying the general periodicity formula
\beq
u(x,t+T_p)=e^{2iT_p+4i\phi_n}u(x-\Delta_x,t)+O(\eps^2|\log\eps |).
\eeq
This periodicity property is a direct consequence of the periodicity property of the $\theta$-function (see section~\ref{sec:sec7}). Therefore we conclude that, {\it whithin $O(\eps^2|\log\eps |)$ corrections, the solution of this Cauchy problem is also periodic in $t$, with period $T_p$, up to the multiplicative phase factor $exp(2iT_p+4i\phi_n)$ and up to the global $x$-translation of the quantity $\Delta_x$}.

Also in this case, it is easy to verify that the two representations (\ref{reg_RW1}) and (\ref{reg_lin2}) of the solution, defined respectively in the $t$ intervals $|t-T_n(|\alpha_n |)|\le O(1)$ and $|t-T_p|\le O(1)$, match successfully in the intermediate region $T_n(|\alpha_n |)\ll t \ll T_p$, as it has to be. 

The time periodicity allows one to infer that {\it the above Cauchy problem leads to an exact recurrence of RWs (of the nonlinear stages of MI), alternating with an exact recurrence of linear stages of MI}. To be precise, we have the following result.

\begin{figure}[H]
\centering
\includegraphics[width=14cm]{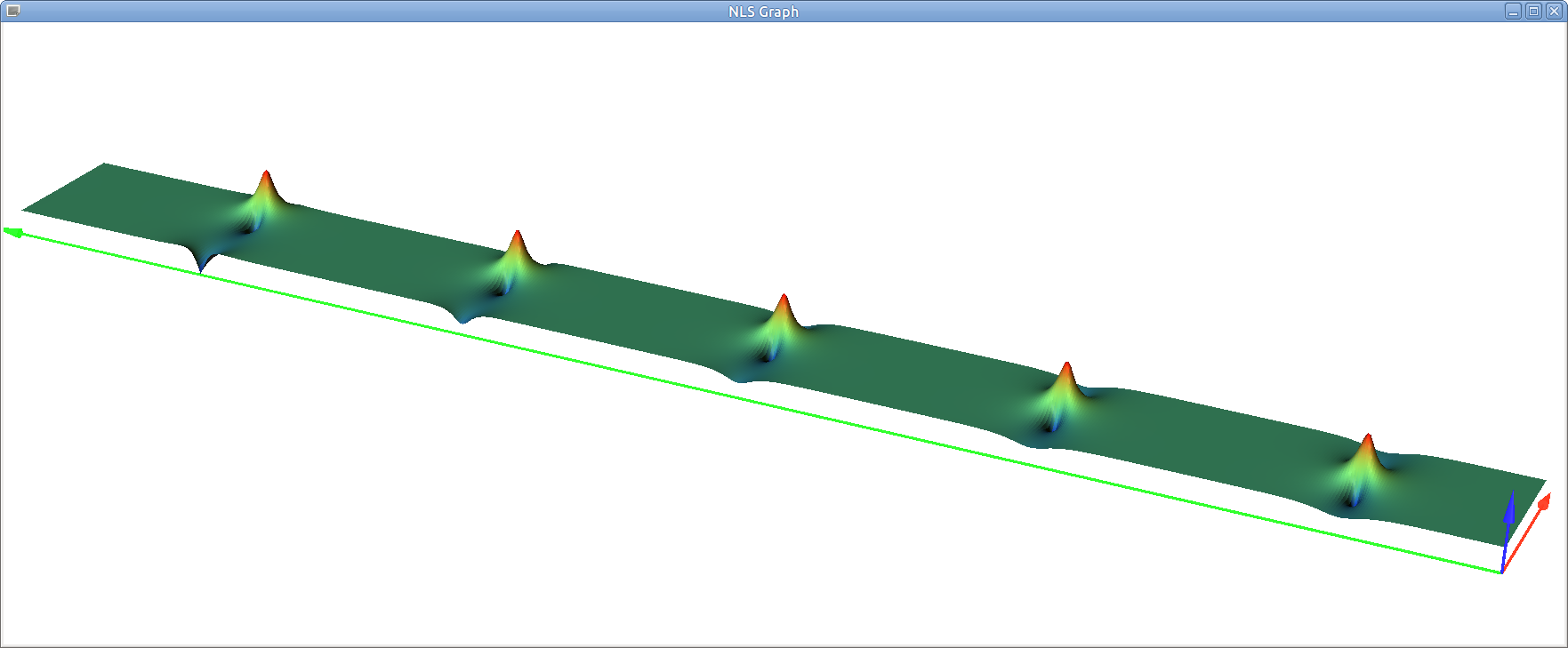} 
\caption{\label{fig:fig2} The 3D plotting of $|u(x,t)|$ describing the RW sequence, obtained through the numerical integration of NLS via the Split Step Fourier Method (SSFM) \cite{SSFM1,SSFM2}. Here $L=6$ ($N=n=1$), with $c_1 = \eps/2,~c_{-1} = \eps (0.3-0.4 i)/2,  ~\epsilon = 10^{-4}$, and the short axis is the $x$-axis, with $x\in [-L/2,L/2]$. The numerical output is in perfect agreement with the theoretical predictions.}
\end{figure}

\begin{figure}[H]
\centering
\includegraphics[width=8cm]{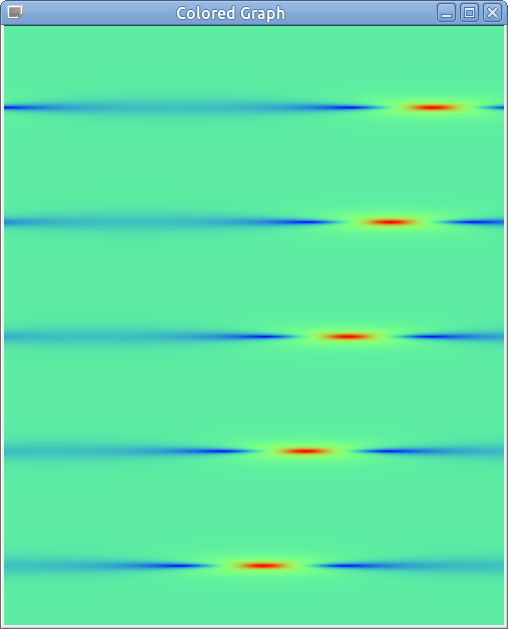}
\caption{\label{fig:fig3} The color level plotting for the numerical experiment of Fig.~\ref{fig:fig2}, in which the periodicity properties of the dynamics are evident.}
\end{figure}

{\it The Cauchy problem (\ref{Cauchy1a}) gives rise to an infinite sequence of RWs, and the $m^{th}$ RW of the sequence ($m\ge 1$) is described, in the time interval $|t-T_n(|\alpha_n|)-(m-1)T_p |\le O(1)$, by the analytic deterministic formula:
\beq\label{RWn_1UM_a}
\ba{l}
u(x,t)=A\Big(x,t;\phi_n,x_n^{(m)},t_n^{(m)},\rho_n^{(m)} \Big)+O(\eps), \ \ m\ge 1 ,
\ea
\eeq
where
\beq\label{parameters_1n_a}
\ba{l}
x_n^{(m)}=X^+_n+(m-1)\Delta^{(x)}, \ \ t_n^{(m)}=T_1(|\alpha_n|)+(m-1)T_p, \\ \rho_n^{(m)}=2\phi_n+(m-1)4\phi_n ,
\ea
\eeq
in terms of the initial data (see Figures~\ref{fig:fig2} and \ref{fig:fig3}). Apart from the first RW appearance, in which the RW contains information only on half of the initial data (the one encoded in the parameter $\alpha_1$), in all the subsequent appearances the RW contains, at the leading order, informations on the full unstable part of the initial datum, encoded in both parameters $\alpha_n$ and $\beta_n$}.

\noindent
Two remarks are important at this point, in addition to the considerations on the instabilities we made at the beginning of this section. 
\vskip 5pt
\noindent
a) If the initial condition (\ref{Cauchy}),(\ref{Cauchy1a}) is replaced by a more general initial condition in which we excite also all the stable modes, then (\ref{reg_lin1}) is replaced by a formula containing also the infinitely many $O(\eps)$ oscillations corresponding to the stable modes. But the behavior of the solution in the overlapping region $1\ll |t|\ll O(\sigma^{-1}_n|\log\eps|)$ is still given by equation (\ref{overlapping1}) and the matching at $O(1)$ is not affected. Therefore the sequence of RWs is still described by equations (\ref{RWn_1UM_a}), (\ref{parameters_1n_a}), and the differences between the two Cauchy problems are hidden in the $O(\eps)$ corrections. {\it As far as the $O(1)$ RW recurrence is concerned, only the part of the initial perturbation $\eps(x)$ exciting the unstable mode is relevant}.
\vskip 5pt
\noindent
b) The above results are valid up to $O(\eps^2 |\log\eps |)$. It means that, in principle, the above RW recurrence formulae may not give a correct description for large times of $O((\eps|\log\eps|)^{-1})$; but since $O((\eps|\log\eps|)^{-1})$ is much larger than the recurrence time $O(|\log\eps|)$, it follows that the above formulae should give an accurate description of the RW recurrence for many consecutive appearances of the RWs.

\subsection{The case of a finite number of open unstable gaps and the first appearance of the RW}
\label{sec:sec2.2}

If the initial perturbation (\ref{Cauchy1a}) excites just the $n^{th}$ unstable mode and the number $\tilde s$ of open unstable gaps is greater than $1$, then a uniform in time description of the RW recurrence requires a detailed study of how the unstable gaps open up due to this initial perturbation, and this study is postponed to a subsequent paper. However it is well-known that only the gaps corresponding to the resonant points $\pm\lambda_{ns}$, $s\in\ZZ^+$ are opened. Moreover it is possible to prove that the length of the corresponding gap is of order $\eps^s$. From the explicit formula 
$\sigma_n=4n\frac{\pi}{L}\sqrt{1-\left(\frac{\pi n}{L}\right)^2}$ it follows that $\sigma_{ns}< s~\sigma_n$. Consequently, the generation time $T_{ns}$ for the mode $ns$, of order $\sigma_{ns}^{-1}|\log(\eps^s)|$, is greater than the generation time $T_n$ for the leading excited mode:
\begin{equation}
\label{eq:theor_est}
T_{ns}-T_n \sim\left[\frac{s}{\sigma_{sn}}-\frac{1}{\sigma_{n}} \right] |\log(\eps)|\gg1 .
\end{equation}
Since the RW is exponentially localized in an $O(1)$ time interval, the other unstable modes $ns$ do not sensibly affect the first nonlinear stage of MI, described by the excited mode: \\ 
{\it the first nonlinear stage of MI (the first appearance of the RW), takes place in the time interval $|t-T_n(|\alpha_n|)|\le O(1)$, and is described again by the Akhmediev 1-breather solution}
\beq\label{first}
u(x,t)\sim A(x,t;\phi_n,X^+_n,T_n(|\alpha_n|),2\phi_n),
\eeq
where $T_n(|\alpha_n|)$ is defined in (\ref{def_period}) (see Figures~\ref{fig:fig4} and \ref{fig:fig5}). According to (\ref{eq:theor_est}), it is a good approximation of the first appearance only if $\eps$ is sufficiently small. We remark that the modulus of the first RW (\ref{first}) has its maxima at $t=T_n(|\alpha_n|)$ in the $n$ positions 
\beq\label{position_max}
X^+_n+\frac{L}{n}j, \ \ 0\le j\le n-1, \ \ \mbox{mod }L , 
\eeq
and the value of such maxima is
\beq\label{maxima_nN}
M(n,L)=1+2\sin\phi_n .
\eeq 

At this point it is important to observe that, when we make a numerical experiment, we have corrections to the gap size due to the numerical noise.  
For standard double-precision calculations, the level of numerical noise can be estimated by the round-off error, of order 
$10^{-18}-10^{-17}$. Therefore, the theoretical estimate (\ref{eq:theor_est}), implying the approximation (\ref{first}), works well only if the theoretical generation time $T_n$ is essentially smaller than the generation time $T^{\tiny{num}}$ due to numerical noise, of the order 
$T^{\tiny{num}}\sim \frac{18\log(10)}{\sigma_j}$, where $j$ is the most unstable mode corresponding to the highest $\sigma_j$.
 
\begin{figure}[H]
\centering
\includegraphics[width=9cm]{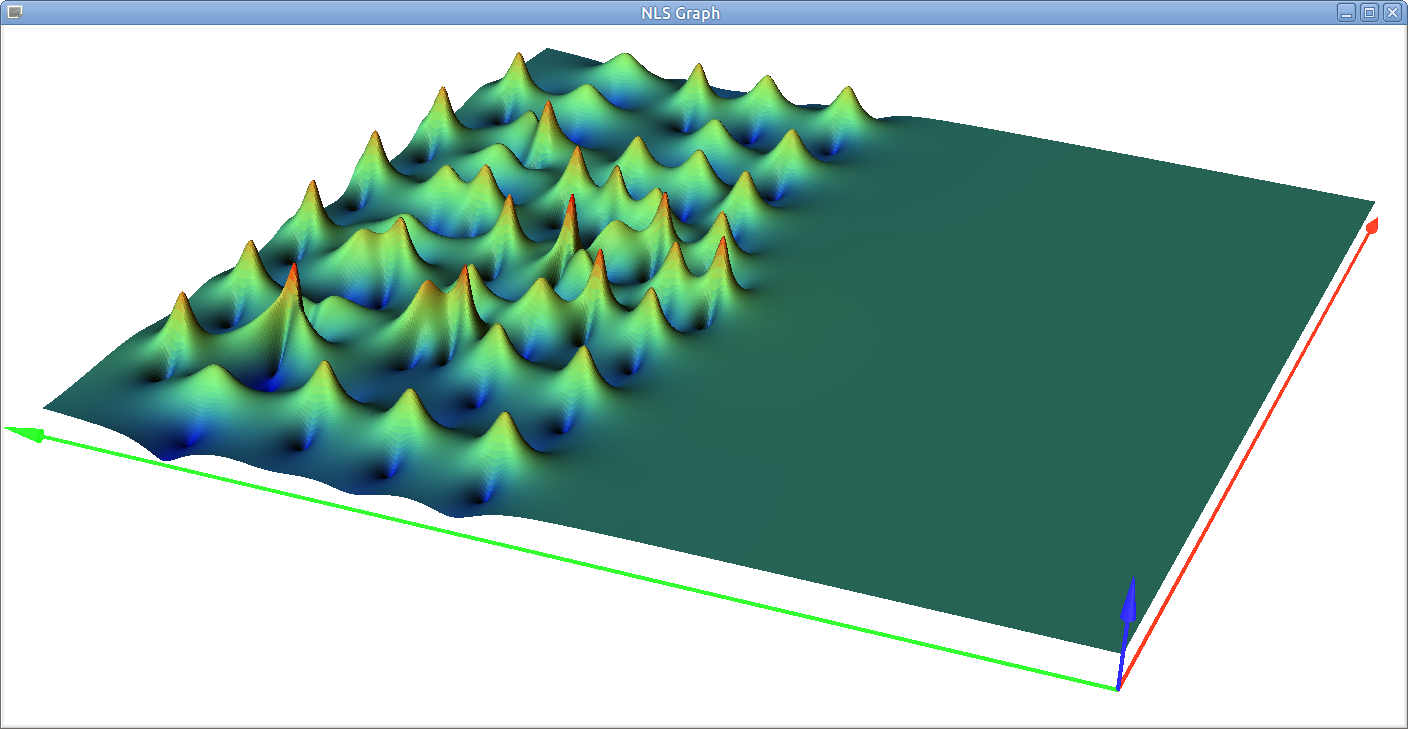}
\includegraphics[width=3.8cm]{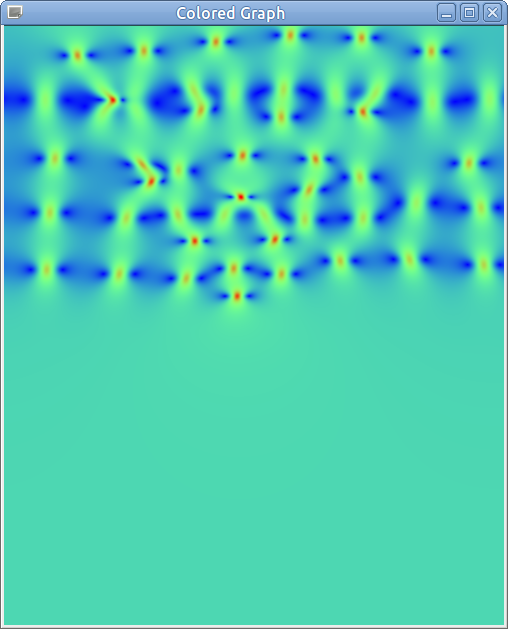}
\caption{\label{fig:fig7} The 3D plotting and the level plotting of $|u(x,t)|$ obtained through the numerical integration of NLS via the SSFM, for $L=36$ ($N=11$ unstable modes), in the case in which only the first mode $k_1$ is excited ($n=1$), with $c_1 = \eps/2,~c_{-1} = \eps (0.3-0.4 i)/2,  ~\epsilon = 10^{-3}$. The first and the $8^{th}$ mode appear almost at the same time due to round off error, as predicted through simple qualitative arguments.}
\end{figure}

For example, if we excite the first harmonic ($n=1$) for $L=36$ (corresponding to $N=11$ unstable modes, and corresponding to the case in which the most unstable mode is 
the 8-th, with $\sigma_8\sim 1.999$), then the choice $\eps=10^{-3}$ first implies that the generation time due to numerical noise is essentially smaller than the theoretical generation time $T_8$: $T^{\tiny{num}}/T_8\sim 18/24$; in addition:  
\beq\label{est1}
\frac{T_1}{T^{\tiny{num}}}\sim \frac{3}{\sigma_1}\left(\frac{18}{\sigma_8}\right)^{-1}\sim\frac{8.62}{9}.
\eeq
Consequently one expects that, in a numerical experiment, the first and the 8-th mode appear almost simultaneously due to numerical noise, and this prediction is confirmed by the experiment (see Fig.~\ref{fig:fig7}). Therefore we see an essential 
correction to the first appearance time, and formula (\ref{first}) does not describe the first appearance well. 

\begin{figure}[H]
\centering
\includegraphics[width=9cm]{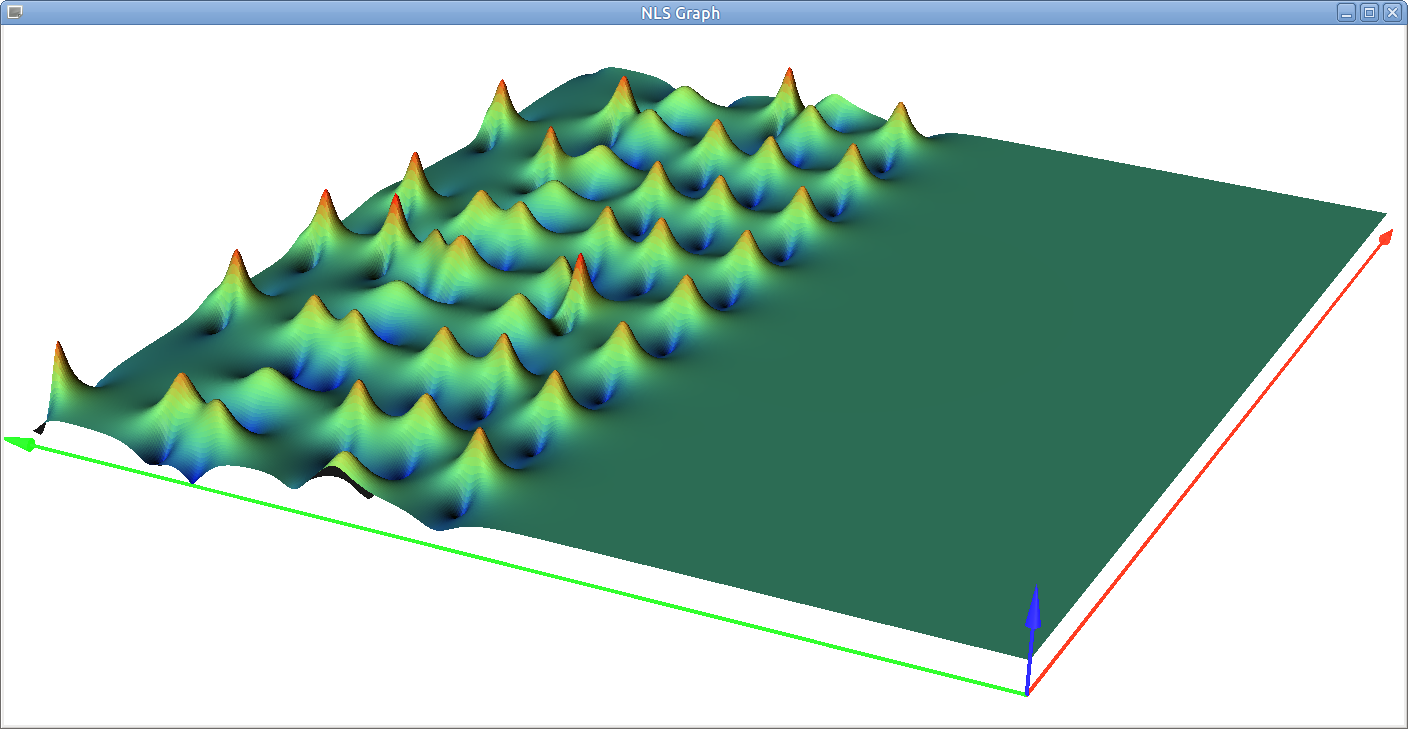}
\includegraphics[width=3.8cm]{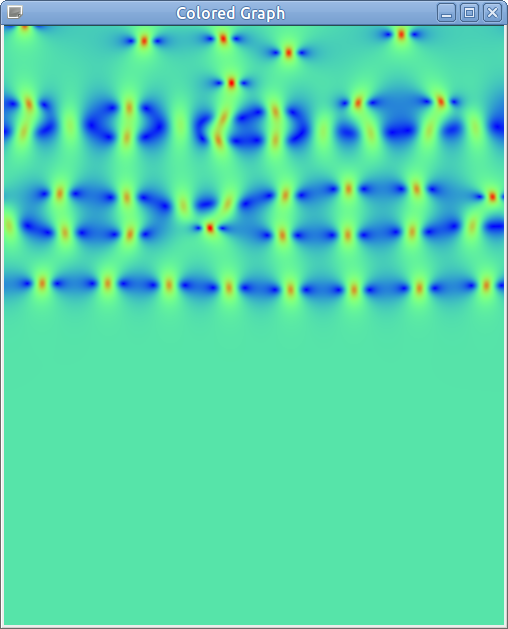}
\caption{\label{fig:fig8} The 3D plotting and the level plotting of $|u(x,t)|$ obtained through the numerical integration of NLS via the SSFM, for $L=36$ ($N=11$ unstable modes), in the case in which only the first mode $k_1$ is excited ($n=1$), with $c_1 = \eps/2,~c_{-1} = \eps (0.3-0.4 i)/2,  ~\epsilon = 10^{-4}$. The $8^{th}$ mode appears before the first, due to round off error, as predicted through simple qualitative arguments.}
\end{figure}

\begin{figure}[H]
\centering
\includegraphics[width=9cm]{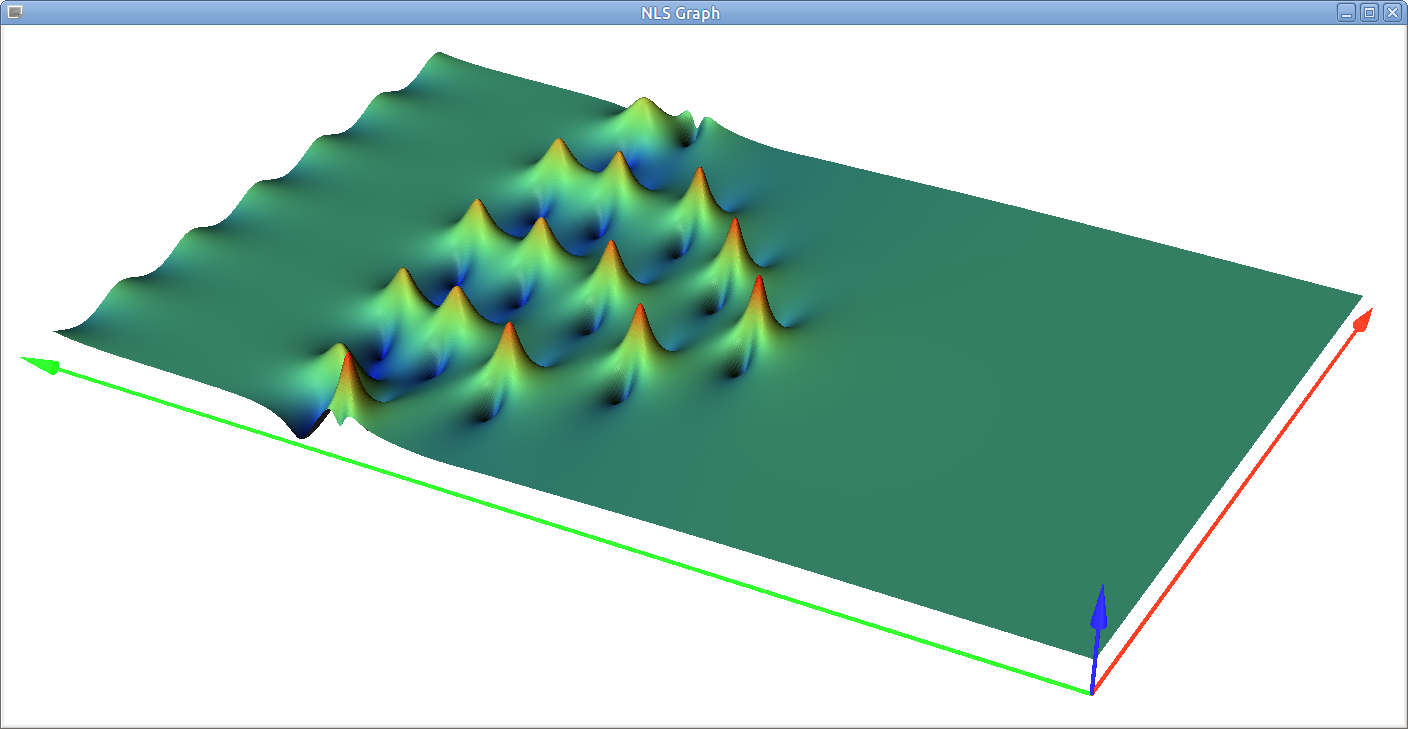} 
\includegraphics[width=3.8cm]{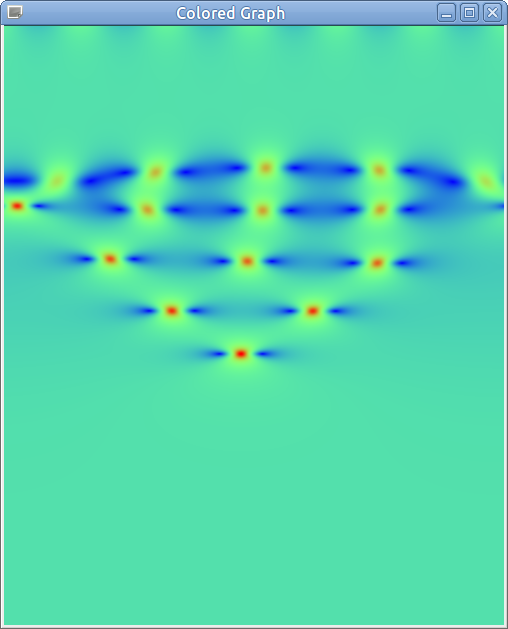}
\caption{\label{fig:fig4}
The 3D plotting and level plotting of $|u(x,t)|$ obtained through the numerical integration of NLS via the SSFM, for $L=20$ ($N=6$ unstable modes), in the case in which only the first mode $k_1$ is excited ($n=1$), with $c_1 = \eps/2,~c_{-1} = \eps (0.3-0.4 i)/2,  ~\epsilon = 10^{-4}$. The agreement with (\ref{first}) is good.}
\end{figure}

\begin{figure}[H]
\centering
\includegraphics[width=9cm]{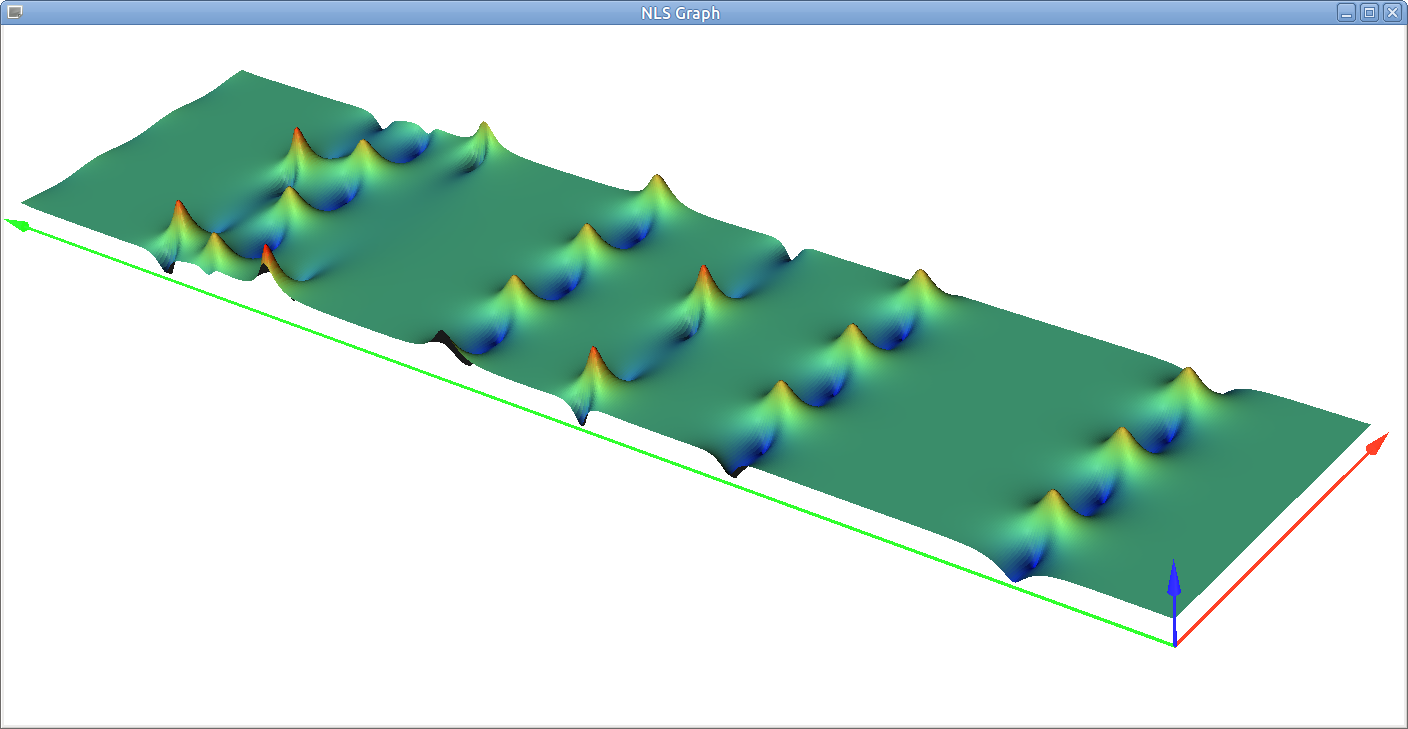}
\includegraphics[width=3.8cm]{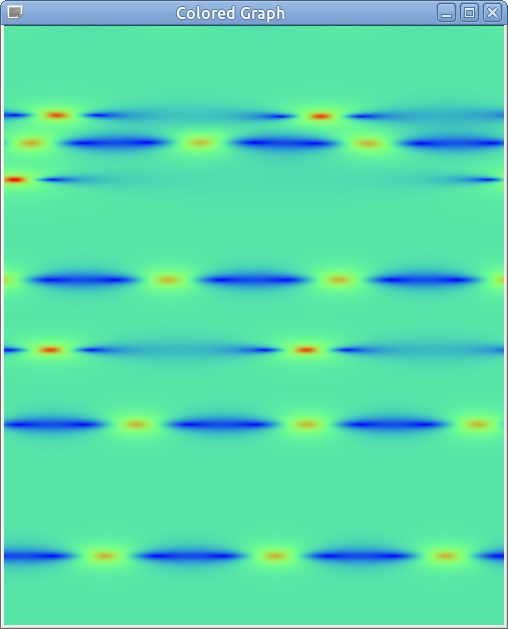}
\caption{\label{fig:fig5}
The 3D plotting of $|u(x,t)|$ obtained through the numerical integration of NLS via the SSFM, for $L=12$ ($N=3$ unstable modes), in the case in which only the third mode $k_3$ is excited ($n=3$), with $c_3 = \eps/2,~c_{-3} = \eps (0.3-0.4 i)/2,  ~\epsilon = 10^{-4}$. The agreement with (\ref{first}) is very good.}
\end{figure}

If we repeat the same experiment with $\eps=10^{-4}$, the 8-th harmonics is the first to appear in the numerical experiment (see Fig~\ref{fig:fig8}), as predicted by the qualitative formula
\beq\label{est2}
\frac{T_1}{T^{\tiny{num}}}\sim \frac{4}{\sigma_1}\left(\frac{18}{\sigma_8}\right)^{-1}\sim\frac{11.5}{9} >1 .
\eeq
The ``theoretical'' generation time $T_8$ of the most unstable 8-th mode is much larger than the appearance time $T^{\tiny{num}}$ due 
to numerical noise, and $T^{\tiny{num}}$ is smaller than $T_1$. The 8-th mode appears first, and the approximation 
for the first appearance (\ref{first}) is completely wrong.

Consider now the case $L=20$ ($N=6$ unstable modes, with the 4-th mode as the most unstable), in the case in which only the first mode $k_1$ is excited ($n=1$), with $c_1 = \eps/2,~c_{-1} = \eps (0.3-0.4 i)/2,  ~\epsilon = 10^{-4}$. In this case we have the following qualitative estimates for the first appearance times:
\begin{align}
\label{est3}
T_1\sim 6.4 \log(10) < T_2\sim 6.7 \log(10) < T_3 \sim  7.2  \log(10) < \\ 
< T_4 = 8.2 \log(10) < T^{\tiny{num}}\sim  9.2 \log(10) 
\end{align}
and therefore we infer that the numerical noise does not affects essentially the first appearance of the first 4 harmonics. Indeed, the 
numerical output confirms this theoretical prediction (see Fig.~\ref{fig:fig4}). In this numerical output, the maximum of the modulus is $2.95$ in the point 
$(x,t)=(-0.351,9.037)$, while formula (\ref{first}) predicts the maximum of the modulus to be $1+2\sin\phi_1=2.93$, in the point $(X^+_1,T_1(|\alpha_1|))=(-0.333,8.980)$. Taking into account how rough are the estimates (\ref{est3}), the agreement is surprisingly good.

If we repeat the same experiment exciting the 3-rd mode only, the first and the second gaps are theoretically close, but they are open  due to numerical noise. The most unstable gap is the second one, and we have the following qualitative estimate:
\beq\label{est4}
\frac{T_3}{T^{\tiny{num}}}\sim \frac{4}{\sigma_3}\left(\frac{18}{\sigma_2}\right)^{-1}\sim\frac{2.06}{10.1}\sim \frac{1}{5} \ll 1 .
\eeq
It means that the generation time of the RW associated with the numerical noise is approximately 5 times larger than the first appearance 
time $T_3$. The numerical output confirms this prediction, but the second harmonics appears a bit earlier $T^{\tiny{num}}\sim 4 T_3$
(see Fig.\ref{fig:fig5}). The first appearance is well described by (\ref{first}): in the numerical output, the maximum of the modulus is $2.2376$ in the point $(x,t)=(0.343,5.7037)$, while formula (\ref{first}) predicts the maximum of the modulus to be $1+2\sin\phi_3=2.2379$ in the point $(X^+_3,T_3(|\alpha_3|))=(0.351,5.7038)$. Therefore first appearance approximation (\ref{first}) works very well.

Reasoning as before, the upper bound for the maximum 
\beq\label{upper_bound_maxima_nN}
M(n,L)<1+2\sqrt{1-\left(\frac{n}{N+1}\right)^2}
\eeq 
follows from the formulas
\beq
\sin\phi_n =\sqrt{1-\left(\frac{\pi}{L}n\right)^2}, \ \ \ \ \pi N<L<\pi(N+1),
\eeq
and is obtained when $L$ is close to $\pi(N+1)$. It follows that, if we have $N$ unstable modes, the first RW has its maximum if the initial condition excites the first unstable mode ($n=1$).
 
If, for instance, we have $N=3$ unstable modes, if the initial perturbation excites the first ($n=1$), and if $L$ is close to the upper bound $\pi (N+1)$, then the maximum value of the amplitude of the RW (\ref{first}) is close to $2.936$ times the background.

In the next sections we derive the results summarized in this section.

\section{The Cauchy problem via the finite gap method}
\label{sec:sec3}

Let us recall the main definitions from the periodic spectral theory of the self-focusing NLS equation (\ref{NLS}). 

We use the following zero-curvature representation \cite{ZakharovShabat}: the function $u(x,t)$ satisfies the NLS equation (\ref{NLS}) if and only if the 
following pair of linear problems is compatible: 
\begin{equation}
\label{eq:lp-x}
\vec\Psi_x(\lambda,x,t)=U(\lambda,x,t)\vec\Psi(\lambda,x,t),
\end{equation}
\begin{equation}
\label{eq:lp-t}
\vec\Psi_t(\lambda,x,t)=V(\lambda,x,t)\vec\Psi(\lambda,x,t),
\end{equation}
$$
U=\left [\begin {array}{cc} -i \lambda & i u(x,t)
\\\noalign{\medskip} i \overline{u(x,t)} & i \lambda\end {array}
\right ],
$$
$$
V(\lambda,x,t)= \left[\begin {array}{cc} -2 i \lambda^2 + i u(x,t)\overline{u(x,t)} & 2 i \lambda u(x,t) - u_x(x,t)
\\\noalign{\medskip} 2 i \lambda \overline{u(x,t)} +\overline{u_x(x,t)} & 2 i \lambda^2- i u(x,t)\overline{u(x,t)} 
\end {array}
\right ],
$$
where 
$$
\vec\Psi(\lambda,x,t)= \left [\begin {array}{c} \Psi_1(\lambda,x,t) \\
\Psi_2(\lambda,x,t) \end {array}\right ].
$$
The linear problem (\ref{eq:lp-x}) can be rewritten as a spectral problem
\begin{equation}
\label{eq:lp-x2}
L\vec\Psi(\lambda,x,t)=\lambda \vec\Psi(\lambda,x,t),
\end{equation}
where 
$$
L=\left[ \begin {array}{cc} i\partial_x & u(x,t) \\ -  \overline{u(x,t)} & -i\partial_x \end {array}\right].
$$
It is essential that $L$ is not self-adjoint, and the spectrum of this problem typically contains complex points. 

In the present text we consider the $x$-periodic problem:
\begin{equation}
\label{eq:per1}
u(x+L,t)=u(x,t).
\end{equation}
In the periodic theory of the NLS equation the following two spectral problems are used to define the spectral data:
\begin{enumerate}
\item The spectral problem on the line, i.e. the spectral problem in $L^2(\RR)$. It is also called the \textbf{main spectrum}.
\item The spectral problem on the interval $[x_0,x_0+L]$ with the following Dirichlet-type boundary conditions:
\begin{equation}
\label{eq:dir1}
\Psi_1(\lambda,x_0,t)=\Psi_1(\lambda,x_0+L,t)=0.
\end{equation}
This spectrum is called the \textbf{auxiliary spectrum}  or \textbf{divisor}. 

\begin{remark}
Let us point out that, in many papers dedicated to the finite-gap NLS solutions, 
a different auxiliary problem is used. Namely, one imposes the following symmetric boundary condition:
$$
\Psi_1(\lambda,x_0,t)+\Psi_2(\lambda,x_0,t) = \Psi_1(\lambda,x_0+L,t)+\Psi_2(\lambda,x_0+L,t)=0.
$$ 
This approach has the following advantage: all divisor points are located in a compact area of the spectral curve, 
but it requires one extra divisor point and increases the complexity of the formulas.
\end{remark}
\end{enumerate}

\begin{enumerate}
\item \textbf{The main spectrum}.
Consider a fixed time point $t=t_0$.
To define the spectrum of the problem on the line, it is convenient to introduce the monodromy matrix. Consider the 
matrix equation
$$
L\hat\Psi(\lambda,x,t_0)=\lambda\hat\Psi(\lambda,x,t_0),
$$
where $\hat\Psi$ is a $2\times2$ matrix with the initial condition
$$
\hat\Psi(\lambda,x_0,t_0)=\left [\begin {array}{cc} 1 & 0
\\ 0 & 1 \end {array}\right].
$$
Then the monodromy matrix $\hat T(\lambda,x_0,t_0)$ is defined by:
$$
\hat T(\lambda,x_0,t_0)=\hat\Psi(\lambda,x_0+L,t_0).
$$
The eigenvalues and eigenvectors of $T(\lambda,x_0,t_0)$ are defined on a two-sheeted covering of the 
$\lambda$-plane. This Riemann surface $\Gamma$ is called the \textbf{spectral curve}. The monodromy matrices 
corresponding to different $x_0$ and $t_0$ coincide up to conjugation, therefore $\Gamma$ is well-defined and does not 
depend on time. The eigenvectors of $T(\lambda,x_0,t_0)$ are the Bloch eigenfunctions of $L$
\begin{eqnarray}
\label{eq:bloch1}
L  \vec\Psi(\gamma,x,t) =\lambda(\gamma) \vec\Psi(\gamma,x,t),\hphantom{aaa\gamma \in \Gamma}\nonumber\\
\vec\Psi(\gamma,x+L,t)=\kappa(\gamma) \vec\Psi(\gamma,x,t),\gamma \in \Gamma,
\end{eqnarray}
$\lambda(\gamma)$ denote the projection of the point $\gamma$ to the $\lambda$-plane.
If $\lambda(\gamma_1)=\lambda(\gamma_2)$, then $\kappa(\gamma_1)\kappa(\gamma_2)=1$.

It is convenient to write:
\begin{equation}
\label{eq:bloch2}
\kappa(\gamma)= e^{iLp(\gamma)}.
\end{equation}
The multivalued function $p(\gamma)$ is called \textbf{quasimomentum.} Its differential $dp(\gamma)$ is well-defined
and meromorphic on $\Gamma$, and all periods of $dp$ are pure real.  

The spectrum of $L$ is exatly the projection of the set $\{\gamma\in\Gamma, |\kappa(\gamma)|=1\}$ to the $\lambda$-plane. 
Equivalently, the spectrum is defined by the condition:
$$
\Im p(\gamma)=0.
$$
For $\lambda\in\RR$ the matrix $U$ is skew-hermitian, and the monodromy matrix is unitary, therefore the whole real 
line lies in the spectrum of $L$ in $L^2(\RR)$. The end points of the spectrum are the branch points of $\Gamma$.
At these points we have:
\begin{equation}
\label{eq:branch1}
\kappa(\gamma)=\pm 1.
\end{equation}
Equation (\ref{eq:branch1}) is also satisfied at the double points, which can be obtained by merging an even number 
of branch points. All real double points in the focusing NLS theory are removable, i.e. they do not arise in the inverse 
spectral transform. But a finite number of non-removable complex double points may be present, and they arise in the 
formulas for the finite-gap solutions.

Equivalently, the branch and double points of $\Gamma$ are exactly the eigenvalues of $L$ on the spaces of periodic 
and antiperiodic functions:
\begin{eqnarray}
\label{eq:branch2}
L  \vec\Psi(\gamma,x,t) =\lambda(\gamma) \vec\Psi(\gamma,x,t), \hphantom{aa \gamma \in \Gamma} \nonumber\\
\vec\Psi(\gamma,x+L,t)=\pm \vec\Psi(\gamma,x,t), \ \  \gamma \in \Gamma.\nonumber
\end{eqnarray}

In our text we calculate the branch points of $\Gamma$ using formula (\ref{eq:branch2}).

If 
$$
\vec\Psi(\lambda,x,t)= \left [\begin {array}{c} \Psi_1(\lambda,x,t) \\
\Psi_2(\lambda,x,t) \end {array}\right ]
$$
satisfies (\ref{eq:lp-x2}), then the function
$$
\vec\Psi^+(\lambda,x,t)= \left [\begin {array}{c} \overline{\Psi_2(\lambda,x,t)} \\
-\overline{\Psi_1(\lambda,x,t)} \end {array}\right ]
$$
also satisfies the same equation with complex conjugate eigenvalue:
\begin{equation}
\label{eq:lp-x3}
L\vec\Psi^+(\lambda,x,t)=\overline{\lambda} \vec\Psi^+(\lambda,x,t).
\end{equation}
It immediately implies that $\Gamma$ is real, i.e. the set of branch points of $\Gamma$ is invariant with respect 
to the complex conjugation.

Potential $u(x,t)$ is called \textbf{finite-gap} if the spectral curve $\Gamma$ is algebraic, i.e., if it can be written 
in the form
\begin{equation}
\label{eq:gamma1}
\nu^2 =\prod\limits_{j=1}^{2g+2} (\lambda-E_j).
\end{equation}
It means that $\Gamma$ has only a finite number of branch points and non-removable double points. Such solutions can be 
written in terms of the Riemann theta-functions. Any smooth, periodic in $x$ solution admits an arbitrarily good finite 
gap approximation, for any fixed time interval.  

\item \textbf{The auxiliary spectrum.} The auxiliary spectrum is defined as the set of points $\gamma\in\Gamma$ such 
that the first component of the Bloch eigenfunction is equal to 0 at the point $x_0$.
\begin{eqnarray}
\label{eq:dir1}
L\vec\Psi(\gamma,x,t)= \lambda(\gamma) \vec\Psi(\gamma,x,t),\nonumber\\
\vec\Psi(\gamma,x+L,t))=\kappa(\gamma) \vec\Psi(\gamma,x,t),\\
\Psi_1(\gamma,x_0,t_0)=0.\nonumber
\end{eqnarray}
Equivalently, the auxiliary spectrum coincides with the set of zeroes of the first component of the Bloch 
eigenfunction:
\begin{equation}
\label{eq:dir2}
\Psi_1(\gamma,x,t)=0;
\end{equation}
therefore it is called \textbf{divisor of zeroes.} The zeroes of $\Psi_1(\gamma,x,t)$ depend on $x$ and $t$. The $x$, 
$t$ dinamics becomes linear after the Abel transform. 

\end{enumerate}

\section{Spectral data for a small perturbation of the constant solution}
\label{sec:sec4}

In our paper we study small periodic perturbations of the spatially constant solution $u_0(x,t)=e^{2i t}$. 

Let us calculate the spectral data associated with such Cauchy data. We assume that $x_0=t_0=0$, and in this Section
we write $u(x)$ instead of $u(x,0)$.

Let $u(x)=1+\epsilon(x)$, where $|\epsilon(x)|\ll 1$, $\epsilon(x+L)=\epsilon(x)$. Then
$$
U=\left [\begin {array}{cc} -i \lambda & i (1+\epsilon(x))
\\\noalign{\medskip} i (1+\bar \epsilon(x)) & i \lambda\end {array}
\right ],
$$
where
$$
\epsilon(x,0) = \sum c_k e^{i\frac{2\pi}{L}kx}, \ \ c_k\ll 1.
$$
It is convenient to write:
$$
L=L_0+L_1, \ \ L_0 = \left[ \begin {array}{cc} i\partial_x & 1 \\ -1  & -i\partial_x \end {array}\right], \ \ 
L_1 = \left[ \begin {array}{cc} 0 & \epsilon(x) \\ -\overline{\epsilon(x)}  & 0 \end {array}\right],
$$
and the spectral data for $L$ will be calculated using the perturbation theory near the spectral data for $L_0$.

\subsection{The spectral data for the unperturbed operator}
\label{sec:sec4.1}

The unperturbed spectral curve $\Gamma_0$ is rational, and a point $\gamma\in\Gamma_0$ is a pair of complex numbers
$\gamma=(\lambda,\mu)$ satisfying the following quadratic equation:
$$
\mu^2=\lambda^2+1.
$$
The Bloch eigenfucntions for operator $L_0$ can be easily calculated explicitly:
\begin{equation}
\label{eq:bloch3}
\psi^{\pm}(\gamma,x)=\left[\begin {array}{c} 1 \\ \lambda(\gamma)\pm \mu(\gamma) \end {array} \right ] e^{\pm i\mu(\gamma) x},
\end{equation}
$$
L \psi^{\pm}(\gamma,x) = \lambda(\gamma)  \psi^{\pm}(\gamma,x).
$$
These eigenfunctions are periodic (antiperiodic) iff $\frac{L}{2\pi}\mu$ is an even (an odd) integer. 
Let us introduce the following enumeration of the periodic and antiperiodic spectral points:
$$
\mu_n = \frac{\pi n}{L}, \ \ \lambda^{\pm}_n=\pm\sqrt{\mu_n^2-1}, \ \ \Re{\lambda^+_n}+\Im{\lambda^+_n}>0, \ \  
n=0,1,2,\ldots\infty.
$$
We also assume that
$$
\lambda_n =\lambda_{-n} = \lambda^+_n.
$$

Taking into account that the squared eigenfunctions provide the proper basis for the linearized theory, we notice that the  
Fourier modes correspond the the points $\mu_n$. These modes are unstable if the corresponding $\lambda_n$ are 
imaginary ($|\mu_n|<1$)  and stable if the  $\lambda_n$ are real ($|\mu_n|\ge 1$), see the Introduction. Therefore the spectral points 
with  $|n|<\frac{L}{\pi}$ are \textbf{unstable}, and spectral points with $|n|\ge\frac{L}{\pi}$  are \textbf{stable}.

We have the following basis of eigenfunctions for the periodic and antiperiodic problems:
\begin{equation}
\label{eq:basis1}
\psi^{\pm}_{n}=\left[\begin {array}{c} 1 \\ \mu_n\pm\lambda_n \end {array} \right ] e^{i\frac{\pi}{L}nx},
\end{equation}
$$
L_0\psi^{\pm}_{n}=\pm\lambda_{n} \psi^{\pm}_{n}.
$$

The curve $\Gamma_0$ has two branch points $E_0=i$, $\overline{E_0}=-i$ corresponding to $n=0$. If $n>0$, there are no 
branching at the points $\lambda^{\pm}_n$, but the monodromy matrix becomes diagonal with coinciding eigenvalues:
$$
T(\lambda^{\pm}_n,0,0)= \left [\begin {array}{cc} (-1)^n &  0
\\ \noalign{\medskip} 0 & (-1)^n \end {array}
\right ].
$$
Such points are called \textbf{resonant points}. They are also the eigenvalues of the Dirichlet problem (\ref{eq:dir1}),
and the Dirichlet eigenfunctions are given by
$$
\psi^{\mbox{Dir}}_{n}(x)=\psi^{+}_{n}(x)-\psi^{-}_{n}(x), \ \ n>0.
$$

Therefore the divisor of the unperturbed problem is located at the resonant points.

After a generic small perturbation such double points split into pairs of branch points, and no other branch points 
are generated. The divisor of the perturbed problem has exactly one point near each resonant point of the unperturbed 
problem.

\subsection{The spectral data for the perturbed operator}
\label{sec:sec4.2}

To calculate the perturbed spectral curve, we develop the perturbation theory for the periodic and antiperiodic problems 
using the basis (\ref{eq:basis1}).

It is also convenient to introduce the following notations:
\begin{align}
\label{def_alpha_beta_2}
\alpha_n =\overline{c_n}-(\mu_n+\lambda_n)^2 c_{-n}, \ \ \beta_n =\overline{c_{-n}}-(\mu_n-\lambda_n)^2 c_{n}, \ \ 
 \hphantom{n\ge 1.}
 \\
\tilde{\alpha_n} =\overline{c_n}-(\mu_n-\lambda_n)^2 c_{-n}, \ \ \tilde{\beta_n} =\overline{c_{-n}}-(\mu_n+\lambda_n)^2 c_{n},\ \  n\ge 1.
\end{align}

If $\epsilon(x)$ has only one harmonic:
$$
\epsilon(x)= c_n e^{i\frac{2\pi}{L}nx},
$$
then the matrix elements for $L_1$ can be easily calculated:
$$
L_1 \psi^{\pm}_m = \left[\begin {array}{c} c_n(\mu_m\pm\lambda_m) \\ 0 \end {array} \right]  e^{i\mu_{m+2n}x}+
\left[\begin {array}{c} 0 \\-\overline{c_n} \end {array} \right] e^{i\mu_{m-2n}x}=
$$
$$
=\frac{c_n(\lambda_{m+2n}-\mu_{m+2n})(\mu_m\pm\lambda_m)   }{2\lambda_{m+2n}}\psi^+_{m+2n}+
\frac{c_n(\lambda_{m+2n}+\mu_{m+2n})(\mu_m\pm\lambda_m)   }{2\lambda_{m+2n}}\psi^-_{m+2n}-
$$
$$
-\frac{\overline{c_n} }{\lambda_{m-2n}}\psi^+_{m-2n}+\frac{\overline{c_n} }{\lambda_{m-2n}}\psi^-_{m-2n}.
$$
Let us introduce the following notations: $|l\pm>$ denotes the basic vector $\psi^{\pm}_l$, $<l\pm|$ denote the adjoint 
basis:
$$
<l+|m+> = \delta_{lm}, \ \ <l-|m-> = \delta_{lm}, \ \ <l+|m-> = <l-|m+>=0.
$$

For an arbitrary periodic perturbation, the matrix elements can be written in the following form: 
$$
<m+|L_1|l+> = \frac{c_{(m-l)/2}(\lambda_{m}-\mu_{m})(\lambda_l+\mu_l)-\overline{c_{-(m-l)/2}}}{2\lambda_{m}}
$$
$$
<m-|L_1|l+> = \frac{c_{(m-l)/2}(\lambda_{m}+\mu_{m})(\lambda_l+\mu_l)+\overline{c_{-(m-l)/2}}}{2\lambda_{m}}
$$
$$
<m+|L_1|l-> = \frac{c_{(m-l)/2}(\lambda_{m}-\mu_{m})(-\lambda_l+\mu_l)-\overline{c_{-(m-l)/2}}}{2\lambda_{m}}
$$
$$
<m-|L_1|l-> = \frac{c_{(m-l)/2}(\lambda_{m}+\mu_{m})(-\lambda_l+\mu_l)+\overline{c_{-(m-l)/2}}}{2\lambda_{m}}.
$$
Here we use a slightly non-standard notations: $<f|L|g>$ denotes the matrix element for both orthogonal and 
non-orthogonal basis. 

Let us recall some basic facts from the perturbation theory. Consider a small perturbation of a diagonal operator 
$\LL=\LL_0+\varepsilon \LL_1$. Let $\lambda_1$ be an eigenvalue of $\LL_0$ of multiplicity $k$. We can write:
$$
\LL_0=\left[\begin{array}{cc} A_0 & 0 \\ 0 & D_0 \end{array} \right],
$$
where $A_0$ denotes the $k\times k$ diagonal matrix with diagonal elements equal to $\lambda_1$
\begin{equation}
\label{eq:a0}
A_0=\left[\begin{array}{ccccc}\lambda_1 & 0 & 0 & \ldots & 0 \\ 
0 & \lambda_1 & 0 & \ldots & 0 \\ 0 & 0 & \lambda_1 & \ldots & 0 \\ 
\vdots & \vdots & \vdots & \ddots & \vdots \\
0 & 0 & 0 & \ldots & \lambda_1
\end{array}\right]
\end{equation}
and the block $D_0$ is diagonal and all diagonal elements are different from $\lambda_1$:
$$ 
D_0=\left[\begin{array}{cccccc}\lambda_2 & 0 & 0 & 0 & \ldots & 0 \\ 
0 & \lambda_3 & 0 & 0 & \ldots & 0 \\ 0 & 0 &   \lambda_4 & 0 & \ldots & 0 \\ 
0 & 0 &  0 & \lambda_5 & \ldots & 0 \\ 
\vdots & \vdots & \vdots & \vdots & \ddots & \vdots \\
0 & 0 & 0 & 0 & \ldots & \lambda_N
\end{array}\right]
$$
Denote the corresponding blocks of $\LL_1$ by $A_1$, $B_1$, $C_1$, $D_1$ respectively:
$$
\LL_1= \left[\begin{array}{cc} A_1 & B_1 \\ C_1 & D_1 \end{array} \right],
$$
and consider the characteristic equation for $\LL$ 
\begin{equation}
\label{eq:char1}
\det(\LL-\lambda E)=0.
\end{equation}
Using the formula for the block matrix determinant
$$
\det\left[\begin{array}{cc} A & B \\ C & D \end{array} \right]=\det D \det(A-BD^{-1}C),
$$
and assuming that $\lambda=\lambda_1+\delta\lambda$,  $|\delta\lambda|\ll1$, (\ref{eq:char1}) implies that
\begin{equation}
\label{eq:char2}
\det(\varepsilon A_1 -\delta\lambda E - \varepsilon^2 B_1 (D_0 -\lambda_1 E +\varepsilon D_1 - \delta\lambda E)^{-1} C_1) =0, 
\end{equation}
where 
$$
(D_0 -\lambda_1 E +\varepsilon D_1 - \delta\lambda E)^{-1}=\sum\limits_{n=0}^{\infty} (D_0 -\lambda_1 E)^{-1}
[( \delta\lambda E - \varepsilon D_1) (D_0 -\lambda_1 E)^{-1}]^n.
$$
Let us point out the procedure works for both hermitian and non-hermitian perturbations.

If the eigenvalues $\tilde\lambda_1$,\ldots,$\tilde\lambda_n$  of $A_1$ are pairwise distinct, then, after perturbation,
the eigenvalue $\lambda_1$ of multiplicity $k$ splits into $k$ simple eigenvalues $\hat\lambda_{1,j}(\varepsilon)$
such that $\hat\lambda_{1,j}(\varepsilon)=\lambda_1+\varepsilon\tilde\lambda_j+O(\varepsilon^2)$, $j=1,\ldots,k$.
Therefore, to calculate the perturbations of the eigenvalue $\lambda_1$ up to $O(\varepsilon^2)$ correction, it is 
sufficient to calculate the eigenvalues of the block $[A_0+\varepsilon A_1]$ in the matrix $\LL$. Equivalently, 
we have the following rule:
\begin{lemma}
\label{lem:l1}
Let $\LL_0$ be a diagonal operator, and one of the eigenvalues of $\LL_0$ is equal to $\lambda_1$ and has multiplicity 
$k$. Denote the positions of $\lambda_1$ in $\LL_0$ by $j_1$,\ldots,$j_n$. Consider a small perturbation 
of $\LL_0$: $\LL=\LL_0+\varepsilon \LL_1$. 

Denote by $A_1$ the $k\times k$ matrix obtained from $\LL$ removing all rows and all columns except $j_1$,\ldots,$j_n$.

Assume that all eigenvalues $\tilde\lambda_1$,\ldots,$\tilde\lambda_n$ of $A_1$ are pairwise distinct. Then, for 
sufficiently small $\varepsilon$, the operator $\LL$ has $k$ simple eigenvalues coinciding with the eigenvalues
of $A_0+\varepsilon A_1$, up to $O(\varepsilon^2)$ corrections.
\end{lemma}
\begin{remark}
The operator $A_0+\varepsilon A_1$ in Lemma~\ref{lem:l1} can be obtained by removing from $\LL$ all columns and all rows 
except $j_1$,\ldots,$j_n$.
\end{remark}

Now we are ready to calculate the eigenvalues of $L$ to the leading order of perturbation theory using Lemma~\ref{lem:l1}.

Let us define
$$
P_{ml}=\left[\begin{array}{cc}
\frac{c_{(m-l)/2}(\lambda_{m}-\mu_{m})(\lambda_l+\mu_l)-\overline{c_{-(m-l)/2}}}{2\lambda_{m}} & 
\frac{c_{(m-l)/2}(\lambda_{m}-\mu_{m})(-\lambda_l+\mu_l)-\overline{c_{-(m-l)/2}}}{2\lambda_{m}} \\
\frac{c_{(m-l)/2}(\lambda_{m}+\mu_{m})(\lambda_l+\mu_l)+\overline{c_{-(m-l)/2}}}{2\lambda_{m}} &
\frac{c_{(m-l)/2}(\lambda_{m}+\mu_{m})(-\lambda_l+\mu_l)+\overline{c_{-(m-l)/2}}}{2\lambda_{m}}
\end{array} \right].
$$

In the basis (\ref{eq:basis1}), operator $L$ has the following form:
$$\hat L=
\begin{array}{c|c|c|c|c|c|c|c|c} \ddots & \ldots & \ldots & \ldots & \ldots & \ldots & \ldots & \ldots & \ldots \\
\hline 
\vdots & \begin{matrix} \lambda_3 & 0 \\ 0 & -\lambda_3 \end{matrix} & \bigzero & \bigP{-3\,-1} & 
\bigzero & \bigP{-3\,1} & \bigzero &\bigP{-3\,3}  & \vdots \\
\hline
\vdots & \bigzero & \begin{matrix} \lambda_2 & 0 \\ 0 & -\lambda_2 \end{matrix} & \bigzero & \bigP{-2\,0} & 
\bigzero & \bigP{-2\,2} & \bigzero & \vdots \\
\hline
\vdots &  \bigP{-1\,-3} & \bigzero & \begin{matrix} \lambda_1 & 0 \\ 0 & -\lambda_1 \end{matrix} & \bigzero & 
\bigP{-1\,1} & \bigzero &  \bigP{-1\,3}  & \vdots \\
\hline
\vdots & \bigzero & \bigP{0\,-2}&   \bigzero & \begin{matrix} \lambda_0 & 0 \\ 0 & -\lambda_0 \end{matrix} & 
\bigzero & \bigP{0\,2} & \bigzero &  \vdots \\
\hline
\vdots & \bigP{1\,-3} & \bigzero &   \bigP{1\,-1} & \bigzero & 
\begin{matrix} \lambda_1 & 0 \\ 0 & -\lambda_1 \end{matrix} & \bigzero & \bigP{1\,3} & \vdots \\
\hline
\vdots & \bigzero & \bigP{2\,-2} & 
\bigzero & \bigP{2\,0} & \bigzero &  \begin{matrix} \lambda_2 & 0 \\ 0 & -\lambda_2 \end{matrix} & \bigzero & \vdots \\
\hline
\vdots &  \bigP{3\,-3} & \bigzero & \bigP{3\,-1} & \bigzero &\bigP{3\,1} & 
\bigzero & \begin{matrix} \lambda_3 & 0 \\ 0 & -\lambda_3 \end{matrix} &\vdots \\
\hline
\vdots & \ldots & \ldots & \ldots & \ldots & \ldots & \ldots & \ldots & \ddots
\end{array}
$$
It is convenient to separate the periodic and antiperiodic subspaces
$$
\hat L_{\mbox{\scriptsize even}}=\begin{array}{c|c|c|c|c|c|c} \ddots & \ldots & \ldots & \ldots & \ldots & \ldots & \ldots \\
\hline 
\vdots & \begin{matrix} \lambda_4 & 0 \\ 0 & -\lambda_4 \end{matrix} & \bigP{-4\,-2} & \bigP{-4\,0} 
& \bigP{-4\,2} &\bigP{-4\,4}  & \vdots \\
\hline
\vdots & \bigP{-2\,-4}  & \begin{matrix} \lambda_2 & 0 \\ 0 & -\lambda_2 \end{matrix} & \bigP{-2\,0} & 
\bigP{-2\,2} & \bigP{-2\,4} & \vdots \\
\hline
\vdots & \bigP{0\,-4} & \bigP{0\,-2} & \begin{matrix} \lambda_0 & 0 \\ 0 & -\lambda_0 \end{matrix} & 
\bigP{0\,2} & \bigP{0\,4} &  \vdots \\
\hline
\vdots & \bigP{2\,-4} & \bigP{2\,-2} & \bigP{2\,0} & 
\begin{matrix} \lambda_2 & 0 \\ 0 & -\lambda_2 \end{matrix} &  \bigP{2\,4}  & \vdots \\
\hline
\vdots &  \bigP{4\,\-4 } & \bigP{4\,-2} & \bigP{4\,0} &\bigP{4\,2} & 
\begin{matrix} \lambda_4 & 0 \\ 0 & -\lambda_4 \end{matrix} &\vdots \\
\hline
\vdots & \ldots & \ldots & \ldots & \ldots & \ldots & \ddots
\end{array}
$$
$$
\hat L_{\mbox{\scriptsize odd}}=\begin{array}{c|c|c|c|c|c} \ddots & \ldots & \ldots & \ldots & \ldots & \ldots \\
\hline 
\vdots & \begin{matrix} \lambda_3 & 0 \\ 0 & -\lambda_3 \end{matrix} & \bigP{-3\,-1} & \bigP{-3\,1} 
& \bigP{-3\,3} & \vdots \\
\hline
\vdots & \bigP{-1\,-3}  & \begin{matrix} \lambda_1 & 0 \\ 0 & -\lambda_1 \end{matrix} & \bigP{-1\,1} & 
\bigP{-1\,3} & \vdots \\
\hline
\vdots & \bigP{1\,-3} & \bigP{1\,-1} & \begin{matrix} \lambda_1 & 0 \\ 0 & -\lambda_1 \end{matrix} & 
\bigP{1\,3} &  \vdots \\
\hline
\vdots & \bigP{3\,-3} & \bigP{3\,-1} & \bigP{3\,1} & 
\begin{matrix} \lambda_3 & 0 \\ 0 & -\lambda_3 \end{matrix} & \vdots \\
\hline
\vdots & \ldots & \ldots & \ldots & \ddots
\end{array}
$$
After this small perturbation, each resonant point of $L_0$  either splits into two branch points, or remains a double 
point. By Lemma~\ref{lem:l1}, to calculate the splitting of the resonant points $\lambda_n$, $-\lambda_n$  to the 
leading order, it is sufficient to consider the following submatrices, corresponding to the linear spans of 
$\{\psi^{+}_n, \psi^{+}_{-n} \}$ and   $\{\psi^{-}_n, \psi^{-}_{-n} \}$:
$$
Q^+_n=\left[\begin{array}{c|c}
\lambda_n & \frac{c_{-n}(\lambda_{-n}-\mu_{-n})(\lambda_{n}+\mu_{n})-\overline{c_{n}}}{2\lambda_{n}} \\
\hline 
\frac{c_{n}(\lambda_{n}-\mu_{n})(\lambda_{-n}+\mu_{-n})-\overline{c_{-n}}}{2\lambda_{n}} & \lambda_n 
\end{array}\right]=
$$
$$
=\left[\begin{array}{c|c}
\lambda_n & \frac{c_{-n}(\lambda_{n}+\mu_{n})^2-\overline{c_{n}}}{2\lambda_{n}} \\
\hline 
\frac{c_{n}(\lambda_{n}-\mu_{n})^2-\overline{c_{-n}}}{2\lambda_{n}} & \lambda_n 
\end{array}\right],
$$
and 
$$
Q^-_n=\left[\begin{array}{c|c}
-\lambda_n & \frac{c_{-n}(\lambda_{-n}+\mu_{-n})(-\lambda_{n}+\mu_{n})+\overline{c_{n}}}{2\lambda_{n}} \\
\hline 
\frac{c_{n}(\lambda_{n}+\mu_{n})(-\lambda_{-n}+\mu_{-n})+\overline{c_{-n}}}{2\lambda_{n}} & -\lambda_n 
\end{array}\right]=
$$
$$
=\left[\begin{array}{c|c}
-\lambda_n & \frac{-c_{-n}(\lambda_{n}-\mu_{n})^2+\overline{c_{n}}}{2\lambda_{n}} \\
\hline 
\frac{-c_{n}(\lambda_{n}+\mu_{n})^2+\overline{c_{-n}}}{2\lambda_{n}} & -\lambda_n 
\end{array}\right],
$$
and we obtain the following approximate formulas for the branch points:
\begin{align}
\label{eq:bps}
E_{l}=\lambda_n\mp\frac{1}{2\lambda_n}\sqrt{\big(\overline{c_{n}}- c_{-n}(\lambda_{n}+\mu_{n})^2\big)
\big(\overline{c_{-n}}-c_{n}(\lambda_{n}-\mu_{n})^2   \big)  }+O(\epsilon^2),  \ \ l=2n-1,2n,\\
\tilde{E_{l}}=-\lambda_n\pm\frac{1}{2\lambda_n}\sqrt{\big(\overline{c_{n}}- c_{-n}(\lambda_{n}-\mu_{n})^2\big)
\big(\overline{c_{-n}}-c_{n}(\lambda_{n}+\mu_{n})^2   \big)  }+O(\epsilon^2). \ \ \hphantom{l=2n-1,2n}
\end{align}
and, finally:
\begin{align}
\label{eq:bps2}
E_{l}=\lambda_n\mp\frac{1}{2\lambda_n}\sqrt{\alpha_n\beta_n}+O(\epsilon^2),  \ \ l=2n-1,2n,\\
\tilde{E_{l}}=-\lambda_n\pm\frac{1}{2\lambda_n}\sqrt{\tilde\alpha_n\tilde\beta_n}+O(\epsilon^2),  \ \ \hphantom{l=2n-1,2n,}\\
\end{align}
where $\alpha_n$, $\beta_n$, $\tilde\alpha_n$, $\tilde\beta_n$, are defined in (\ref {def_alpha_beta_2}). 
Here we assume that $\Re\sqrt{\alpha_n\beta_n}\ge0$, $\Re\sqrt{\tilde\alpha_n\tilde\beta_n}\ge0$ for unstable points, and  
$\Im\sqrt{\alpha_n\beta_n}<0$, $\Im\sqrt{\tilde\alpha_n\tilde\beta_n}<0$ for stable points.
For 
perturbations of the unstable points we have:
\begin{equation}
\label{eq:bps3}
\tilde{E_{l}}=\overline{E_{l}}.
\end{equation}

\begin{figure}
\centering
\mbox{\epsfxsize=10cm\epsffile{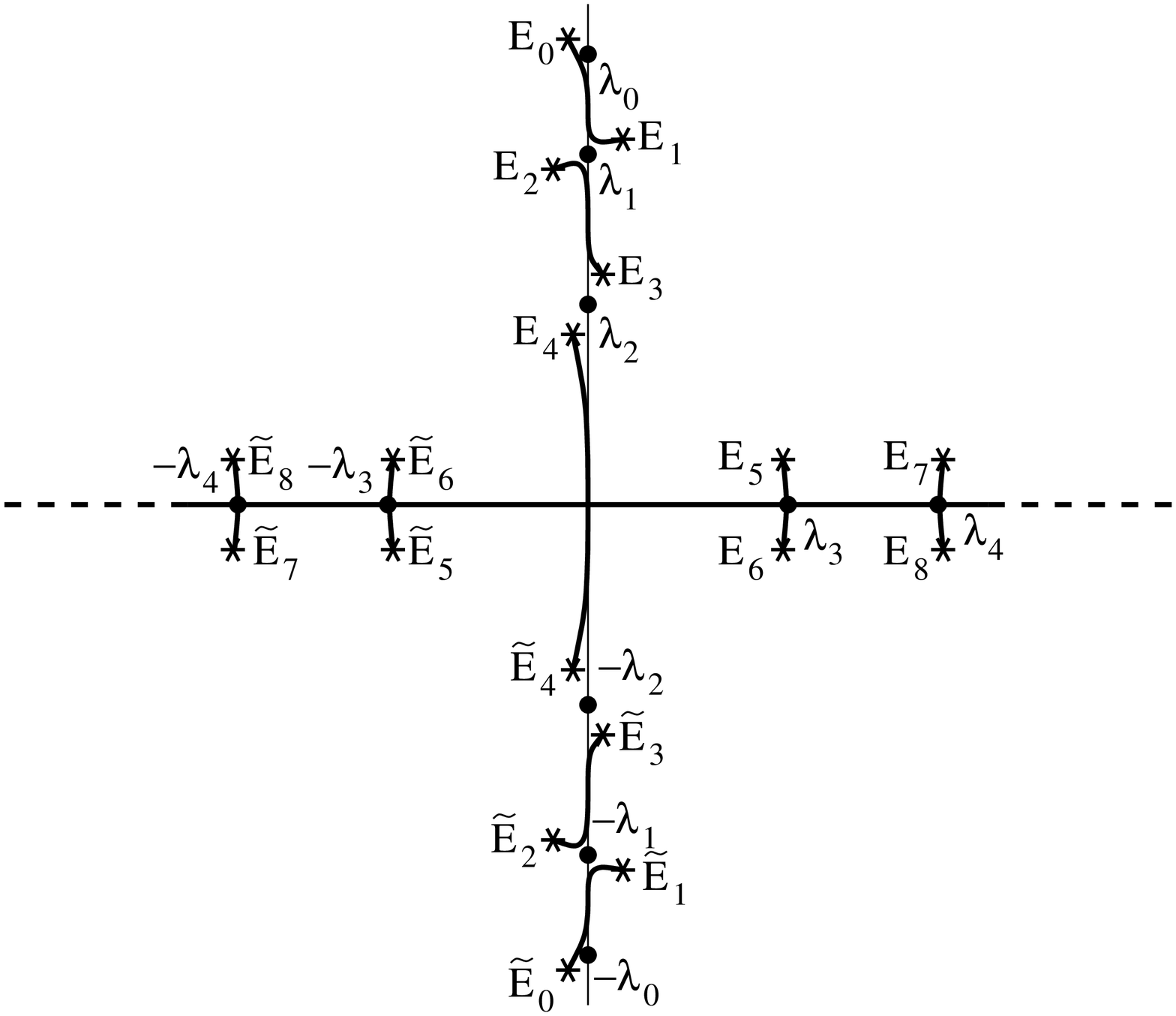}}
\end{figure}

Let us define the following enumeration for the  unperturbed divisor points 
$\gamma_n=(\lambda^{\mbox{\small div}}_n, \mu^{\mbox{\small div}}_n)$,  $n\ne 0$:
$$
\lambda^{\mbox{\small div}}_n=\left\{\begin{array}{ll} \lambda_n, & n>0, \\  -\lambda_n, & n<0, \end{array}\right. \ \ 
\mu^{\mbox{\small div}}_n=\mu_n.
$$
Let us calculate now the divisor positions up to $O(\epsilon^2)$ corrections. In contrast with the branch points, the 
Bloch multipliers for the Dirichlet spectrum are generically different from $\pm 1$; moreover their absolute values 
do not have to be equal to 1. Nevertheless they are sufficiently close to $\pm 1$ if the perturbation is small enough. 

Consider the restrictions of the operators $L_0$, $L_1$ to the space with the Bloch boundary conditions:
\begin{equation}
\label{eq:bloch4}
\vec\Psi(x+L) = \pm e^{iL\delta\mu }\vec\Psi(x), \ \ |\delta\mu|\ll 1.
\end{equation}
The eigenfunctions of $L_0$ in the space (\ref{eq:bloch4}) are the functions $\psi^{\pm}(\gamma,x)$, where 
$\gamma\in\Gamma_0$ are the following points:
$$
\mu_n(\delta\mu) = \frac{\pi n}{L}+\delta\mu, \ \ \lambda^{\pm}_n(\delta\mu)=\pm\sqrt{[\mu_n(\delta\mu)]^2-1},
$$
$$
\Re{\lambda^+_n(\delta\mu)}+\Im{\lambda^+_n(\delta\mu)}>0, \ \ 
\lambda_n(\delta\mu) = \lambda^+_n(\delta\mu).
$$
In contrast with the previous calculation, now
$$
\lambda_{-n}(\delta\mu)\ne\lambda_n(\delta\mu).
$$
Moreover
$$
\lambda_n(\delta\mu)=\lambda_n +\frac{\mu_n}{\lambda_n}\delta\mu+O\left((\delta\mu)^2\right).
$$
Let us denote the eigenfunctions of $L_0$ in (\ref{eq:bloch4}) by: 
\begin{equation}
\label{eq:basis2}
\psi^{\pm}_{n}(\delta\mu)=\left[\begin {array}{c} 1 \\ \mu_n(\delta\mu)\pm\lambda_n(\delta\mu) \end {array} \right ] 
e^{i\left[\frac{\pi}{L}n+\delta\mu\right]x},
\end{equation}
$$
L_0 \psi^{\pm}_{n}(\delta\mu,x) = \pm \lambda_{-n}(\delta\mu) \psi^{\pm}_{n}(\delta\mu,x). 
$$
Denote the matrix of $L$ in this basis by  $\hat L(\delta\mu_n)$.

To calculate  the divisor point $\gamma_n$ after perturbation we have to find a pair of numbers $\delta\lambda_n$,
$\delta\mu_n$ such that the Dirichlet boundary condition (\ref{eq:dir1}) is fulfilled. In the  basis (\ref{eq:basis2}) this condition 
can be reformulated as:

\textbf{Matrix $\hat L(\delta\mu_n)$ has eigenvalue $\lambda^{\pm}_n+\delta\lambda_n$ with the 
following property: the sum of all components of the eigenvector is equal to 0.}

We have the following formula for $\hat L(\delta\mu)$:

\noindent
\resizebox{\linewidth}{!}{%
$
$$
\hat L(\delta\mu)=\begin{array}{c|c|c|c|c|c|c} \ddots & \ldots & \ldots & \ldots & \ldots & \ldots & \ldots \\
\hline 
\vdots & \begin{matrix} \lambda_{-2}(\delta\mu) & 0 \\ 0 & -\lambda_{-2}(\delta\mu) \end{matrix} & \bigzero & \bigP{-2\,0}(\delta\mu) & 
\bigzero & \bigP{-2\,2}(\delta\mu) & \vdots  \\
\hline
\vdots & \bigzero & \begin{matrix} \lambda_{-1}(\delta\mu) & 0 \\ 0 & -\lambda_{-1}(\delta\mu) \end{matrix} & \bigzero & 
\bigP{-1\,1}(\delta\mu) & \bigzero & \vdots \\
\hline
\vdots &  \bigP{0\,-2}(\delta\mu) & \bigzero & \begin{matrix} \lambda_0(\delta\mu) & 0 \\ 0 & -\lambda_0(\delta\mu) \end{matrix} & \bigzero & 
\bigP{0\,2}(\delta\mu) & \vdots \\
\hline
\vdots & \bigzero & \bigP{1\,-1}(\delta\mu)&   \bigzero & \begin{matrix} \lambda_1(\delta\mu) & 0 \\ 0 & -\lambda_1(\delta\mu) \end{matrix} & 
\bigzero & \vdots \\
\hline
\vdots & \bigP{2\,-2}(\delta\mu) & \bigzero &   \bigP{2\,0}(\delta\mu) & \bigzero & 
\begin{matrix} \lambda_2(\delta\mu) & 0 \\ 0 & -\lambda_2(\delta\mu) \end{matrix} & \vdots \\
\hline
\vdots & \ldots & \ldots & \ldots & \ldots & \ldots & \ldots 
\end{array}
$$
$
}
where

\noindent
\resizebox{\linewidth}{!}{%
$
$$
P_{ml}(\delta\mu)=\left[\begin{array}{cc}
\frac{c_{(m-l)/2}(\lambda_{m}(\delta\mu)-\mu_{m}(\delta\mu))(\lambda_l(\delta\mu)+
\mu_l(\delta\mu))-\overline{c_{-(m-l)/2}}}{2\lambda_{m}(\delta\mu)} & 
\frac{c_{(m-l)/2}(\lambda_{m}(\delta\mu)-\mu_{m}(\delta\mu))(-\lambda_l(\delta\mu)+\mu_l(\delta\mu))-
\overline{c_{-(m-l)/2}}}{2\lambda_{m}(\delta\mu)} \\
\frac{c_{(m-l)/2}(\lambda_{m}(\delta\mu)+\mu_{m}(\delta\mu))(\lambda_l(\delta\mu)+\mu_l(\delta\mu))+
\overline{c_{-(m-l)/2}}}{2\lambda_{m}(\delta\mu)} &
\frac{c_{(m-l)/2}(\lambda_{m}(\delta\mu)+\mu_{m}(\delta\mu))(-\lambda_l(\delta\mu)+\mu_l(\delta\mu))+
\overline{c_{-(m-l)/2}}}{2\lambda_{m}(\delta\mu)}
\end{array} \right]
$$
$
}

Assuming $\delta\mu=O(\epsilon)$, we obtain
$$
P_{ml}(\delta\mu)=P_{ml}+O(\epsilon^2).
$$

Now we are ready to calculate the divisor positions to the leading order. 

If $n>0$ we consider the following block:

$$
Q^{d+}_n=\left[\begin{array}{c|c}
\lambda_n-\frac{\mu_n}{\lambda_n}\delta\mu & \frac{c_{-n}(\lambda_{n}+\mu_{n})^2-\overline{c_{n}}}{2\lambda_{n}} \\
\hline 
\frac{c_{n}(\lambda_{n}-\mu_{n})^2-\overline{c_{-n}}}{2\lambda_{n}} & \lambda_n+\frac{\mu_n}{\lambda_n}\delta\mu 
\end{array}\right],
$$
and for $n<0$ we consider the following block:
$$
Q^{d-}_n=\left[\begin{array}{c|c}
-\lambda_{-n} +\frac{\mu_{-n}}{\lambda_{-n}}\delta\mu & \frac{-c_{n}(\lambda_{-n}-\mu_{-n})^2+\overline{c_{-n}}}{2\lambda_{n}} \\
\hline 
\frac{-c_{-n}(\lambda_{-n}+\mu_{-n})^2+\overline{c_{n}}}{2\lambda_{-n}} & -\lambda_{-n} -\frac{\mu_{-n}}{\lambda_{-n}}\delta\mu
\end{array}\right],
$$
Let $n>0$. The vector 
$$
\left[\begin{array}{c} 1 \\ -1
\end{array}\right]
$$
is an eigenvector for $Q^+_n$ iff 
$$
-\frac{\mu_{n}}{\lambda_{n}}\delta\mu  -\frac{c_{-n}(\lambda_{n}+\mu_{n})^2-\overline{c_{n}}}{2\lambda_{n}}
=\frac{\mu_{n}}{\lambda_{n}}\delta\mu  -\frac{c_{n}(\lambda_{n}-\mu_{n})^2-\overline{c_{-n}}}{2\lambda_{n}};
$$
therefore, up to $O(\epsilon^2)$ corrections, we have:
$$
\delta\mu_n = \frac{1}{4\mu_n}\left[\big(
c_{n}(\lambda_{n}-\mu_{n})^2-\overline{c_{-n}}\big) -\big(c_{-n}(\lambda_{n}+\mu_{n})^2-\overline{c_{n}}\big)\right],
$$
and 
\begin{equation}
\label{eq:div1}
\delta\lambda_n=-\frac{\mu_{n}}{\lambda_{n}}\delta\mu  -\frac{c_{-n}(\lambda_{n}+\mu_{n})^2-\overline{c_{n}}}{2\lambda_{n}}=
\end{equation}
$$
=\frac{1}{4\lambda_n}\left[-\big(
c_{n}(\lambda_{n}-\mu_{n})^2-\overline{c_{-n}}\big) -\big(c_{-n}(\lambda_{n}+\mu_{n})^2-\overline{c_{n}}\big)\right]=
$$
$$
=\frac{1}{4\lambda_n}\left[\big(\overline{c_{n}}+\overline{c_{-n}}\big)-
c_{n}(\lambda_{n}-\mu_{n})^2-c_{-n}(\lambda_{n}+\mu_{n})^2\right].
$$ 

From the definition of the quasimomentum we immediately obtain:
\begin{equation}
\label{eq:div2}
p(\gamma_{n})=\frac{1}{4\mu_n}\left[\big(
c_{n}(\lambda_{n}-\mu_{n})^2-\overline{c_{-n}}\big) -\big(c_{-n}(\lambda_{n}+\mu_{n})^2-\overline{c_{n}}\big)\right]
+O(\epsilon^2)\ \  (\!\!\!\!\mod\pi/L).
\end{equation}

Finally we obtain:

\begin{equation}
\label{eq:div_alpha1}
\lambda(\gamma_n) = \lambda_n+\frac{1}{4\lambda_n}\left[\alpha_n+\beta_n\right]+O(\epsilon^2), \ \
p(\gamma_{n})= \frac{1}{4\mu_n}\left[\alpha_n-\beta_n\right]+O(\epsilon^2)
\end{equation}

Analogously, this vector is an eigenvector for $Q^-_{-n}$ iff
$$
\frac{\mu_{n}}{\lambda_{n}}\delta\mu- \frac{-c_{-n}(\lambda_{n}-\mu_{n})^2+\overline{c_{n}}}{2\lambda_{n}}=
-\frac{\mu_{n}}{\lambda_{n}}\delta\mu-\frac{-c_{n}(\lambda_{n}+\mu_{n})^2+\overline{c_{-n}}}{2\lambda_{n}},
$$
and, up to $O(\epsilon^2)$ correction:
$$
\delta\mu=\frac{1}{4\mu_n}\left[\big(
c_{n}(\lambda_{n}+\mu_{n})^2-\overline{c_{-n}}\big) -\big(c_{-n}(\lambda_{n}-\mu_{n})^2-\overline{c_{n}}\big)\right],
$$
\begin{equation}
\label{eq:div3}
\delta\lambda_{-n}=\frac{\mu_{n}}{\lambda_{n}}\delta\mu +
\frac{c_{-n}(\lambda_{n}-\mu_{n})^2-\overline{c_{n}}}{2\lambda_{n}}=
\end{equation}
$$
=\frac{1}{4\lambda_n}\left[\big(
c_{n}(\lambda_{n}+\mu_{n})^2-\overline{c_{-n}}\big) +\big(c_{-n}(\lambda_{n}-\mu_{n})^2-\overline{c_{n}}\big)\right]=
$$
$$
=\frac{1}{4\lambda_n}\left[-\big(\overline{c_{-n}}+\overline{c_{n}}\big)+
c_{n}(\lambda_{n}+\mu_{n})^2 + c_{-n}(\lambda_{n}-\mu_{n})^2\right].
$$
\begin{equation}
\label{eq:div4}
p(\gamma_{-n})=\frac{1}{4\mu_n}\left[\big(
c_{n}(\lambda_{n}+\mu_{n})^2-\overline{c_{-n}}\big) -\big(c_{-n}(\lambda_{n}-\mu_{n})^2-\overline{c_{n}}\big)\right]
+O(\epsilon^2)\ \  (\!\!\!\!\mod\pi/L),
\end{equation}
and
\begin{equation}
\label{eq:div_alpha2}
\lambda(\gamma_{-n}) =-\lambda_n-\frac{1}{4\lambda_n}\left[\tilde\alpha_n+\tilde\beta_n\right]+O(\epsilon^2), \ \
p(\gamma_{-n})= \frac{1}{4\mu_n}\left[\tilde\alpha_n-\tilde\beta_n\right]+O(\epsilon^2)
\end{equation}

A generic small perturbation of the constant solution generates an infinite genus spectral curve. Of course, it is
natural to approximate it by a finite-genus curve. The perturbations corresponding to stable resonant points remain 
of order $\epsilon$ for all $t$ and can be well-described by the linear perturbation theory. Therefore we can close 
all gaps associated with stable points, obtaining a finite-gap approximation of the spectral curve. 

Since the first nonlinear stage of modulation instability arose at order $|\log{\epsilon}|$ time,

\section{The spectral curve for a single harmonic perturbation}
\label{sec:sec5}

Assume now that we perturb the constant solution by one-frequency perturbation:
\begin{equation}
\label{pert1}
\epsilon(x) = c_{-n} e^{-i\frac{2\pi}{L}n x}+c_n e^{i\frac{2\pi}{L}n x}, \ \ |c_{-n}|\ll 1, \ \ |c_n|\ll 1.
\end{equation}
Starting from this point, we assume that the perturbation (\ref{pert1}) corresponds to an unstable resonant 
point; then $\lambda^{\pm}_n$ are pure imaginary.

As we discussed in Section~\ref{sec:sec2.2}, the behavior of solutions, at least up to the first appearance, can be described using 
the genus 2 approximation, in which only the 2 gaps associated with the excited harmonic are open. Moreover, if all other non-zero
gaps are stable, this approximation provides a good model for the long-time behavior. This is true, in particular, if $L/\pi>n>L/2\pi$. 

To the leading order of perturbation theory, the curve $\Gamma$ has genus 2 and six branch points $E_0$,  $\overline{E_0}$,
$E_{2n-1}$,  $\overline{E_{2n-1}}$, $E_{2n}$,  $\overline{E_{2n}}$. We use the following enumeration: $E_0=i+O(\epsilon^2)$,   
$E_{1,2}=\lambda_n+O(\epsilon)$, and $\Gamma$ is defined by:
\begin{equation}
\label{eq:curve1}
\mu^2=(\lambda-E_0) (\lambda-E_{2n-1}) (\lambda-E_{2n}) 
(\lambda-\overline{E_0})(\lambda-\overline{E_{2n-1}}) (\lambda-\overline{E_{2n-1}}).
\end{equation}

Let us denote:
$$
\mu^{(0)}(\lambda)=\sqrt{\lambda^2+1},
$$
$$
z_n=\frac{E_{2n-1}+E_{2n}}{2}=\lambda_n+O\left(\epsilon^2 \right), \ \  \mu^{(0)}_n=\mu^{(0)}(\lambda_n)=\sqrt{\lambda_n^2+1}>0.
$$

The natural compactification of $\Gamma$ has two infinite points:
$$
\infty_{+}: \mu\sim-\lambda^3, \ \  \infty_{-}: \mu\sim \lambda^3,
$$
We have the following system of cuts (marked by the dashed lines): $[E_0,i\infty]$, $[\overline{E_0},-i\infty]$, 
$[E_{2n-1},E_{2n}]$, $[\overline{E_{2n-1}},\overline{E_{2n}}]$. We use a slightly non-standard agreement: $\infty_+$ is located on the 
Sheet~2 (dashes blue lines), and  $\infty_{-}$ is located on the Sheet~1 (solid blue lines). 

To calculate finite-gap solutions we need the periods of basic holomorphic differentials and some meromorphic 
differentials, the vector of Riemann constants 
and the Abel transform of the Dirichlet spectrum. Let us calculate them to the leading order.

\subsection{The Riemann matrix}
\label{sec:sec5.1}

To start with, let us fix the following system of cycles: 

\begin{figure}
\centering
\mbox{\epsfxsize=8cm\epsffile{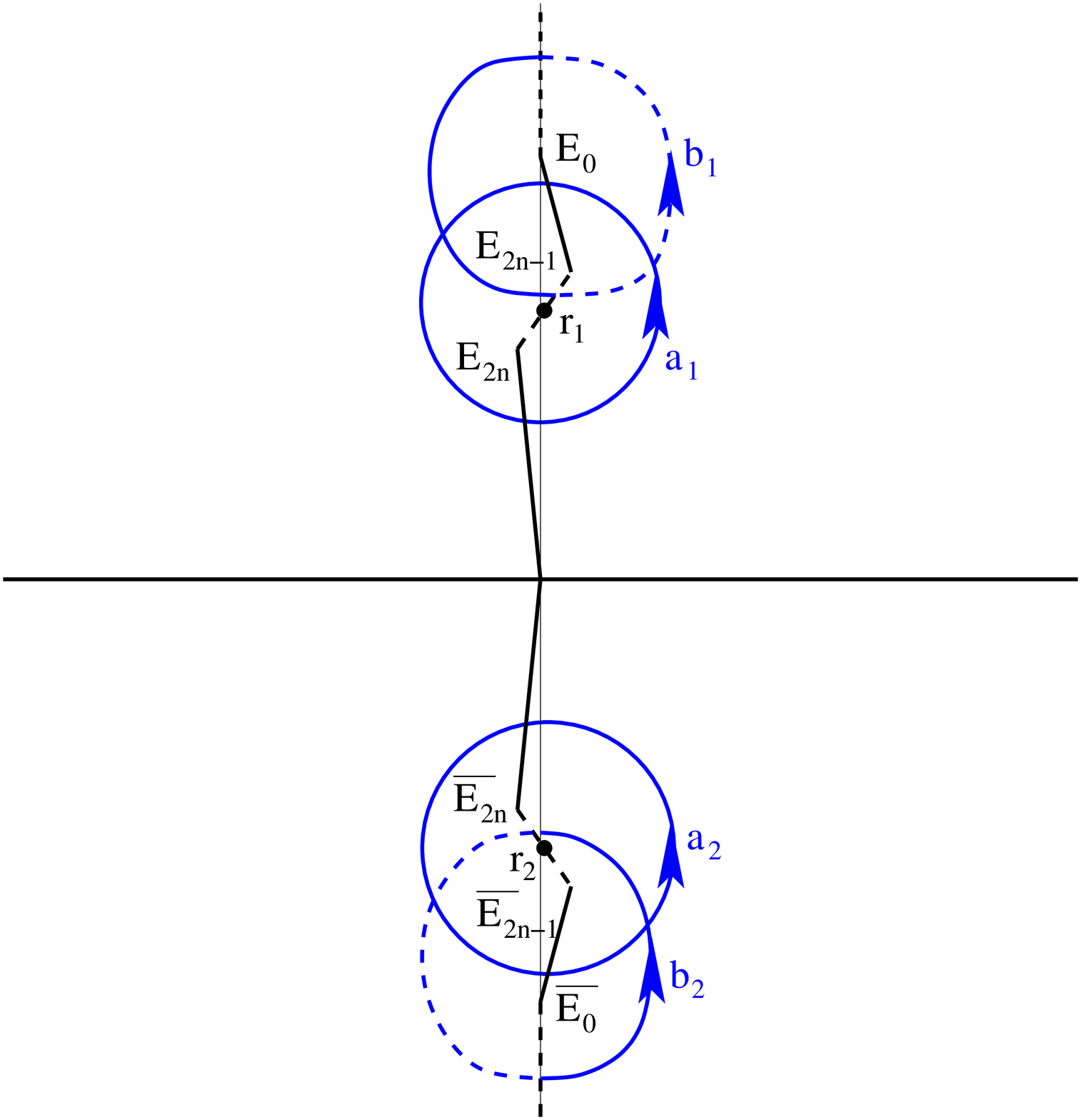}}
\end{figure}

The cycles $a_1$ and $a_2$ are on the Sheet~1.

On the Sheet~1:
\begin{enumerate}
\item If $\Im\lambda>\Im{\overline{E_0}}$, then $0\le\arg\sqrt{\lambda-\overline{E_0}}\le \pi/2$. 
\item If $\Im\lambda<\Im{E_0}$, then $-\pi/2\le\arg\sqrt{\lambda-E_0 }\le 0$.
\item If $|\lambda-z_n |\gg|E_{2n}-E_{2n-1}|$, then $\sqrt{(\lambda-E_{2n-1})(\lambda-E_{2n})}\sim \lambda-z_n$.
\item If $|\lambda-\overline{z_n}|\gg|E_{2n}-E_{2n-1}|$, then $\sqrt{(\lambda-\overline{E_{2n-1}})(\lambda-\overline{E_{2n}})}\sim \lambda-\overline{z_n}$. 
\end{enumerate}
Therefore on the cycle $a_1$: 
$$
\mu\sim \mu^{(0)}_n (z_n-\overline{z_n})  (\lambda-z_n).
$$
The basic holomorphic differentials are defined by the conditions:
\begin{equation}
\label{eq:norm1}
\oint\limits_{a_m} \omega_l =\delta_{ml},
\end{equation}
therefore
$$
\omega_1 =\frac{\mu^{(0)}_n}{2\pi i}\frac{\lambda-\overline{r_1}}{\mu}d\lambda, \ \ 
\omega_2 =\frac{\mu^{(0)}_n}{2\pi i}\frac{\lambda-r_1}{\mu}d\lambda,
$$
where the coefficients $\frac{\mu^{(0)}_n}{2\pi i}$  are {\bf exact}, and
$$
r_1=z_n-\delta_1, 
$$
where $\delta_1$ is computed in the following way.
On the cycle $a_1$:
$$
\omega_1 \sim \frac{1}{2\pi i}\frac{z_n-\overline{z_n}}{ (z_n-\overline{z_n})  (\lambda-z_n) }d\lambda\sim
\frac{1}{2\pi i}\frac{d\lambda }{\lambda-z_n }, 
$$
$$
\omega_2 \sim \frac{1}{2\pi i}\frac{(\lambda-z_n)}{(\lambda-\overline{z_n})(\lambda-z_n)} d\lambda
\sim \frac{1}{2\pi i (\lambda-\overline{z_n}) } d\lambda.
$$
Let us calculate $\delta_1$. We us the following standard formulas (for $R_0> |E|$):
\begin{align*}
\oint\limits_{|\nu|=R_0} \frac{d\nu}{\sqrt{\nu^2-E^2}}&= 2\pi i, \\ 
\oint\limits_{|\nu|=R_0} \frac{\nu d\nu}{\sqrt{\nu^2-E^2}}&= 0, \\
\oint\limits_{|\nu|=R_0} \frac{\nu^2 d\nu}{\sqrt{\nu^2-E^2}}&= \pi i E^2. \\
\end{align*}
Let us denote:
$$
\lambda=\nu+z_n
$$
$$
\tilde\sigma_1=\frac{1}{z_n-E_0}+\frac{1}{z_n-\overline{E_0}}+\frac{1}{z_n-\overline{E_{2n-1}}}+\frac{1}{z_n-\overline{E_{2n}}}
$$
$$
\tilde\sigma_2=\frac{1}{(z_n-E_0)(z_n-\overline{E_0})} +\frac{1}{(z_n-E_0)(z_n-\overline{E_{2n-1}})}+\frac{1}{(z_n-E_0)(z_n-\overline{E_{2n}})}+
$$
$$
+\frac{1}{(z_n-\overline{E_0})(z_n-\overline{E_{2n-1}})} + \frac{1}{(z_n-\overline{E_0})(z_n-\overline{E_{2n}})} +
\frac{1}{(z_n-\overline{E_{2n-1}})(z_n-\overline{E_{2n}})} 
$$
$$
\tilde\mu= \sqrt{(\lambda-E_0)(\lambda-\overline{E_0})(\lambda-\overline{E_{2n-1}}) (\lambda-\overline{E_{2n}})}
$$
$$
\mu=\tilde\mu\sqrt{(\nu+(z_n-E_{2n-1})) (\nu-(z_n-E_{2n-1}))}
$$
$$
\tilde\mu=\mu^{(0)}_n (z_n-\overline{z_n})\left[1+\tilde\sigma_1\nu+\tilde\sigma_2\nu^2+O(\nu^3)\right],
$$
$$
\frac{1}{\tilde\mu}=\frac{1}{\mu^{(0)}_n (z_n-\overline{z_n})  } \left[1-\frac{1}{2}\tilde\sigma_1\nu+\left[\frac{3}{8}\tilde\sigma_1^2 
-\frac{1}{2}\tilde\sigma_2\right]\nu^2+O(\nu^3)\right],
$$
At $a_1$:
$$
\omega_2 =\frac{\mu^{(0)}_n}{2\pi \mu^{(0)}_n (z_n-\overline{z_n})} 
\frac{\left[1-\frac{1}{2}\tilde\sigma_1\nu+O(\nu^2)\right][\nu+\delta_1+O(\nu^3)]}{\sqrt{(\nu-(E_{2n-1}-z_n))( \nu-(E_{2n}-z_n))}}d\nu =
$$
$$
=\frac{\left[\delta_1+\nu-\frac{1}{2}\tilde\sigma_1\nu^2   +O(\nu^3)\right]}{(z_n-\overline{z_n})}
\frac{d\nu}{2\pi i \sqrt{(\nu-(E_{2n-1}-z_n))( \nu-(E_{2n}-z_n))}}
$$
$$
0=(z_n-\overline{z_n}) \oint\limits_{a_1} \omega_2=\delta_1-\frac{1}{16}\tilde\sigma_1\left(E_{2n}-E_{2n-1}\right)^2+O(c^3).
$$
Therefore
$$
\delta_1=\frac{1}{16}\tilde\sigma_1\left(E_{2n}-E_{2n-1}\right)^2+O(c^3).
$$

On the cycles $a_1$, $b_1$ we have: 
$$
\frac{\lambda-\overline{r_1}}{\sqrt{(\lambda-\overline{E_{2n-1}}) (\lambda-\overline{E_{2n}})}}=1+
\frac{\overline{\delta_1}}{\lambda-\overline{z_n}}
+\frac{1}{8}\frac{\left(\overline{E_{2n}}-\overline{E_{2n-1}} \right)^2}{(\lambda-\overline{z_n})^2}+O\left(E_{2n}-E_{2n-1}\right)^4,
$$
therefore
$$
\oint\limits_{a_1} \omega_1=\left[1+O\left(E_{2n}-E_{2n-1}\right)^2\right] \cdot\frac{\mu^{(0)}_n}{2\pi i}\cdot
\oint\limits_{a_1} \frac{d\lambda}{ \sqrt{(\lambda-E_0)(\lambda-E_{2n-1})(\lambda-E_{2n}) (\lambda-\overline{E_0})}},
$$
$$
\oint\limits_{b_1} \omega_1=\left[1+O\left(E_{2n}-E_{2n-1}\right)^2\right] \cdot\frac{\mu^{(0)}_n}{2\pi i}\cdot
\oint\limits_{b_1} \frac{d\lambda}{ \sqrt{(\lambda-E_0)(\lambda-E_{2n-1})(\lambda-E_{2n}) (\lambda-\overline{E_0})}},
$$
and 
$$
b_{11}=\left[1+O\left(\epsilon^2\right)\right] \hat b ,
$$
where $\hat b$ is the $b$-period of the elliptic curve:
$$
\hat\mu^2=(\lambda-E_0)(\lambda-E_{2n-1})(\lambda-E_{2n}) (\lambda-\overline{E_0}).
$$
From \cite{AS1972} the answer is known and it is expressed in terms of the projective cross-ratio
$d$:
$$
d=\frac{(E_{2n-1}-E_0)}{(E_{2n-1}-\overline{E_0})} \frac{(E_{2n}-\overline{E_0})} {(E_{2n}-E_0)} ,
$$
$$
m=\left(\frac{\sqrt{d}-1}{\sqrt{d}+1}\right)^2 .
$$
We can also write:
$$
\sqrt{m}=\frac{(E_{2n-1}-E_{2n})}{2i\left(\mu^{(0)}_n\right)^2}+O(\epsilon^2) .
$$
The sign of the square root was chosen to comply the following rule: if $E_{2n-1}$, $E_{2n}$ are pure imaginary and 
$\Im E_{2n-1}>\Im E_{2n}$, then $b_{11}$ is pure imaginary. Therefore
$$
\sqrt{m}=\frac{i\sqrt{\alpha_n\beta_n}}{2 \lambda_n^{0}\left(\mu^{(0)}_n\right)^2}+O(\epsilon^2).
$$
\begin{equation}
\label{eq:riemann1}
b_{11}\sim\hat b=\frac{1}{2\pi i} \log\left(\frac{m}{16}+ 8\left(\frac{m}{16} \right)^2 +\ldots \right).
\end{equation}
We have the following approximation:
$$
\exp{\left(\pi i b_{11}\right)}=\frac{\sqrt{m}}{4}+O(\epsilon^2)=
\frac{i\sqrt{\alpha_n\beta_n}}{8 \lambda_n^{0}\left(\mu^{(0)}_n\right)^2}+O(\epsilon^2).
$$

To calculate $b_{21}$ we use the following formula: for a regular function $f(\lambda)$
$$
\int\limits_{E}^1\frac{\lambda f(\lambda)d\lambda}{\sqrt{\lambda^2-E^2}}=
\int\limits_{0}^1 f(\lambda)d\lambda + O\left(E^2 \right)
$$
$$
b_{21}=\frac{2\mu^{(0)}_n}{2\pi i}\int\limits_{i \Im z_n}^{i} \frac{d\lambda}{\sqrt{(\lambda^2+1)}\ \ (\lambda+i\Im z_n)}+
O\left(\log(\epsilon)\cdot (\epsilon)^2\right),
$$
where  $\sqrt{(\lambda^2+1)}<0$ (dashed line on the contour). Equivalently,
\begin{equation}
\label{eq:riemann2}
b_{12}=b_{21}=\frac{-i}{\pi}\log(\Im\lambda_n)+O(\log(\epsilon)\cdot \left(\epsilon)^2\right)=
\end{equation}
The Riemann matrix is symmetric, therefore  
\begin{equation}
\label{eq:riemann3}
b_{12}=b_{21}.
\end{equation}
It is easy to check that 
\begin{equation}
\label{eq:riemann4}
b_{22}=-\overline{b_{11}}.
\end{equation}

\subsection{The vector of Riemann constants}
\label{sec:sec5.2}

In the calculation of the Riemann constants we use the following basis of cycles (here the orientations of $a_2$, $b_2$ are 
the opposite with respect to the previous pictures. But this does not affect the answer, modulo periods):

\begin{figure}
\centering
\mbox{\epsfxsize=8cm\epsffile{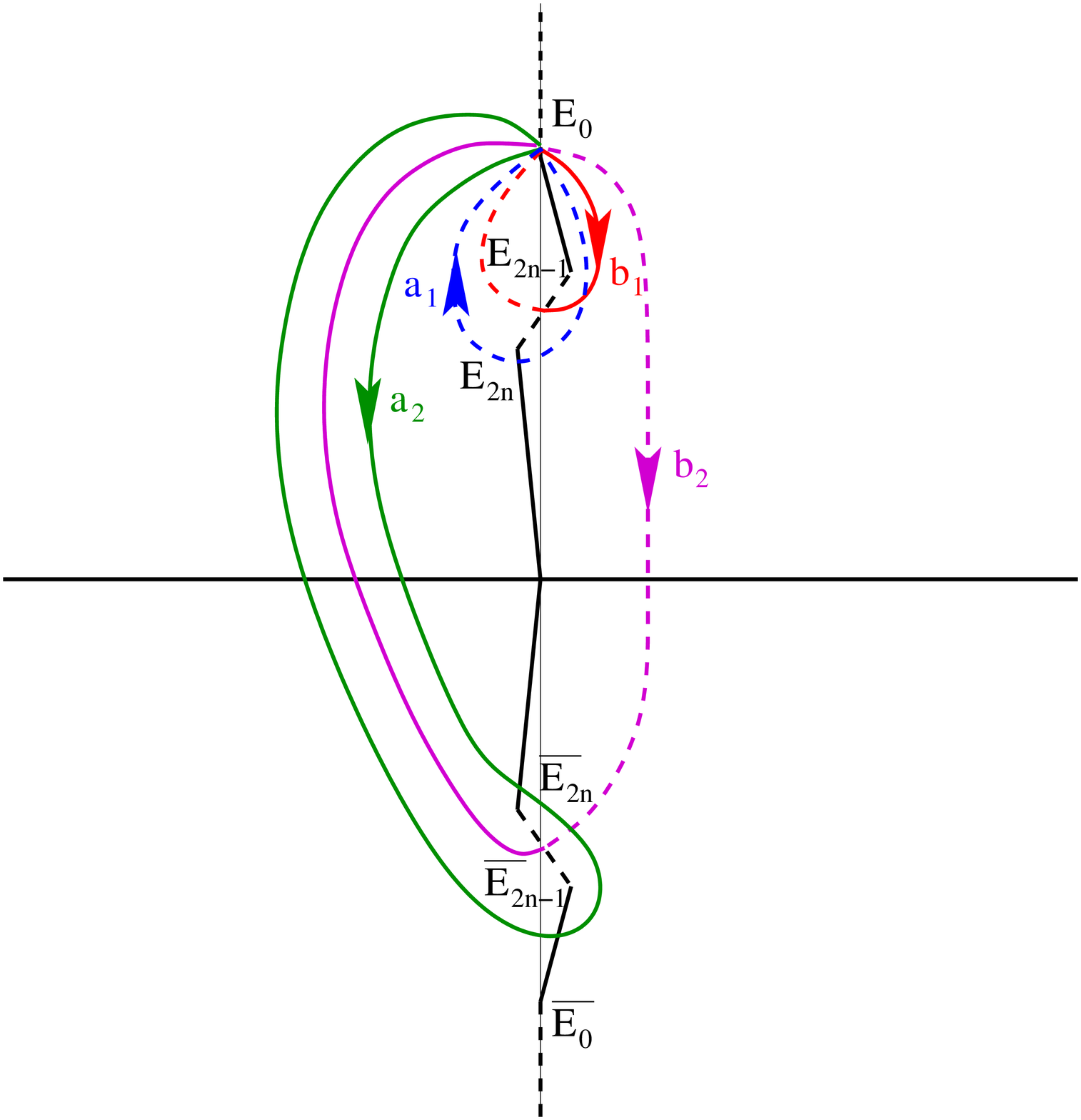}}
\end{figure}

Here we assume that $E_0$ is the starting point. Consider the standard octagon:
\begin{figure}
\centering
\mbox{\epsfxsize=6cm\epsffile{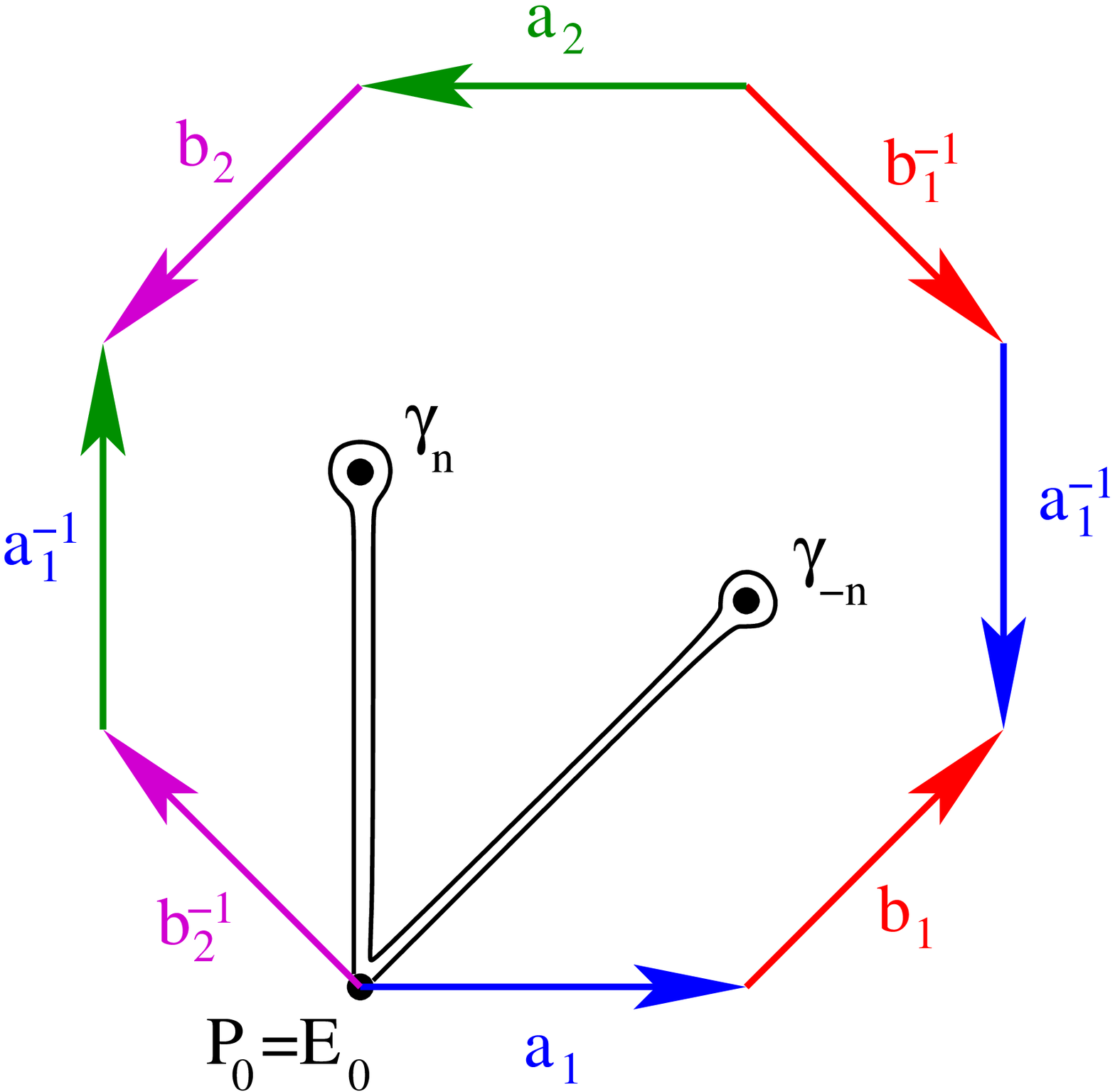}}
\end{figure}

Let us denote 
$$
\vec a_1=\left[\begin{array}{c} 1 \\ 0 \end{array}\right], \ \ \vec a_2=\left[\begin{array}{c} 0 \\ 1 \end{array}\right], \ \
\vec b_1=\left[\begin{array}{c} b_{11} \\ b_{12} \end{array}\right], \ \ \vec b_2=\left[\begin{array}{c} b_{21} \\ b_{22} \end{array}\right].
$$

The vector of Rienann constants is defined by (see \cite{Mum1983}):
$$
\vec K = \left[\begin{array}{c} K_1 \\ K_2 \end{array}\right]
$$
$$
K_1 = \frac{b_{11}}{2} + \oint\limits_{a_1}  \omega_1(\gamma) A_1(\gamma) +
\oint\limits_{a_2} \omega_1(\gamma) A_2(\gamma),
$$
$$
K_2 = \frac{b_{22}}{2} + \oint\limits_{a_1} \omega_2(\gamma) A_1(\gamma) +
\oint\limits_{a_2} \omega_2(\gamma) A_2(\gamma).
$$
Here the cycles $a_1$, $a_2$ start from the point $E_0$! Therefore 
$$
\oint\limits_{a_2} \omega_1(\gamma) A_2(\gamma) + \oint\limits_{a_2} \omega_2(\gamma) A_1(\gamma)= \oint\limits_{a_2} d(A_1(\gamma) 
A_2(\gamma))=\left. A_1(\gamma)A_2(\gamma)\vphantom{\int}\right|_{\gamma=E_0}^{\gamma=E_0+a_2}=
$$
$$
=A_1(E_0+a_2)A_2(E_0+a_2)-A_1(E_0)A_2(E_0)= 0\cdot 1-0\cdot 0=0 ,
$$
having used (\ref{eq:norm1}), and
\begin{align}
A_1(E_0) &=0,&  A_1(E_0+a_1) &=1,&  A_1(E_0+a_2) &=0,  \\
A_2(E_0) &=0,&  A_2(E_0+a_1) &=0,&  A_2(E_0+a_2) &=1.\nonumber
\end{align}
Therefore:
$$
\oint\limits_{a_2} \omega_1(\gamma) A_2(\gamma) =- \oint\limits_{a_2} \omega_2(\gamma) A_1(\gamma),
$$
$$
\oint\limits_{a_1} \omega_2(\gamma) A_1(\gamma) =- \oint\limits_{a_1} \omega_1(\gamma) A_2(\gamma),
$$
and
$$
K_1 = \frac{b_{11}}{2} + \oint\limits_{a_1}  \omega_1(\gamma) A_1(\gamma) -
\oint\limits_{a_2} \omega_2(\gamma) A_1(\gamma),
$$
$$
K_2 = \frac{b_{22}}{2} - \oint\limits_{a_1} \omega_2(\gamma) A_1(\gamma) +
\oint\limits_{a_2} \omega_2(\gamma) A_2(\gamma).
$$
The last formulas are more convenient. In addition:
$$
\oint\limits_{a_1}  \omega_1(\gamma) A_1(\gamma)=\frac{1}{2}\oint\limits_{a_1}d(A_1^2(\gamma))=\frac{1}{2}[A_1^2(E_0+a_1)-A_1^2(E_0)]=
\frac{1}{2}[1^2-0^2]=\frac{1}{2},
$$
$$
\oint\limits_{a_2}  \omega_2(\gamma) A_2(\gamma)=\frac{1}{2}\oint\limits_{a_2}d(A_2^2(\gamma))=\frac{1}{2}[A_2^2(E_0+a_2)-A_2^2(E_0)]=
\frac{1}{2}[1^2-0^2]=\frac{1}{2}.
$$
Therefore we obtain
$$
K_1 = \frac{b_{11}}{2} + \frac{1}{2} - \oint\limits_{a_2} \omega_2(\gamma) A_1(\gamma),
$$
$$
K_2 = \frac{b_{22}}{2} - \oint\limits_{a_1} \omega_1(\gamma) A_2(\gamma) +\frac{1}{2}.
$$

To complete the calculation we use the following fact: on both sides of the path connecting $E_0$ with $E_{2n-1}$ the function
$\omega_1(\gamma)A_2(\gamma)$ has the same values; therefore the answer does not depend on the initial point of $a_1$, and
$a_1$ can be replaced by any homotopic contour. Analogously, on both sides of the path connecting $E_0$ with 
$\overline {E_{2n}}$, the function $\omega_2(\gamma)A_1(\gamma)$ has the same values; therefore the answer does not 
depend on the initial point of $a_2$, and $a_2$ can be replaced by any homotopic contour. Let us remark that 
$$
A_2(E_{2n-1})=A_2(E_{2n}), \ \ A_1(\overline{E_{2n-1}})=A_1(\overline{E_{2n}}),
$$ 
and, on the cycles around the intervals $[E_{2n-1},E_{2n}]$ and $[\overline{E_{2n-1}},\overline{E_{2n}}]$, one can write respectively:
$$
A_2(\gamma)=A_2(E_{2n-1})+\tilde A_2(\gamma), \ \ A_1(\gamma)=A_1(\overline{E_{2n}})+\tilde A_1(\gamma),
$$
where $\tilde A_2(\gamma)$, $\tilde A_1(\gamma)$ have opposite values on opposite sides of the cuts. Therefore
$$
\oint\limits_{a_2} \omega_2(\gamma) \tilde A_1(\gamma)=  \oint\limits_{a_1} \omega_1(\gamma) \tilde A_2(\gamma)=0,
$$

\begin{figure}
\centering
\mbox{\epsfxsize=14cm\epsffile{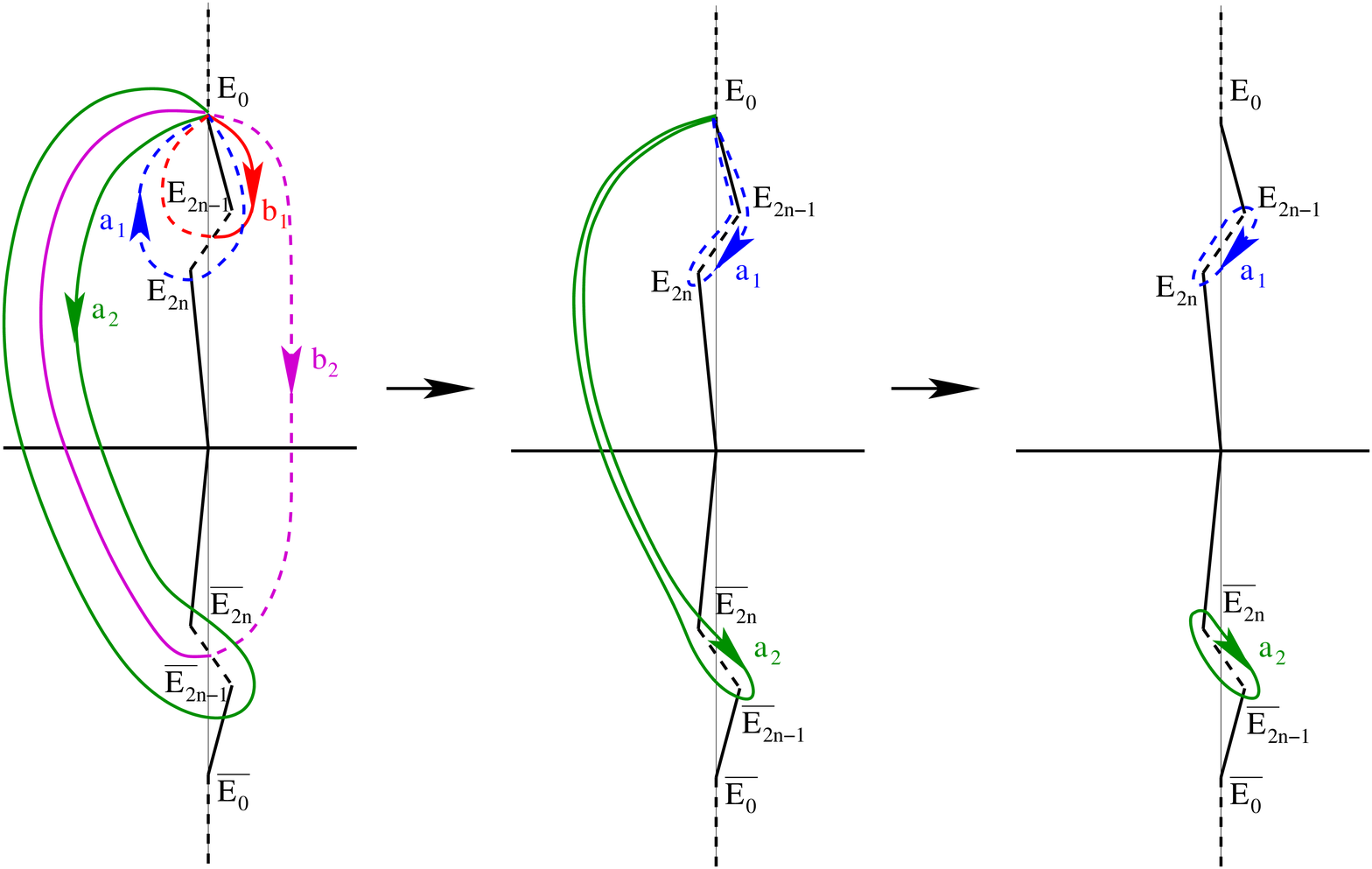}}
\end{figure}
and
$$
K_1 = \frac{b_{11}}{2} + \frac{1}{2} - \oint\limits_{a_2} \omega_2(\gamma) A_1(\overline{E_{2n}}),
$$
$$
K_2 = \frac{b_{22}}{2} - \oint\limits_{a_1} \omega_1(\gamma) A_2(E_{2n-1}) +\frac{1}{2};
$$
therefore
$$
K_1 = \frac{b_{11}}{2} + \frac{1}{2} - A_1(\overline{E_{2n}}),
$$
$$
K_2 = \frac{b_{22}}{2} +\frac{1}{2}- A_2(E_{2n-1}).
$$
It is easy to check that:
$$
A_1(\overline{E_{2n}})=-\frac{b_{12}}{2}+\frac{1}{2}, \ \ A_2(E_{2n-1})=-\frac{b_{21}}{2}.
$$
Finally we obtain:
\begin{equation}
\label{eq:rc1}
K_1 = \frac{b_{11}}{2} + \frac{b_{12}}{2},
\end{equation}
\begin{equation}
\label{eq:rc2}
K_2 = \frac{b_{22}}{2} + \frac{b_{21}}{2}+ \frac{1}{2} .
\end{equation}

\subsection{Periods of some meromorphic differentials}
\label{sec:sec5.3}

The formulas for finite-gap solutions include the following meromorphic differentials:
\begin{enumerate}
\item A 3-rd kind meromorphic differential $\Omega$ with zero $a$-periods and first order poles at $\infty_+$,  
$\infty_-$:
\begin{align*}
\Omega &= -\frac{d\lambda}{\lambda}+O(1)& &\mbox{at} \ \ \infty_+\\
\Omega &= \frac{d\lambda}{\lambda}+O(1)& &\mbox{at} \ \ \infty_-\\ 
\oint\limits_{a_1}\Omega&=\oint\limits_{a_2}\Omega =0.&
\end{align*}    
From the Riemann bilinear relations (see \cite{Spr1957}), it follows that:
$$
A_j(\infty_-)-A_j(\infty_+)=Z_j=\frac{1}{2\pi i}\oint\limits_{b_j}\Omega.
$$
Again we assume $E_0$ to be the starting point of the Abel transform. Then:
$$
\vec A(\infty_+)= - \vec A(\infty_-).
$$
Let  us calculate $\vec A(\infty_-)$ up to $O(\epsilon^2)$ corrections. On the
contour connecting $E_0$ with the infinite points we have
$$
\omega_1 =\frac{\mu^{(0)}_n}{2\pi i}\frac{\lambda-\overline{r_1}}{\mu}d\lambda=
\frac{\mu^{(0)}_n}{2\pi i}\frac{d\lambda}{\mu^{(0)}(\lambda)\cdot(\lambda-\lambda_n)} + 
O(\epsilon)^2,
$$
$$
\omega_2 =\frac{\mu^{(0)}_n}{2\pi i}\frac{\lambda-r_1}{\mu}d\lambda=
\frac{\mu^{(0)}_n}{2\pi i}\frac{d\lambda}{\mu^{(0)}(\lambda)\cdot(\lambda+\lambda_n)} + O(\epsilon)^2.
$$
We know that
$$
\int\frac{d\lambda}{\mu^{(0)}(\lambda)\cdot(\lambda-\lambda_n)}=
\frac{1}{\mu^{(0)}_n}
\log\left[\frac{\mu^{(0)}(\lambda)\mu^{(0)}_n-\lambda\lambda_n-1}{\lambda-\lambda_n} \right],
$$
$$
\int\frac{d\lambda}{\mu^{(0)}(\lambda)\cdot(\lambda+\lambda_n)}=
\frac{1}{\mu^{(0)}_n}
\log\left[\frac{\mu^{(0)}(\lambda)\mu^{(0)}_n+\lambda\lambda_n-1}{\lambda+\lambda_n} \right].
$$
At $\infty_-$ we have $\mu^{(0)}(\lambda)=\lambda+O(1/\lambda)$; therefore, up to $O(\epsilon^2)$ correction, we have:
$$
A_1(\infty_-)=\frac{1}{2\pi i} \left\{\log\left[\mu^{(0)}_n-\lambda_n \right]-
\log\left[\frac{-i\lambda_n-1}{i-\lambda_n} \right]\right\}=-\frac{1}{4}+
\frac{\arg(\mu^{(0)}_n-\lambda_n)}{2\pi}=
$$
$$
=-\frac{1}{4}-\frac{\arg(\mu^{(0)}_n+\lambda_n)}{2\pi}.
$$
Analogously, up to $O(\epsilon^2)$ correction:
$$
A_2(\infty_-)=-\frac{1}{4}+\frac{\arg(\mu^{(0)}_n+\lambda_n)}{2\pi}.
$$
Finally we obtain:
\begin{align}
\label{eq:omega_per1}
\vec A(\infty_-) &=  \left[\begin{array}{c} 
-\frac{1}{4}-\frac{\arg(\mu^{(0)}_n+\lambda_n)}{2\pi} \\ 
-\frac{1}{4}+\frac{\arg(\mu^{(0)}_n+\lambda_n)}{2\pi}  \end{array}\right] 
+ O(\epsilon^2), \\
\vec A(\infty_+) &=  \left[\begin{array}{c} 
\frac{1}{4}+\frac{\arg(\mu^{(0)}_n+\lambda_n)}{2\pi} \\ 
\frac{1}{4}-\frac{\arg(\mu^{(0)}_n+\lambda_n)}{2\pi}  \end{array}\right]
+ O(\epsilon^2), \nonumber \\
\vec Z &=  \left[\begin{array}{c} 
-\frac{1}{2}-\frac{\arg(\mu^{(0)}_n+\lambda_n)}{\pi} \\ 
-\frac{1}{2}+\frac{\arg(\mu^{(0)}_n+\lambda_n)}{\pi}  \end{array}\right]
+ O(\epsilon^2),\nonumber \\
\vec A(\infty_+) &= -\vec A(\infty_-), \nonumber \\ 
\vec Z &= \vec A(\infty_-)- \vec A(\infty_+). \nonumber
\end{align}

\item A 2-nd kind meromorphic differential $dp$ such that
\begin{align*}
dp &= -d\lambda +O(1)& &\mbox{at} \ \ \infty_+\\
dp &=  d\lambda +O(1)& &\mbox{at} \ \ \infty_-\\ 
\oint\limits_{a_1}dp &=\oint\limits_{a_2}dp =0.&
\end{align*} 
It is convenient to use the Riemann bilinear relations:
$$
\oint\limits_{a_j}dp=-2\pi i \left[\res\nolimits_{\infty_+} [p\,\omega_j]+ \res\nolimits_{\infty_-} [p\, \omega_j]  \right].
$$
\begin{equation}
\label{eq:dp_per1}
\oint\limits_{a_1}dp=-2\pi i\frac{\mu^{(0)}_n}{2\pi i} \left[ \res\nolimits_{\infty_+} 
\left[-\frac{\lambda(\lambda-\overline{r_1})}{\mu}d\lambda\right] + \res\nolimits_{\infty_-} 
\left[\frac{\lambda(\lambda-\overline{r_1})}{\mu}d\lambda\right]  \right]=2\mu^{(0)}_n,
\end{equation}
\begin{equation}
\label{eq:dp_per2}
\oint\limits_{a_2}dp=-2\pi i\frac{\mu^{(0)}_n}{2\pi i} \left[ \res\nolimits_{\infty_+}\left[ 
-\frac{\lambda(\lambda-r_1)}{\mu}d\lambda\right] + \res\nolimits_{\infty_-} 
\left[\frac{\lambda(\lambda-r_1)}{\mu}d\lambda\right]  \right]=2\mu^{(0)}_n,
\end{equation}
Let us point out that formulas (\ref{eq:dp_per1}), (\ref{eq:dp_per2}) are exact.

\item A 2-nd kind meromorphic differential $dq$ such that
\begin{align*}
dq &= -d(\lambda)^2 +O(1)& &\mbox{at} \ \ \infty_+\\
dq &=  d(\lambda)^2 +O(1)& &\mbox{at} \ \ \infty_-\\ 
\int\limits_{a_1}dq &=\int\limits_{a_2}dq =0.&
\end{align*} 
Again it is convenient to use the Riemann bilinear relations:
$$
\oint\limits_{a_j}dq=-2\pi i \left[\res\nolimits_{\infty_+} [q\, \omega_j]+ \res\nolimits_{\infty_-} [q\, \omega_j]  \right].
$$
$$
\oint\limits_{a_1}dq=-2\pi i\frac{\mu^{(0)}_n}{2\pi i} \left[ \res\nolimits_{\infty_+} 
\left[-\frac{\lambda^2(\lambda-\overline{r_1})}{\mu}d\lambda\right] + \res\nolimits_{\infty_-} 
\left[\frac{\lambda^2(\lambda-\overline{r_1})}{\mu}d\lambda\right]  \right],
$$
$$
\oint\limits_{a_2}dq=-2\pi i\frac{\mu^{(0)}_n}{2\pi i} \left[ \res\nolimits_{\infty_+} 
\left[-\frac{\lambda^2(\lambda-r_1)}{\mu}d\lambda\right] + \res\nolimits_{\infty_-} 
\left[\frac{\lambda^2(\lambda-r_1)}{\mu}d\lambda\right]  \right].
$$
Taking into account that, for $\lambda\rightarrow\infty_{\pm}$,
$$
\mu = \mp\lambda^3 + O(\epsilon^2)\lambda^2 + O(\lambda),
$$
we obtain
\begin{equation}
\label{eq:dq_per1}
\oint\limits_{a_1}dq=2\lambda_n \mu^{(0)}_n +O(\epsilon^2),
\end{equation}
\begin{equation}
\label{eq:dq_per2}
\oint\limits_{a_2}dq=-2\lambda_n \mu^{(0)}_n +O(\epsilon^2).
\end{equation}
\end{enumerate}

Let us also provide formulas for the differentials: $\Omega$, $dp$, $dq$.
Let us define the following differentials on $\Gamma$:
$$
\Omega_{j}=\frac{\lambda^j (\lambda-\rr_{j})(\lambda-\overline{\rr_{j}})   d\lambda}{\mu}
$$
where the zeroes $\rr_{j}$,  $\overline{\rr_{j}}$ are determined by the normalization condition: 
$$
\int\limits_{a_1} \Omega_{j}=\int\limits_{a_2} \Omega_{j}=0.
$$
We see that
$$
\rr_{j}=z_n+O(\epsilon^2).
$$
It is easy to check that
$$
\Omega =\Omega_0, \ \ dp = \Omega_1 + O(\epsilon^2)\Omega_0, \ \ 
dq = 2 \Omega_2 + O(\epsilon^2)\Omega_1 + (1+O(\epsilon^2))\Omega_0.
$$

Let us write the corresponding formulas for the unperturbed curve $\Gamma_0: \mu^{(0)}(\lambda)=\sqrt{\lambda^2+1}$. 
The corresponding set of basic differentials has the following form:
$$
\Omega^{(0)}_{j}=\frac{\lambda^j d\lambda}{\mu^{(0)}} .
$$
We have:
\begin{equation}
\label{eq:integr1}
\int\limits_{i}^{\gamma} \Omega^{(0)}_{0}=\log\left[\frac{\lambda(\gamma)+\mu^{(0)}(\gamma)}{i}\right],
\end{equation}
\begin{equation}
\label{eq:integr2}
\int\limits_{i}^{\gamma}d\Omega^{(0)}_{1}=\mu^{(0)}(\gamma),
\end{equation}
\begin{equation}
\label{eq:integr3}
\int\limits_{i}^{\gamma} \Omega^{(0)}_{2}=\frac{\lambda(\gamma)\mu^{(0)}(\gamma)}{2} -
\frac{1}{2}\int\limits_{i}^{\gamma}d\Omega^{(0)}_{0} = 
\frac{\lambda(\gamma)\mu^{(0)}(\gamma)}{2} -\log\left[\frac{\lambda(\gamma)+\mu^{(0)}(\gamma)}{i}\right],
\end{equation}
$$
p^{(0)}(\gamma)= \mu^{(0)}(\gamma), \ \ q^{(0)}(\gamma)= \lambda(\gamma)\mu^{(0)}(\gamma),
$$
$$
d p^{(0)} =  \Omega^{(0)}_{1}, \ \ d q^{(0)} = 2 \Omega^{(0)}_{2} + \Omega^{(0)}_{0}
$$

\subsection{Abel transform of the divisor}
\label{sec:sec5.4}

To calculate the Abel transform of the divisor point we use the following representation:
$$
dp=\frac{(\lambda-{\cal P}_0)(\lambda-{\cal P}_1)(\lambda-\overline{{\cal P}_1})d\lambda}{\mu}
$$
$$
{\cal P}_0=O(\epsilon^2), \ \ {\cal P}_1=\lambda_0+O(\epsilon^2).
$$

Near the point $\lambda_n$:
$$
dp=\left[\frac{\lambda_n}{\mu^{(0)}_j}+O(\epsilon^2)\right] 
\frac{(\lambda-{\cal P}_1) d\lambda}{\sqrt{(\lambda-E_{2n-1})(\lambda-E_{2n})}}=
\left[\frac{\lambda_n}{\mu^{(0)}_n}+O(\epsilon^2)\right] d\sqrt{(\lambda-E_{2n-1})(\lambda-E_{2n})}, 
$$
and
$$
p=\left[\frac{\lambda_n}{\mu^{(0)}_n}+O(\epsilon^2)\right] \sqrt{(\lambda-E_{2n-1})(\lambda-E_{2n})}, \ \  \!\!\!\!\mod\pi/L.
$$
Therefore, if $\gamma_n$ denotes the divisor point near $\lambda_n$, then, from (\ref{eq:div1}), (\ref{eq:div2}), we obtain:
$$
\lambda(\gamma_n)-\lambda_n=\frac{1}{4\lambda_n}\left[\alpha_n+\beta_n\right]+O(\epsilon^2),
$$ 
$$
\sqrt{(\lambda(\gamma_n)-E_{2n-1})(\lambda(\gamma_n)-E_{2n})}  =\frac{1}{4\lambda_n}\left[\alpha_n-\beta_n\right]
+O(\epsilon^2).
$$
Let us expand the holomorphic differentials $\omega_1$, $\omega_2$ near this point.
$$
\omega_1 =\frac{\mu^{(0)}_n}{2\pi i}\frac{\lambda-\overline{r_1}}{\mu}d\lambda=
\frac{\mu^{(0)}_n}{2\pi i} \frac{1}{\sqrt{\lambda^2+1}} 
\frac{\lambda-\overline{r_1}}{\sqrt{(\lambda-\overline{E_{2n-1}})(\lambda-\overline{E_{2n}})}} 
\frac{d\lambda}{\sqrt{(\lambda-E_{2n-1})(\lambda-E_{2n})} } + O(\epsilon^2)=
$$
$$
=\frac{\mu^{(0)}_n}{2\pi i} \frac{1}{\sqrt{\lambda^2+1}} 
\frac{d\lambda}{\sqrt{(\lambda-E_{2n-1})(\lambda-E_{2n})} } + O(\epsilon^2)=
$$
$$
=\frac{\mu^{(0)}_n}{2\pi i} \frac{1}{\sqrt{(\mu^{(0)}_n)^2+2(\lambda-\lambda_n) \lambda_n+(\lambda-\lambda_n)^2)}}
\frac{d\lambda}{\sqrt{(\lambda-E_{2n-1})(\lambda-E_{2n})} } + O(\epsilon^2)=
$$
$$
=\frac{\mu^{(0)}_n}{2\pi i} \frac{1}{\mu^{(0)}_n}\left[1-\frac{\lambda_n}{(\mu^{(0)}_n)^2}(\lambda-\lambda_n) \right]
\frac{d\lambda}{\sqrt{(\lambda-E_{2n-1})(\lambda-E_{2n})} } + O(\epsilon^2)+O((\lambda-\lambda_n)^2) =
$$
$$
=\frac{1}{2\pi i}\frac{d\lambda}{\sqrt{(\lambda-E_{2n-1})(\lambda-E_{2n})} }-
\frac{1}{2\pi i}\frac{\lambda_n}{(\mu^{(0)}_n)^2}
d\sqrt{(\lambda-E_{2n-1})(\lambda-E_{2n})} + O(\epsilon^2)+O((\lambda-\lambda_n)^2)
$$
$$
\omega_2=\frac{\mu_0}{4\pi i \lambda_n^2}dp(1+O(\epsilon))\sim \frac{1}{4\pi i\lambda_n}d\sqrt{(\lambda-E_{2n-1})(\lambda-E_{2n})}
$$
Therefore:
$$
\int \omega_1= \frac{1}{2\pi i}\log{\left[(\lambda-\lambda_n)+
\sqrt{(\lambda-E_{2n-1})(\lambda-E_{2n})}\right]} -\frac{1}{2\pi i}\frac{\lambda_n}{(\mu^{(0)}_n)^2}
\sqrt{(\lambda-E_{2n-1})(\lambda-E_{2n})}   + O(\epsilon^2) 
$$
$$
\int \omega_2=\frac{1}{4\pi i\lambda_n} \sqrt{(\lambda-E_{2n-1})(\lambda-E_{2n})} + O(\epsilon^2),
$$
and we obtain:
$$
A_{1\,(E_{2n-1})}(\gamma_n)=\frac{1}{2\pi i}\log\left[\frac{(\lambda(\gamma_n)-\lambda_n)+
\sqrt{(\lambda(\gamma_n)-E_{2n-1})(\lambda(\gamma_n)-E_{2n})}}{E_{2n-1}-\lambda_n} \right]-
$$
$$
-\frac{\lambda_n}{2\pi i (\mu^{(0)}_n)^2}
\sqrt{(\lambda(\gamma_n)-E_{2n-1})(\lambda(\gamma_n)-E_{2n})} +O(\epsilon^2),
$$
$$
A_{2\,(E_{2n-1})}(\gamma_n)=\frac{1}{4\pi i \lambda_n}\sqrt{(\lambda(\gamma_n)-E_{2n-1})(\lambda(\gamma_n)-E_{2n})} +O(\epsilon^2).
$$
Finally we obtain:
\begin{equation}
\label{eq:abel1}
A_{1\, (E_{2n-1})}(\gamma_n)=\frac{1}{2\pi i}
\log\left[\frac{\alpha_n}{2\lambda_n (E_{2n-1}-\lambda_n) }
\right] -\frac{1}{8\pi i (\mu^{(0)}_n)^2} \left[\alpha_n-\beta_n\right]+\ldots=
\end{equation}
$$
\frac{1}{2\pi i}
\log\left[-\sqrt{\frac{\alpha_n}{\beta_n}}
\right] -\frac{1}{8\pi i (\mu^{(0)}_n)^2} \left[\alpha_n-\beta_n\right]+\ldots,
$$
\begin{equation}
\label{eq:abel2}
A_{2\,(E_{2n-1})}(\gamma_n)=\frac{1}{16 \pi i\lambda_n^2}\left[\alpha_n-\beta_n \right] +O(\epsilon^2).
\end{equation}

Analogously, near the point $-\lambda_n$ we have:
$$
\omega_1=\frac{\mu^{(0)}_n}{4\pi i \lambda_n^2}dp(1+O(\epsilon))\sim -\frac{1}{4\pi i\lambda_n}d\sqrt{(\lambda-\overline{E_{2n-1}})
(\lambda-\overline{E_{2n}})},
$$
$$
\omega_2 =\frac{\mu^{(0)}_n}{2\pi i}\frac{\lambda-r_1}{\mu}d\lambda=
$$
$$
=\frac{1}{2\pi i}\frac{d\lambda}{\sqrt{(\lambda-\overline{E_{2n-1}})(\lambda-\overline{E_{2n}})} }+
\frac{1}{2\pi i}\frac{\lambda_n}{(\mu^{(0)}_n)^2}
d\sqrt{(\lambda-\overline{E_{2n-1}})(\lambda-\overline{E_{2n}})} + O(\epsilon^2)+O((\lambda+\lambda_n)^2).
$$
Therefore:
$$
\int \omega_1=-\frac{1}{4\pi i\lambda_n}\sqrt{(\lambda-\overline{E_{2n-1}})(\lambda-\overline{E_{2n}})} + O(\epsilon^2),
$$
$$
\int \omega_2= \frac{1}{2\pi i}\log{\left[(\lambda+\lambda_n)+
\sqrt{(\lambda-\overline{E_{2n-1}})(\lambda-\overline{E_{2n}})}\right]} +\frac{1}{2\pi i}\frac{\lambda_n}{(\mu^{(0)}_n)^2}
\sqrt{(\lambda-\overline{E_{2n-1}})(\lambda-\overline{E_{2n}})}   + \ldots 
$$
$$
A_{1\,(\overline{E_{2n-1}})}(\gamma_{-n})=-\frac{1}{4\pi i \lambda_n}\sqrt{(\lambda(\gamma_{-n})-\overline{E_{2n-1}})(\lambda(\gamma_{-n})-\overline{E_{2n}})} +O(\epsilon^2),
$$
$$
A_{2\,(\overline{E_{2n-1}})}(\gamma_{-n})=\frac{1}{2\pi i}\log\left[\frac{(\lambda(\gamma_{-n})+\lambda_n)+
\sqrt{(\lambda(\gamma_{-n})-\overline{E_{2n-1}})(\lambda(\gamma_{-n})-\overline{E_{2n}})}}{\overline{E_{2n-1}}+\lambda_n} \right]+ 
$$
$$
+\frac{\lambda_n}{2\pi i (\mu^{(0)}_n)^2}
\sqrt{(\lambda(\gamma_{-n})-\overline{E_{2n-1}})(\lambda(\gamma_{-n})-\overline{E_{2n}})} +\ldots
$$
But
$$
\lambda(\gamma_{-n})+\lambda_n =- \frac{1}{4\lambda_n}\left[\tilde\alpha_n+\tilde\beta_n\right]+\ldots
$$
$$
\sqrt{(\lambda(\gamma_{-n})-\overline{E_{2n-1}})(\lambda(\gamma_{-n})-\overline{E_{2n}})}=-\frac{1}{4\lambda_n}
\left[\tilde\alpha_n-\tilde\beta_n \right] +O(\epsilon^2);
$$
therefore
\begin{equation}
\label{eq:abel3}
A_{1\,(\overline{E_{2n-1}})}(\gamma_{-n})=\frac{1}{16 \pi i\lambda_n^2}
\left[\tilde\alpha_n-\tilde\beta_n \right]
+O(\epsilon^2),
\end{equation}
\begin{equation}
\label{eq:abel4}
A_{2\,(\overline{E_{2n-1}})}(\gamma_{-n})=\frac{1}{2\pi i}\log\left[\frac{-\tilde\alpha_n }{2\lambda_n(\overline{E_{2n-1}}+\lambda_n)} \right]-\frac{1}{8\pi i (\mu^{(0)}_n)^2} \left[\tilde\alpha_n-\tilde\beta_n \right] +\ldots=
\end{equation}
$$
=\frac{1}{2\pi i}\log\left[-\sqrt{\frac{\tilde\alpha_n}{\tilde\beta_n}} \right]-\frac{1}{8\pi i (\mu^{(0)}_n)^2} \left[\tilde\alpha_n-\tilde\beta_n \right] +\ldots.
$$
\section{Theta-functional formulas for wave function and potentials}
\label{sec:sec6}

Let us recall the basic formulas from the finite-gap NLS theory.

We use the following normalization for the Riemann theta-functions:
\begin{equation}
\label{eq:theta1}
\theta(z|B)=\sum\limits_{n_1,n_2}
\exp{\Big[2 \pi i (n_1 z_1 +n_2 z_2) + \pi i (b_{11}n_1^2 + 2 b_{12}n_1 n_2 + b_{22}n_2^2  )  \Big] }.
\end{equation}
Normalization (\ref{eq:theta1}) implies the following periodicity properties:
\begin{align*}
\theta\left(\vec z+ \vec a_1 \right)&=\theta\left(\vec z\right)\\ 
\theta\left(\vec z+ \vec a_2 \right)&=\theta\left(\vec z\right)\\
\theta\left(\vec z+ \vec b_1 \right)&=\theta\left(\vec z\right) \exp\left(-\pi i b_{11}-2\pi i z_1 \right)  \\ 
\theta\left(\vec z+ \vec b_2 \right)&=\theta\left(\vec z\right) \exp\left(-\pi i b_{22}-2\pi i z_2 \right).
\end{align*}

The wave functions are defined by the Its formula \cite{Its2}:
$$
\Psi_1(\gamma,x,t)=\kappa_1(x,t) \exp\left(ix\int\limits_{E_0}^{\gamma}dp + 2it \int\limits_{E_0}^{\gamma}dq  \right)\times
$$
$$
\times\frac{\theta(\vec A(\gamma) -\vec U_1 x - \vec U_2 t -\vec A(\gamma_n)-\vec A(\gamma_{-n}) 
-\vec K)}
{\theta(\vec A(\gamma)-\vec A(\gamma_n)-\vec A(\gamma_{-n})-\vec K)},
$$
$$
\Psi_2(\gamma,x,t)=\kappa_2(x,t) \exp\left(\int\limits_{E_0}^{\gamma}\Omega+ ix\int\limits_{E_0}^{\gamma}dp + 2it \int\limits_{E_0}^{\gamma}dq  \right)\times
$$
$$
\times\frac{\theta(\vec A(\gamma)+\vec Z- \vec U_1 x - \vec U_2 t -\vec A(\gamma_n)-\vec A(\gamma_{-n}) 
-\vec K)}
{\theta(\vec A(\gamma)-\vec A(\gamma_n)-\vec A(\gamma_{-n})-\vec K)}.
$$
Here:
\begin{equation}
\label{eq:u1}
\vec U_1=-\frac{1}{2\pi}\left[\begin{array}{c}\oint_{b_1}dp \\ \oint_{b_2}dp \end{array}\right]= 
\left[\begin{array}{c}-\frac{\mu^{(0)}_n}{\pi} \\ -\frac{\mu^{(0)}_n}{\pi} \end{array}\right] ,
\end{equation}
\begin{equation}
\label{eq:u2}
\vec U_2=-\frac{2}{2\pi}\left[\begin{array}{c}\oint_{b_1}dq \\ \oint_{b_2}dq \end{array}\right]=
\left[\begin{array}{c}-\frac{2\lambda_n\mu^{(0)}_n}{\pi} \\ \frac{2\lambda_n\mu^{(0)}_n}{\pi} \end{array}\right]+ O(\epsilon^2).
\end{equation}

To calculate the normalization functions $\kappa_1(x,t)$, $\kappa_2(x,t)$ we require the large $\gamma$ expansions for $\Psi_1(\gamma,x,t)$,
$\Psi_2(\gamma,x,t)$.

\begin{lemma}
Let $C$ be a positive constant, the distance between $\gamma$ and the interval $[-i,i]$ is greater then $C$ and the
integration path does not intersect the filled area in the picture.
 
\begin{figure}
\centering
\mbox{\epsfxsize=4cm\epsffile{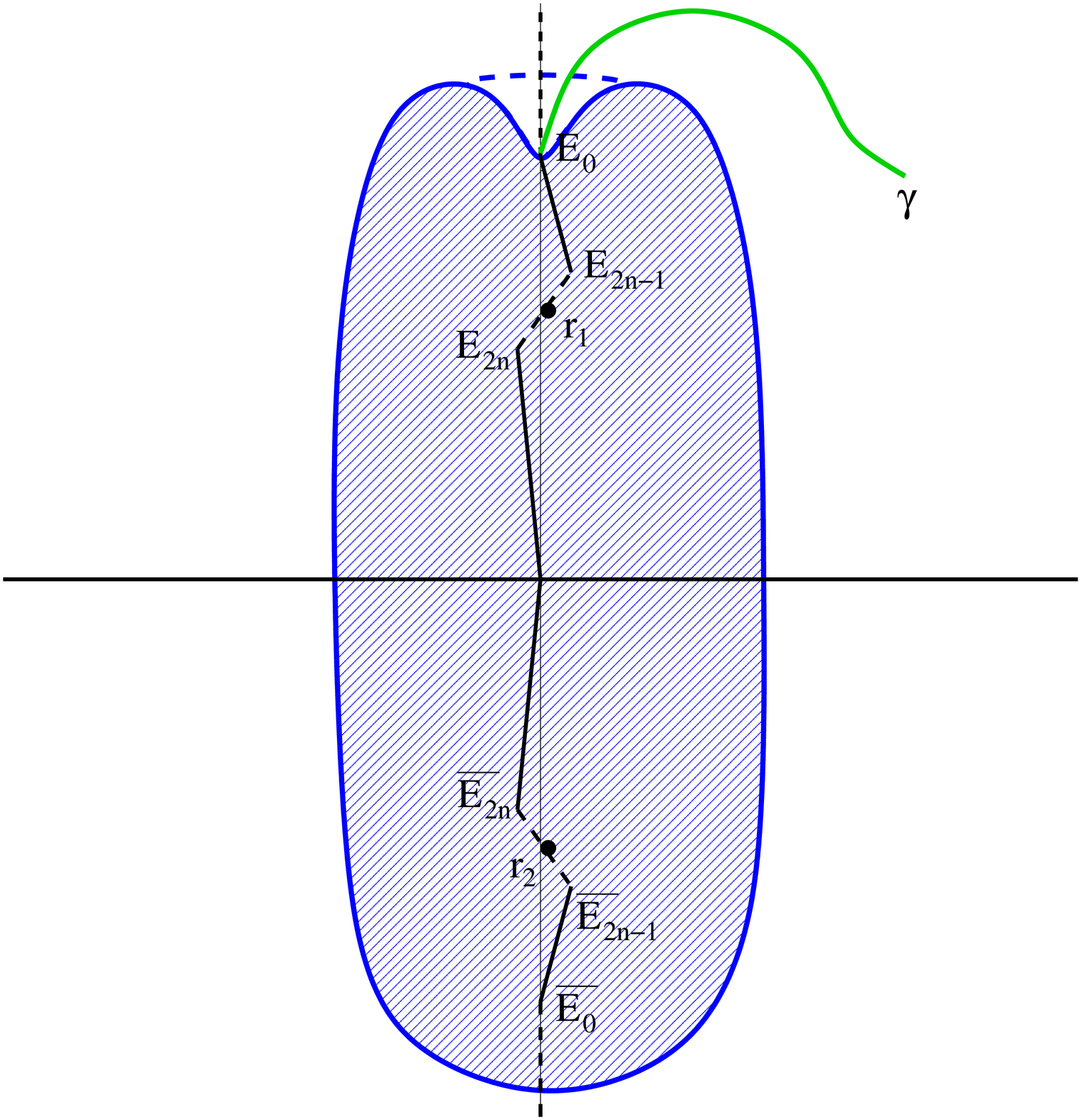}}
\end{figure}

Then
$$
\int\limits_{E_0}^{\gamma}\Omega_{j}-\int\limits_{i}^{\gamma}\Omega^{(0)}_{j}=O(\epsilon^2).
$$
\end{lemma}

Applying (\ref{eq:integr1})-(\ref{eq:integr3}) for large $\gamma$, we obtain:
$$
\Psi_1(\gamma,x,t)=(1+O(\epsilon^2))\cdot\kappa_1(x,t)\cdot \exp\left(i\mu^{(0)} x + 2i\lambda\mu^{(0)} t  \right)\times
$$
$$
\times\frac{\theta(\vec A(\gamma) -\vec U_1 x - \vec U_2 t -\vec A(\gamma_n)-\vec A(\gamma_{-n}) 
-\vec K)}
{\theta(\vec A(\gamma)-\vec A(\gamma_n)-\vec A(\gamma_{-n})-\vec K)},
$$
$$
\Psi_2(\gamma,x,t)=(1+O(\epsilon^2))\cdot\kappa_2(x,t)\cdot\frac{\lambda+\mu^{(0)}}{i}  \cdot \exp\left(i\mu^{(0)} x + 2i\lambda\mu^{(0)} t  \right) \times
$$
$$
\times\frac{\theta(\vec A(\gamma)+\vec Z- \vec U_1 x - \vec U_2 t -\vec A(\gamma_n)-\vec A(\gamma_{-n}) 
-\vec K)}
{\theta(\vec A(\gamma)-\vec A(\gamma_n)-\vec A(\gamma_{-n})-\vec K)}.
$$

Let us recall how the expansions of the wave functions near the infinite points are connected with the potential.

The properly normalized wave function satisfy the following linear spectral problem 
\begin{equation}
\label{eq:lp-x4}
\Psi_x=U\Psi, \ \ 
U=\left [\begin {array}{cc} -i\, \lambda & i\, u(x,t) \\ 
i\, \overline{u(x,t)} & i\, \lambda\end {array}
\right ].
\end{equation}
Near the point $\infty_+$, $\mu\sim-\lambda^{3}$ we have:
$$
\Psi=\left[ \begin{array}{c} 1 + \frac{\alpha^{+}_{1}}{\lambda}+  \frac{\alpha^{+}_{2}}{\lambda^2}+
O\left(\frac{1}{\lambda^3} \right) \\ 
\frac{\beta^{+}_{1}}{\lambda}+ \frac{\beta^{+}_{2}}{\lambda^2} +O\left(\frac{1}{\lambda^3}\right)
 \end{array} \right] e^{-i\lambda x}
$$
Substituting this expansion into (\ref{eq:lp-x4}) we immediately obtain:
\begin{align}
\label{eq:exp1}
&\beta^{+}_{1}=-\frac{1}{2} \bar u ,&\\
&{\alpha^{+}_{1}}'=-\frac{i}{2} u \bar u ,& \nonumber\\
&\beta^{+}_{2}=\frac{i}{4} \bar u'-\frac{1}{2}\bar u  \alpha^{+}_{1} .&\nonumber
\end{align}

Analogously, near the point $\infty_-$, $\mu\sim \lambda^{3}$ we have 
$$
\Psi=\left[ \begin{array}{c} \alpha^{-}_{0}  + \frac{\alpha^{-}_{1}}{\lambda}+  \frac{\alpha^{-}_{2}}{\lambda^2}+
O\left(\frac{1}{\lambda^3} \right) \\ 
\beta^{-}_{-1} \lambda + \beta^{-}_{0} + \frac{\beta^{-}_{1}}{\lambda}+ \frac{\beta^{-}_{2}}{\lambda^2} +
O\left(\frac{1}{\lambda^3}\right) \end{array} \right] e^{i\lambda x},
$$
and
\begin{align}
\label{eq:exp2}
&\partial_x{\beta^{-}_{-1}}= 0 ,&\\
&\alpha^{-}_{0}=\frac{\beta^{-}_{-1}}{2} u ,&\nonumber\\
&{\beta^{-}_{0}}'= \frac{i\beta^{-}_{-1}}{2} u \bar u ,&\nonumber\\
&\alpha^{-}_{1}= \frac{i\beta^{-}_{-1}}{4} u'+\frac{1}{2} u \beta^{-}_{0} .&\nonumber
\end{align}

Moreover, using (\ref{eq:lp-t}), one can easily check that 
$$
\partial_t \beta^{-}_{-1}= 0.
$$

\begin{enumerate}
\item To calculate $\kappa_1(x,t)$, assume that $\gamma\rightarrow\infty_+$:
$$
\mu^{(0)}=-\lambda -\frac{1}{2\lambda}+O\left(\frac{1}{\lambda^2} \right),
$$
$$
\exp\left(i\mu^{(0)} x + 2i\lambda\mu^{(0)} t  \right)=\exp(-i\lambda x -2i\lambda^2 t)\cdot \exp(-i t)\cdot \left(1+ 
O\left(\frac{1}{\lambda}\right)\right)\cdot(1+O(\epsilon^2)),
$$
and
$$
\kappa_1(x,t)=\frac{\theta(\vec A(\infty_+)-\vec A(\gamma_n)-\vec A(\gamma_{-n})-\vec K)}
{\theta(\vec A(\infty_+) -\vec U_1 x - \vec U_2 t -\vec A(\gamma_n)-\vec A(\gamma_{-n}) 
-\vec K)}\cdot \exp(it)\cdot(1+O(\epsilon^2)).
$$
\item Assume now that $\gamma\rightarrow\infty_-$:
$$
\mu^{(0)}=\lambda +\frac{1}{2\lambda}+O\left(\frac{1}{\lambda^2} \right),
$$
$$
\exp\left(i\mu^{(0)} x + 2i\lambda\mu^{(0)} t  \right)=\exp(i\lambda x +2i\lambda^2 t) \exp(i t) \left(1+ 
O\left(\frac{1}{\lambda}\right)\right)(1+O(\epsilon^2))
$$
and
$$
\Psi_1(\gamma,x,t)=(1+O(\epsilon^2)) \frac{\theta(\vec A(\infty_+)-\vec A(\gamma_n)-\vec A(\gamma_{-n})-\vec K)}
{\theta(\vec A(\infty_+) -\vec U_1 x - \vec U_2 t -\vec A(\gamma_n)-\vec A(\gamma_{-n}) 
-\vec K)}\exp(2it)\times 
$$
$$
\times\frac{\theta(\vec A(\infty_-) -\vec U_1 x - \vec U_2 t -\vec A(\gamma_n)-\vec A(\gamma_{-n}) 
-\vec K)}
{\theta(\vec A(\infty_-)-\vec A(\gamma_n)-\vec A(\gamma_{-n})-\vec K)}\exp(i\lambda x +2i\lambda^2 t)\left(1+ 
O\left(\frac{1}{\lambda}\right)\right)
$$
$$
\alpha^{-}_{0}=(1+O(\epsilon^2))\exp(2it)
\frac{\theta(\vec A(\infty_-) -\vec U_1 x - \vec U_2 t -\vec A(\gamma_n)-\vec A(\gamma_{-n})-\vec K)}
{\theta(\vec A(\infty_+) -\vec U_1 x - \vec U_2 t -\vec A(\gamma_n)-\vec A(\gamma_{-n})-\vec K)}\times
$$
$$
\times\frac{\theta(\vec A(\infty_+)-\vec A(\gamma_n)-\vec A(\gamma_{-n})-\vec K)}
{\theta(\vec A(\infty_-)-\vec A(\gamma_n)-\vec A(\gamma_{-n})-\vec K)}.
$$
\item
Expanding $\Psi_2(x,t)$ near $\infty_-$, we obtain:
$$
\beta^{-}_{-1}(x,t)= (1+O(\epsilon^2))\kappa_2(x,t)\frac{2}{i}e^{it}
\frac{\theta(\vec A(\infty_-)+\vec Z- \vec U_1 x - \vec U_2 t -\vec A(\gamma_n)-\vec A(\gamma_{-n}) 
-\vec K)}
{\theta(\vec A(\infty_-)-\vec A(\gamma_n)-\vec A(\gamma_{-n})-\vec K)}.
$$
But 
$$
\beta^{-}_{-1}(x,t)=\beta^{-}_{-1}(0,0)=\frac{2}{u(0,0)};
$$
therefore
$$
\kappa_2(x,t)=(1+O(\epsilon^2))\cdot\frac{ie^{-it}}{u(0,0)} 
\frac{\theta(\vec A(\infty_-)-\vec A(\gamma_n)-\vec A(\gamma_{-n})-\vec K)}
{\theta(\vec A(\infty_-)+\vec Z- \vec U_1 x - \vec U_2 t -\vec A(\gamma_n)-\vec A(\gamma_{-n}) -\vec K)}.
$$
\item
Expanding $\Psi_2(x,t)$ near $\infty_+$, we obtain:
$$
\beta^{+}_{1}(x,t)=-(1+O(\epsilon^2))\frac{2e^{-2it}}{u(0,0)}\frac{\theta(\vec A(\infty_-)-\vec A(\gamma_n)-\vec A(\gamma_{-n})-\vec K)}
{\theta(\vec A(\infty_-)+\vec Z- \vec U_1 x - \vec U_2 t -\vec A(\gamma_n)-\vec A(\gamma_{-n}) -\vec K)}\times  
$$
$$
\times\frac{\theta(\vec A(\infty_+)+\vec Z -\vec U_1 x - \vec U_2 t -\vec A(\gamma_n)-\vec A(\gamma_{-n}) 
-\vec K)}
{\theta(\vec A(\infty_+)-\vec A(\gamma_n)-\vec A(\gamma_{-n})-\vec K)}.
$$
\end{enumerate}
Finally, using (\ref{eq:exp1}), (\ref{eq:exp2}), we obtain:
\begin{equation}
\label{eq:pot1}
u(x,t)=(1+O(\epsilon^2))\exp(2it)
\frac{\theta(\vec A(\infty_-) -\vec U_1 x - \vec U_2 t -\vec A(\gamma_n)-\vec A(\gamma_{-n})-\vec K)}
{\theta(\vec A(\infty_+) -\vec U_1 x - \vec U_2 t -\vec A(\gamma_n)-\vec A(\gamma_{-n})-\vec K)}\times
\end{equation}
$$
\times\frac{\theta(\vec A(\infty_+)-\vec A(\gamma_n)-\vec A(\gamma_{-n})-\vec K)}
{\theta(\vec A(\infty_-)-\vec A(\gamma_n)-\vec A(\gamma_{-n})-\vec K)}  u(0,0)=
$$
$$
=(1+O(\epsilon^2))\exp(2it)\cdot u(0,0)
\frac{\theta(\vec A(\infty_-) -\vec U_1 x - \vec U_2 t -\vec A(\gamma_n)-\vec A(\gamma_{-n})-\vec K)}
{\theta(\vec A(\infty_-)-\vec Z -\vec U_1 x - \vec U_2 t -\vec A(\gamma_n)-\vec A(\gamma_{-n})-\vec K)}\times
$$
$$
\times\frac{\theta(\vec A(\infty_-)-\vec Z-\vec A(\gamma_n)-\vec A(\gamma_{-n})-\vec K)}
{\theta(\vec A(\infty_-)-\vec A(\gamma_n)-\vec A(\gamma_{-n})-\vec K)},
$$
\begin{equation}
\label{eq:pot2}
\overline{u(x,t)}=(1+O(\epsilon^2))\frac{\exp(-2it)}{u(0,0)}
\frac{\theta(\vec A(\infty_+)+\vec Z -\vec U_1 x - \vec U_2 t -\vec A(\gamma_n)-\vec A(\gamma_{-n})-\vec K)}
{\theta(\vec A(\infty_-)+\vec Z -\vec U_1 x - \vec U_2 t -\vec A(\gamma_n)-\vec A(\gamma_{-n})-\vec K)}\times
\end{equation}
$$
\times\frac{\theta(\vec A(\infty_-)-\vec A(\gamma_n)-\vec A(\gamma_{-n})-\vec K)}
{\theta(\vec A(\infty_+)-\vec A(\gamma_n)-\vec A(\gamma_{-n})-\vec K)}=
$$
$$
=(1+O(\epsilon^2))\frac{\exp(-2it)}{u(0,0)}
\frac{\theta(\vec A(\infty_-)-\vec U_1 x - \vec U_2 t -\vec A(\gamma_n)-\vec A(\gamma_{-n})-\vec K)}
{\theta(\vec A(\infty_-)+\vec Z -\vec U_1 x - \vec U_2 t -\vec A(\gamma_n)-\vec A(\gamma_{-n})-\vec K)}\times
$$
$$
\times\frac{\theta(\vec A(\infty_-)-\vec A(\gamma_n)-\vec A(\gamma_{-n})-\vec K)}
{\theta(\vec A(\infty_-)-\vec Z -\vec A(\gamma_n)-\vec A(\gamma_{-n})-\vec K)}.
$$

These solutions are expressed in terms of the genus 2 $\theta$-functions. If the gaps are horizontal ($E_{2n}-E_{2n-1}$ is pure real), 
then these solutions can be expressed in terms of elliptic functions \cite{Smirnov1}.

\section{The solution of the Cauchy problem in terms of elementary functions}
\label{sec:sec7}

We want to evaluate the $\theta$-function representation (\ref{eq:pot1}) of the solution of our Cauchy problem
\beq\label{u_theta}
\ba{l}
u(x,t)=e^{2it}u(0,0)~\frac{\theta({Z}^{+}(0,0),B)}{\theta({Z}^{-}(0,0),B)}~\frac{\theta({Z}^{-}(x,t),B)}{\theta({Z}^{+}(x,t),B)}(1+O(\eps^2)), \\
\ \ \\
u(0,0)=1+c_n +c_{-n}, \ \ |c_n|,|c_{-n}|=O(\eps),
\ea
\eeq
where
\beq\label{def_Z}
{Z}^{\pm}(x,t)=\vec{A}(\infty_{\pm})-\vec{U_1}x-\vec{U_2}t-\vec{A}(\gamma_1)-\vec{A}(\gamma_2), \ \ {Z}^{\pm}=\left(
\ba{c}
z_1^{\pm} \\
z_2^{\pm}
\ea
\right).
\eeq
We recall that the components $b_{ij}$ of the B matrix are expanded as follows 
\beq\label{b_asympt}
\ba{l}
\pi i b_{11}\sim \log(\frac{\sqrt{\alpha_n \beta_n}}{k_n \sigma_n}), \ \ 
\pi i b_{22}\sim \log(\frac{\sqrt{\tilde\alpha_n\tilde\beta_n}}{k_n \sigma_n}), \\ 2\pi i b_{12}\sim \log(\sin\phi_n)^2 ,
\ea
\eeq
where
\beq\label{def_alpha}
\ba{l}
\alpha_n=\overline{c_n}-e^{2i\phi_n}c_{-n}=O(\eps), \\ \beta_n=\overline{c_{-n}}-e^{-2i\phi_n}c_{n}=O(\eps), \\
\tilde\alpha_n=\overline{c_n}-e^{-2i\phi_n}c_{-n}=-e^{-2i\phi_n}\overline{\beta_n}=O(\eps), \\
\tilde\beta_n=\overline{c_{-n}}-e^{2i\phi_n}c_{n}=-e^{2i\phi_n}\overline{\alpha_n}=O(\eps), 
\ea
\eeq
so that 
\beq\label{exp_b}
\ba{l}
e^{\pi i b_{11}}\sim \frac{\sqrt{\alpha_n\beta_n}}{k_n \sigma_n}=O(\eps), \ \ e^{\pi i b_{22}}\sim \frac{\sqrt{\tilde\alpha_n\tilde\beta_n}}{k_n \sigma_n}=O(\eps) , \\ e^{2\pi i b_{12}}\sim (\sin\phi_n)^2 
\ea
\eeq
and 
\beq
\ba{l}
e^{2\pi i z_1^{\pm}(x,t)}\sim \mp i\sqrt{\frac{\beta_n}{\alpha_n}}e^{\varphi_-(x,t)\pm i\phi_n}, \ \
e^{2\pi i z_2^{\pm}(x,t)}\sim \pm i\sqrt{\frac{\tilde\beta_n}{\tilde\alpha_n}}e^{\varphi_+(x,t)\mp i\phi_n}, 
\ea
\eeq
where
\beq
\varphi_{\pm}(x,t)=i k_n x \pm \sigma_n t ,
\eeq
and $k_n=2\cos\phi_n$, $\sigma_n=2\sin(2\phi_n)$.
\vskip 5pt
If $0\le t\le O(1)$, we have that $exp(2\pi i z_1^{\pm}(x,t)), exp(2\pi i z_2^{\pm}(x,t))=O(1)$; therefore
\beq
e^{\pm 2\pi i z_j^{\pm}(x,t)+n \pi i b_{kk}}=O(\eps^n), \ \ j=1,2, \ k=1,2 .
\eeq
It follows that the $\theta$ functions appearing in (\ref{u_theta}) acquire the contribution $1$ from the lattice point $(0,0)$ and $O(\eps)$ contributions from the lattice points $(-1,0)$,$(1,0)$,$(0,1)$,$(0,-1)$; while the contributions of the other lattice points are $O(\eps^2)$ and can be neglected:
\beq
\ba{l}
\theta({Z}^{\pm}(x,t),B)=  1+\Delta^{\pm}(x,t)+O(\eps^2|\log\eps |) \\
\Delta^{\pm}(x,t)=(e^{2\pi i z_1^{\pm}}+e^{-2\pi i z_1^{\pm}})e^{\pi i b_{11}}+ (e^{2\pi i z_2^{\pm}}+e^{-2\pi i z_2^{\pm}})e^{\pi i b_{22}}\sim \\
\pm\frac{i}{k_1\sigma_1}\left(\beta e^{\varphi_-(x,t)\pm i\phi}-\alpha e^{-\varphi_-(x,t)\mp i\phi}-\tilde\beta e^{\varphi_+(x,t)\mp i\phi}+\tilde\alpha e^{-\varphi_+(x,t)\pm i\phi} \right).
\ea
\eeq
Therefore
\beq\label{u_nonconst}
\ba{l}
\frac{\theta({Z}^{-}(x,t),B)}{\theta({Z}^{+}(x,t),B)}\sim \frac{1+\Delta^{-}(x,t)}{1+\Delta^{+}(x,t)}\sim 1+\Delta^{-}(x,t)-\Delta^{+}(x,t) \\
= 1+\frac{i}{\sigma_1}\Big[e^{\sigma_1 t}(\tilde\beta e^{i k_1 x}+\alpha e^{-i k_1 x})- e^{-\sigma_1 t}(\beta e^{i k_1 x}+\tilde\alpha e^{-i k_1 x} ) \Big]+O(\eps^2|\log\eps |).
\ea
\eeq 
Consequently:
\beq
\ba{l}
\frac{\theta({Z}^{+}(0,0),B)}{\theta({Z}^{-}(0,0),B)}=1-\frac{i}{\sigma_1}\Big[\tilde\beta +\alpha - \beta +\tilde\alpha \Big]+O(\eps^2|\log\eps |)= \\
1-(c_1 +c_{-1})+O(\eps^2|\log\eps |),
\ea
\eeq 
and 
\beq\label{u_const}
u(0,0)\frac{\theta({Z}^{+}(0,0),B)}{\theta({Z}^{-}(0,0),B)}=1+O(\eps^2|\log\eps |).
\eeq
Using (\ref{u_nonconst}) and (\ref{u_const}), we conclude that 
\beq\label{u_lin1}
\ba{l}
u(x,t)=e^{2it}\Big\{1+\frac{1}{\sigma_n}\Big[ 
e^{\sigma_n t}(\tilde\beta_n e^{i k_n x}+\alpha_n e^{-ik_n x})-e^{-\sigma_n t}(\beta_n e^{ik_n x}+\tilde\alpha_n e^{-ik_n x})\Big] \Big\}\\ +O(\eps^2|\log\eps |)=e^{2it}\Big\{1+\frac{2}{\sigma_n}\Big[|\alpha_n |\cos\Big(k_n  (x-X^+_n)\Big)e^{\sigma_n t+i\phi_n}+ \\
|\beta_n|\cos\Big(k_n (x-X^-_n)\Big)e^{-\sigma_n t-i\phi_n}  \Big] \Big\}+O(\eps^2|\log\eps |),
\ea
\eeq
where
\beq\label{def_X}
X^{+}_n=\frac{\arg(\alpha_n)-\phi_n+\pi/2}{k_n}, \ \ X^{-}_n=\frac{-\arg(\beta_n)-\phi_n+\pi/2}{k_n}.
\eeq

\vskip 10pt

If $t=(\sigma_n)^{-1}O(|\log\eps |)$, the balance changes and the $\theta$-function is $O(1)$ at the four points $(0,0)$, $(0,1)$, $(-1,0)$, $(-1,1)$, and smaller in the remaining points; we are in the first nonlinear stage of modulation instability (the first appearence of the RW). More precisely, if $|t-T_n(|\alpha_n |)|\le O(1)$, where
\beq
T_n(\zeta)=\frac{1}{\sigma_n}\log\left(\frac{(\sigma_n)^2}{2\zeta} \right) ,
\eeq
then 
\beq
\ba{l}
\theta({Z}^{\pm}(x,t),B)\sim 1+e^{-2\pi i z^{\pm}_1+\pi i b_{11}}+e^{2\pi i z^{\pm}_2+\pi i b_{22}}+
e^{2\pi i (z^{\pm}_2-z^{\pm}_1)+\pi i (b_{11}+b_{22}-2b_{12})} \\
\sim 2 e^{\sigma_n (t-T_n(|\alpha_n |))+i\phi_n\mp\phi_n}
\Big(\cosh[\sigma_n (t-T_n(|\alpha_n |))+i\phi_n \mp\phi_n]\mp \\
\sin\phi_n\cos[k_n(x-X^{+}_n)] \Big),
\ea
\eeq
implying that the solution is described, to the leading $O(1)$, by the Akhmediev 1-breather solution:
\beq
u(x,t)=e^{2it+2i\phi_n}\frac{\cosh[\sigma_n (t-T_n(|\alpha_n |))+2i\phi_n ]+\sin(\phi_n) \cos[k_n(x-X^{+}_n)]}{\cosh[\sigma_n (t-T_n(|\alpha_n |))]-\sin(\phi_n) \cos[k_n(x-X^{+}_n)]}.
\eeq

\vskip 10pt

The above two balances are obtained starting from the five-point scheme centered at the origin and moving upward on the secondary
diagonal of the 2-D parameter lattice. Therefore the next balance involves five points centered at $(-1,1)$, corresponding to
the shift of the original summation area along the vector $(-1,1)$. It means that we come back to the same point of the 
Liouville torus and, since $g=2$, the potential $u(x,t)$ is periodic up to a phase multiplier:
$$
u(x+\Delta_x,t+T_p)=exp(i\rho) u(x,t).
$$    

In order to compute the period $T_p$, the $x$-phase shift $\Delta_x$ and the phase increment $\rho$ of  
the solution, we use the finite-gap formula (\ref{u_theta}). 

We return to the starting point of the phase shift if the argument shift in the theta-functions coincide with a period:
\begin{equation}
\label{eq:per2}
-\vec U_1 \Delta_x - \vec U_2 T_p = \left(\begin{array}{c} b_{11} \\ b_{12} \end{array}\right) n_1 + 
\left( \begin{array}{c} b_{21} \\ b_{22} \end{array} \right) n_2 + \left(\begin{array}{c} 1 \\ 0 \end{array}\right) m_1 + 
\left(\begin{array}{c} 0 \\ 1 \end{array}\right) m_2.
\end{equation}
$n_1,n_2,m_1,m_2\in\ZZ$.

Taking into account that
$$
\vec U_1=\left[\begin{array}{c}-\frac{k_n}{2\pi} \\ -\frac{k_n}{2\pi} \end{array}\right], \ \
\vec U_2= \left[\begin{array}{c}\frac{-i\sigma_n}{2\pi} \\ \frac{i\sigma_n}{2\pi} \end{array}\right]+ O(\epsilon^2).
$$
we see that the first repetition corresponds to $n_1=1$, $n_2=-1$ as expected. Separating the real and the imaginary parts of (\ref{eq:per2}), we obtain, modulo $\epsilon^2$ corrections:
\begin{equation}
\label{eq:per3}
\left[\begin{array}{c}\frac{\sigma_n}{2\pi} \\ -\frac{\sigma_n}{2\pi} \end{array}\right] T_p = 
\left[\begin{array}{c} \Im(b_{11}-b_{21}) \\ -\Im (b_{22}-b_{12})\end{array} \right],
\end{equation}
\begin{equation}
\label{eq:per4}
\left[\begin{array}{c}\frac{k_n}{2\pi} \\ \frac{k_n}{2\pi} \end{array}\right] \Delta_x = 
\left[\begin{array}{c} \Re(b_{11}-b_{21}) \\ -\Re (b_{22}-b_{12}) \end{array}\right]=
\left[\begin{array}{c} \Re(b_{11}) \\ -\Re (b_{22}) \end{array}\right] 
\end{equation}
(we used here the fact that, in the leading approximation, $b_{21}=b_{21}$ are pure imaginary).
From formulas (\ref{eq:riemann1}), (\ref{eq:riemann2})  we obtain, to the leading order:
$$
b_{11}-b_{12}=\frac{1}{2\pi i}\log\left[\frac{4\alpha_n\beta_n}{\sigma_n^4} \right]=\frac{i}{2\pi}
\log\left[\frac{\sigma_n^4}{4|\alpha_n\beta_n|} \right]+\frac{\arg(\alpha_n\beta_n)}{2\pi}.
$$ 
Therefore
$$
T_p=\frac{1}{\sigma_n} \log\left[\frac{\sigma_n^4}{4|\alpha_n\beta_n|} \right],
$$
$$
\Delta_x=\frac{\arg(\alpha_n\beta_n)}{k_n}.
$$
The theta-functions have the following periodicity property:
$$
\theta(\vec z+ \vec b_1 -\vec b_2) = \exp(-\pi i(b_{11}+b_{22}-2b_{12}))\exp(-2\pi i (z_1-z_2))\theta(\vec z);
$$
therefore
$$
\frac{\theta({Z}^{-}(\Delta_x,T_p),B)}{\theta({Z}^{+}(\Delta_x,T_p),B)}=
$$
$$
=\exp\left(2\pi i \big[
({Z}^{+}(0,0)- {Z}^{-}(0,0))_1 -({Z}^{+}(0,0)- {Z}^{-}(0,0))_2\big] \right)
\frac{\theta({Z}^{-}(0,0),B)}{\theta({Z}^{+}(0,0),B)}=
$$
$$
=\exp\left(2\pi i \big[
(\vec A(\infty_{+})-\vec A(\infty_{-}))_1 -(\vec A(\infty_{+})-\vec A(\infty_{-}))_2\big] \right)
\frac{\theta({Z}^{-}(0,0),B)}{\theta({Z}^{+}(0,0),B)}=
$$
$$
=\exp\left(2\pi i \left[\left(\frac{\phi_n}{\pi}+\frac{1}{2}\right) -\left(-\frac{\phi_n}{\pi}+\frac{1}{2}\right) \right] \right)
\frac{\theta({Z}^{-}(0,0),B)}{\theta({Z}^{+}(0,0),B)}=
$$
$$
=\exp\left(4 i \phi_n\right)
\frac{\theta({Z}^{-}(0,0),B)}{\theta({Z}^{+}(0,0),B)}.
$$
We conclude that $\rho=2 T_p +4\phi_n$.

\section{Acknowledgments} Two visits of P. G. Grinevich to Roma were supported by the University of Roma ``La Sapienza''. P. G. Grinevich and P. M. Santini acknowledge the warm hospitality and the local support of CIC, Cuernavaca, Mexico, in December 2016. P.G. Grinevich was also partially supported by RFBR grant 17-51-150001.

We are grateful to M. Sommacal for introducing us to the Split Step Fourier Method and for showing us his personalized MatLab code. We acknowledge useful discussions with F. Briscese, F. Calogero, C. Conti, E. DelRe, A. Degasperis, A. Gelash, I. Krichever, A. Its, S. Lombardo, A. Mikhailov, D. Pierangeli, M. Sommacal and V. Zakharov.

\end{document}